\documentclass[
 pra,
 amsmath,amssymb,
 aps,
 twocolumn,
 notitlepage,
 superscriptaddress
]{revtex4-2}

\usepackage{float}
\makeatletter
\let\newfloat\newfloat@ltx
\makeatother

\usepackage{algorithm}
\usepackage{algpseudocode}

\usepackage[colorlinks=true]{hyperref}

\usepackage{ragged2e}
\usepackage{caption}       
\usepackage{subcaption}
\DeclareCaptionJustification{justified}{\justifying}
\usepackage[capitalise]{cleveref}

\usepackage{styles}
\usepackage{fancyhdr}
\usepackage{txfonts}
\usepackage{xurl} 
\usepackage{bm} 
\usepackage{physics} 
\usepackage{tikz} 
\usetikzlibrary{quantikz} 
\usepackage{tikzscale}
\usepackage[shortlabels]{enumitem}

\usepackage{bm} 

\begin{document}

\title{Quantum Algorithm Exploration using Application-Oriented Performance Benchmarks}

\author{Thomas Lubinski}
\affiliation{Quantum Circuits Inc, 25 Science Park, New Haven, CT 06511}
\affiliation{QED-C Technical Advisory Committee on Standards and Performance Benchmarks}

\author{Joshua J. Goings}
\affiliation{IonQ Inc, 4505 Campus Dr, College Park, MD 20740, USA}

\author{Karl Mayer}
\affiliation{Quantinuum, 303 S. Technology Ct, Broomfield, CO 80021, USA}

\author{Sonika Johri}
\affiliation{Coherent Computing, Cupertino, CA, USA}

\author{Nithin Reddy}
\affiliation{San Jose State University, San Jose, USA}

\author{Aman Mehta}
\affiliation{University of California at Los Angeles, USA}

\author{Niranjan Bhatia}
\affiliation{University of California, Berkeley, USA}

\author{Sonny Rappaport}
\affiliation{IonQ Inc, 4505 Campus Dr, College Park, MD 20740, USA}

\author{Daniel Mills}
\affiliation{Quantinuum, Terrington House, 13-15 Hills Road, Cambridge CB2 1NL, UK}

\author{Charles H. Baldwin}
\affiliation{Quantinuum, 303 S. Technology Ct, Broomfield, CO 80021, USA}

\author{Luning Zhao}
\affiliation{IonQ Inc, 4505 Campus Dr, College Park, MD 20740, USA}

\author{Aaron Barbosa}
\affiliation{Q-CTRL, Los Angeles, CA USA}

\author{Smarak Maity}
\affiliation{Q-CTRL, Los Angeles, CA USA}

\author{Pranav S. Mundada}
\affiliation{Q-CTRL, Los Angeles, CA USA}

\collaboration{Quantum Economic Development Consortium (QED-C) collaboration} 

\thanks{This work was sponsored by the Quantum Economic Development Consortium (QED-C) and was performed under the auspices of the QED-C Technical Advisory Committee on Standards and Performance Benchmarks. The authors acknowledge many committee members for their input to and feedback on the project and this manuscript.}

\date{\rule[15pt]{0pt}{0pt}\today}
             
\begin{abstract}

\vspace{0.0cm}
The QED-C suite of Application-Oriented Benchmarks provides the ability to gauge performance characteristics of quantum computers as applied to real-world applications. Its benchmark programs sweep over a range of problem sizes and inputs, capturing key performance metrics related to the quality of results, total time of execution, and quantum gate resources consumed.
The work described in this manuscript investigates challenges in broadening the relevance of this benchmarking methodology to applications of greater complexity.
First, we introduce a method for improving landscape coverage by varying algorithm parameters systematically, exemplifying this functionality in a new scalable HHL linear equation solver benchmark. 
Second, we add a VQE implementation of a Hydrogen Lattice simulation to the QED-C suite, and introduce a methodology for analyzing the result quality and run-time cost trade-off. We observe a decrease in accuracy with increased number of qubits, but only a mild increase in the execution time.
Third, unique characteristics of a supervised machine-learning classification application are explored as a benchmark to gauge the extensibility of the framework to new classes of application. Applying this to a binary classification problem revealed the increase in training time required for larger anzatz circuits, and the significant classical overhead.
Fourth, we add methods to include optimization and error mitigation in the benchmarking workflow which allows us to: identify a favourable trade off between approximate gate synthesis and gate noise; observe the benefits of measurement error mitigation and a form of deterministic error mitigation algorithm; and to contrast the improvement with the resulting time overhead.
Looking to the future, we discuss how the benchmark framework can be instrumental in facilitating the exploration of algorithmic options and their impact on performance.

\end{abstract}


\maketitle

\tableofcontents


\pagestyle{fancy}

\renewcommand{\headrulewidth}{0.0pt}
\lhead{}
\rhead{\thepage}

\renewcommand{\footrulewidth}{0.4pt}
\cfoot{}
\lfoot{Quantum Algorithm Exploration using Application-Oriented Performance Benchmarks}
\rfoot{\today}
\vspace{2cm}


\section{Introduction}
\label{sec:introduction}

Quantum computing is in the early stages of its introduction to a broad community of potential users in multiple application areas, with new technologies and supporting software emerging regularly.
For both providers and users to obtain optimal results, it is important they be able to gauge the evolution of these systems and evaluate available options.
In this context, a variety of methodologies for benchmarking performance and characterizing improvements have become available to the community.

Performance benchmarking tools span various levels of the computing stack. 
At the component level, protocols such as randomized benchmarking~\cite{PhysRevA.77.012307, PhysRevLett.106.180504}, and gate set tomography~\cite{Blume-Kohout2017-no}, are used to characterize individual components within the system.  
System-level benchmarks, such as quantum volume~\cite{Cross_2019}, cross-entropy benchmarking~\cite{Boixo_2018}, mirror circuit benchmarks~\cite{proctor2020measuring}, and CLOPS (Circuit-Layer Operations per Second)~\cite{wack_clops_2021}, provide a single metric measure of quality or speed of execution of quantum programs.
In combination, these two classes of benchmarks complement each other. The former precisely measures the performance of key elements, while the latter aggregates system behavior, providing a summary view.

At the algorithm or application level, a number of benchmarking suites have emerged that provide measures in several dimensions of the performance experienced by users when running complete programs.
These attempt to take into account all factors contributing to the overall quality and run-time efficiency of a quantum computing system. Factors include qubit noise characteristics, program preparation and execution time, classical processing of results, and data transfer to and from the system~\cite{Quetschlich2023mqtbench,tomesh2022supermarq,Mesman2022,Donkers2022,Finzgar_2022,kiwit2023applicationoriented,lubinski2023_10061574,lubinski2023optimization}.
Such benchmarks are valuable for users as they provide measures that may be closer to their actual experience and in the context of recognizable application scenarios.

One of these, the QED-C suite of Application-Oriented Performance Benchmarks for Quantum Computing~\cite{lubinski2023_10061574,lubinski2023optimization}, takes an approach similar to that of the SPEC benchmarks for classical computers~\cite{spec_org, hennessy_patterson_2019_all}, providing algorithms and simple applications structured as benchmarks that sweep over a range of problem sizes and complexity.  
This enables a broad characterization of the overall system performance to be expected for many classes of application, in the dimensions of quality and execution run-time, across different target quantum computing systems.

\vspace{0.3cm}

However, creating benchmarks from applications comes with several challenges. One requirement is to select methods that scalably quantify performance. For example, in the QED-C suite, most applications used simple problem instances producing computational basis states to ease testing. Moreover, quantum algorithms come in many variants, and benchmark results can be impacted dramatically by the particular implementation and execution options. For example, there are multiple options for the ansatz used in a VQE algorithm~\cite{peruzzoVariationalEigenvalueSolver2014}, the number of rounds (repetitions of a group of gates, resulting in deeper circuit) in a QAOA algorithm~\cite{farhi2014quantum}, or the type of encoding of data in a machine learning application~\cite{larose2020robust}. Additionally, with variational algorithms, a classical optimizer must be selected along with the freedom to choose hyperparameters, which may impact the time to solution and the quality of results. Another challenge is selecting appropriate datasets for use with a particular algorithm. For example, in HHL, discussed below, some datasets are efficiently verifiable while others are not. 

The choices made in converting an algorithm to a scalable and verifiable benchmark will have consequences on how the benchmark performs on quantum hardware. Many of these choices impact the number of two-qubit gates, which is the dominant source of error in most current generation systems, and therefore strongly correlates with the performance. Evidence of this can be seen in the QED-C suite where decisions made to simplify verification allow for effective compilation that substantially reduces the number of two-qubit gates as discussed below. Additionally, the choice to have some circuits ideally generate computational basis states means results are greatly improved with plurality voting error mitigation techniques~\cite{chen2023benchmarking}, which are known to scale poorly. Both of these options are unique to the benchmark version of the application and are not typical for an actual use case. It is important to have benchmarks that both represent application use cases and run with comparable resource requirements.

\vspace{0.3cm}

In this work, we describe several enhancements that have been made to the QED-C suite.
These were selected to better match users' experience in running algorithms.
Several new benchmark algorithms have been added, specifically to increase the breadth and complexity of the applications included, add coverage where gaps exist in areas of un-tested qubit number and depth, and to illustrate various execution methods and metric collection options.
We also examine the impact of program optimization and error mitigation techniques and review the enhancements made to enable the inclusion of these.

Our benchmark suite provides a platform in which new applications can readily be developed, studied, and measured without the need to build anew the standardized data collection and reporting mechanisms required for each instance.
By varying the algorithm or its hyperparameters, while keeping constant the problem definition, our approach can provide useful information to developers, who can use the results to analyze solution trade-offs in the context of various target environments.

\vspace{0.3cm}

The remainder of this paper is structured as follows.
Background on the fundamentals guiding this work is provided in~\autoref{sec:background}.
A new benchmark, based on a scalable version of the HHL linear equation solver~\cite{Harrow2009}, is introduced in~\autoref{sec:scalable_hhl_solver}. This new benchmark illustrates how variations on the algorithm impact its volumetric profile and extends the coverage of the benchmark suite.
In~\autoref{sec:hydrogen_lattice}, we analyze the run-time costs associated with finding the ground state energy of a hydrogen lattice using a Variational Quantum Eigensolver algorithm along with a methodology for analyzing the trade-off between the quality of the result and the run-time cost to achieve that quality. 
In~\autoref{program_optimization_techniques}, a set of program optimization techniques is described along with a discussion of their operation, the level of performance improvement obtained, the cost associated with the technique, and the types of applications that exhibit the greatest impact.
Lastly, in~\autoref{machine_learning_benchmark}, we extend the iterative benchmarking framework by including a machine learning algorithm, a simple image classification problem, that defines another application-specific measure and requires more circuits iteratively than the VQE benchmark.


\section{Background}
\label{sec:background}

In this section, we review concepts and definitions from prior benchmarking work that we reference throughout the remainder of this manuscript.
We focus primarily on the application-oriented level of performance evaluation, as there is already a large body of reference material on component and system-level benchmarks~\cite{PhysRevA.77.012307, PhysRevLett.106.180504,Blume-Kohout2017-no,gambetta2012characterization,sarovar2019detecting,Proctor_2022}.


\subsection{System-Level Performance Benchmarks}
\label{sec:system_level_benchmarks}

We briefly describe here two well-known system-level performance benchmarks, Quantum Volume (QV)~\cite{Cross_2019,qiskit_measuring_quantum_volume} and Volumetric Benchmarking (VB)~\cite{BlumeKohout2020volumetricframework,proctor2020measuring}.
As a single-number metric of a quantum computer's general capability, QV captures the combined effects of qubit number and result fidelity for gate model computers~\cite{Cross_2019,Baldwin2022reexaminingquantum, Pelofske_2022}.
VB was proposed to address limitations of QV, using a method by which the observed performance of the quantum computing system is plotted in a two-dimensional depth $\times$ width ``volumetric'' grid.
Our benchmarks build upon the VB method, which highlights the ability of a specific device to execute successfully both wide/shallow and deep/narrow circuits, extending the square circuits represented by QV~\cite{proctor2022trust}.

These two methods (QV and VB) characterize quantum circuit execution quality and scale, but neither provides information about the time it takes a program to run, which is a key factor in evaluating the total cost of any computing solution. 
Circuit Layer Operations per Second (CLOPS) was introduced~\cite{wack_clops_2021} as a measure of run-time performance computed by executing a sequence of quantum volume circuits.
While CLOPS characterizes execution speed in a single number metric that includes data transfer, circuit operations, and other latencies, it is not specific to any application. It also does not capture compilation time or the trade-off between quality and run-time, or true time-to-solution, which is crucial to understand when selecting machines for running applications.
    

\subsection{Application-Oriented Benchmarks}
\label{sec:application_oriented_benchmarks}

System-level metrics are valuable as general measures of system capability, but it can be challenging to predict how well a machine with a certain level of general performance would be able to address a specific class of application. 
Application-focused benchmarks execute well-defined programs that aim to directly yield performance metrics specific to an application type.

\begin{figure}[t!]
\includegraphics[width=0.94\columnwidth]{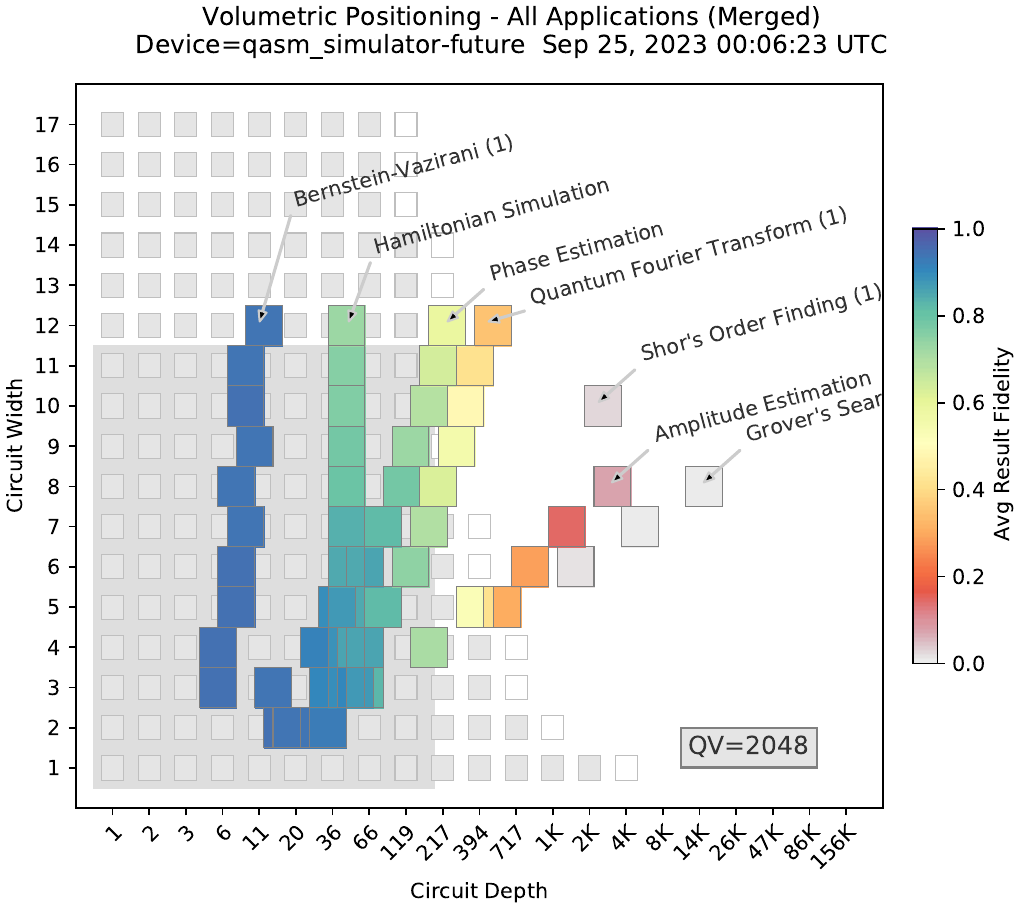}
\caption{\textbf{ A Volumetric Positioning Plot.} Results from executing several of the application-oriented benchmark programs on a classically implemented noisy simulator of a device that exhibits a quantum volume of 2048.
The result quality of the application circuits (shown on a color scale) degrades as the width and depth of the circuits range outside of the quantum volume and volumetric benchmark regions (shown in dark and light grey rectangles). Each application has a unique profile and the success or failure in its execution can be predicted (approximately) from the system-level benchmarks.
}
\label{fig:volumetric_application_profiles}
\end{figure}

The QED-C suite offers an effective methodology to evaluate the performance of a wide variety of quantum programs across a range of quantum hardware and simulator systems~\cite{lubinski2023_10061574,qc-proto-benchmarks,lubinski2023optimization}.
Its benchmark programs are structured to sweep over a range of problem sizes and input characteristics, while systematically capturing key performance metrics, such as quality of result, execution run-time, and resources consumed (i.e. gate count). Supporting infrastructure and abstractions make them accessible to a broad audience of users.
The framework supports benchmarking of individual circuit execution as well as iterative algorithms that repeatedly execute quantum circuits as part of an algorithm such as QAOA~\cite{farhi2014quantum} or VQE~\cite{peruzzoVariationalEigenvalueSolver2014}.

Throughout this manuscript, 
the figure of merit used to represent the quality of result for individual circuits is the ``Normalized Hellinger Fidelity'', a modification of the standard ``Hellinger Fidelity'' that scales the metric to account for the uniform distribution produced by a completely noisy device. This is described in detail in our prior work~\cite{lubinski2023_10061574}.
For iterative algorithms that address problems such as combinatorial optimization, we calculate figures of merit that are application-specific as described in Ref.~\cite{lubinski2023optimization}.
We use the new Hydrogen Lattice and Machine Learning benchmarks to explore application-specific figures of merit derived from functions of observables and how these are impacted by various algorithm options.

\vspace{0.3cm}

\autoref{fig:volumetric_application_profiles} illustrates a visualization technique, referred to as ``volumetric positioning'', that shows the quality of the result obtained from executing each application circuit type, at increasing width and depth, on a classically implemented quantum simulator measured to have a quantum volume of 2048 (depolarizing error 0.05\% for one qubit gates and 0.5\% for two qubit gates, used throughout this manuscript).
Results are plotted over a background defined by the quantum volume of the target machine (dark grey rectangle) and the success region estimated for volumetric circuits (light grey rectangles).
Each application is characterized by its ``profile'', a visual representation of the relationship between circuit depth and the number of qubits used as the problem size grows.
Several of the benchmark programs have multiple versions implemented, as indicated by a number that can follow its name, e.g. Bernstein-Vazirani (1).

A correlation can be seen in~\autoref{fig:volumetric_application_profiles} between the execution fidelity of the application-oriented benchmark circuits at specific qubit widths and circuit depths and the results predicted by QV or VB.
Circuits with fidelity greater than $\sim0.5$ lie within the region defined by the QV of the target system or in the success regions predicted by VB.
This approach helps to validate that results from the application-oriented benchmarks align with those of the system-level benchmarks.

\vspace{0.3cm}

The QED-C benchmark framework incorporates multiple metrics for assessing execution run-time performance.
This is important because total execution time significantly influences the overall number of applications that can be processed within a defined time window on a quantum computer.
Quantum algorithms such as QAOA and VQE are iterative in nature and any run-time overhead is magnified by the number of circuit executions performed.
The framework implements a standard mechanism for the collection of the total elapsed time $t_{\rm elapsed}$ and the quantum execution time on a backend system $t_{\rm quantum}$ (reported differently by each service provider).
In this manuscript, these times are represented in various plots with the labels ``Elapsed'' and ``Quantum''.

Execution time for quantum circuits is closely related to the gate depth of the circuit. In the plots in this manuscript, we use the label ``Algorithmic Depth'' to refer to the number of gate operations in the longest path of the circuit, as defined by a user at the quantum programming API level. The label ``Normalized Depth'' refers to an estimate of the gate depth that is expected after transpiling the program to the gate set of a representative backend system.
For comparison across different backend systems, we approximate the transpiled circuit depth by transpiling circuits with Qiskit at the default optimization level 1, using a default gate set denoted by [`rx', `ry', `rz', `cx'].
More detail about execution time and circuit depth can be found in our prior work~\cite{lubinski2023optimization}.


\section{Benchmarking An HHL Linear Solver}
\label{sec:scalable_hhl_solver}

The Harrow-Hassidim-Lloyd (HHL) algorithm~\cite{Harrow2009,Aaronson2015} is theoretically capable of solving linear systems of equations with exponential speedup over classical solutions (apart from data loading and extraction).
We selected this algorithm as a benchmark for two reasons.
First, the algorithm is constructed from multiple instances of smaller subroutines that we previously configured as benchmarks, together with a particular quantum controlled-rotation implementation.
This combination enables us to reasonably predict where the benchmark profile would lie in the volumetric spectrum.

Secondly, the HHL algorithm employs two independent qubit registers: one for input data, which scales as the size of the problem matrix, and another for performing phase estimation, which scales as the bit precision of the problem matrix elements.
These registers provide a mechanism by which to vary the circuit complexity, impacting the algorithm's profile and extending its coverage across the volumetric spectrum.

In this section, we review the HHL algorithm and our benchmark implementation.
The HHL algorithm has been studied extensively, and there are numerous tutorials and circuit examples available~\cite{Liu_Xie_Liu_Zhao_2022,dervovic2018quantum,morrell2023stepbystep,Cao_2012,Lee2019_HHL_IBMQ}.
However, most of these have been constrained to a fixed number of qubits or a limited set of problem instances, primarily to simplify the implementation's complexity. 
For our benchmark, we developed an implementation that achieves a modest level of scaling, closely resembling one that is described in a recent study~\cite{Martin_2023}.


\subsection{The HHL Algorithm}
\label{subsec:hhl_algorithm}

The HHL algorithm prepares a quantum state that encodes the solution, $x$, for the linear equation shown here:
\begin{equation}
    \label{eq:hhl_linear_equation}
    Ax=b
\end{equation}
The transformation represented by the $N$-by-$N$ matrix $A$ is applied to an unknown $N$-dimensional vector $x$, resulting in the vector $b$.
Finding the value of $x$ is equivalent to determining the inverse of $A$ and applying it to $b$.
The matrix $A$ is assumed to be Hermitian without loss of generality~\cite{Harrow2009}.

When the HHL algorithm is executed, it generates an $n$-qubit quantum state whose amplitudes are proportional to the $N=2^n$ components of the solution vector $x$ in~\autoref{eq:hhl_linear_equation}.
It is important to note that the HHL algorithm does not directly provide the complete solution
vector $x$, since a computational basis measurement
on the output state only returns one random bitstring with a probability given by the square of the amplitude,
rather than the full list of amplitudes.
However, if one is interested in a property of the solution vector, such as the expectation value of an observable in the solution state, the HHL algorithm can be used to extract that information.

Furthermore, there are several caveats that must be satisfied for
the HHL algorithm to provide an exponential 
speedup over classical matrix inversion~\cite{Aaronson2015,Liu_Xie_Liu_Zhao_2022}.
First, the row sparsity of $A$, which is the number of non-zero elements per row, must be at most polynomial in $n$.
Second, the condition number $\kappa$,
which is the ratio between the largest and smallest eigenvalues of $A$, must also be at most polynomial in $n$.

\vspace{0.3cm}

Here, we describe how the HHL algorithm functions.
First, an $n$-qubit ``input'' register is initialized in the state $\ket{b}$, representing the vector $b$.
A Quantum Phase Estimation (QPE) routine is used to encode this input state in the eigenbasis of the matrix $A$.
To realize this, a second group of $n_p$ qubits called the ``phase'' register, is placed into an equal superposition.
(This register is sometimes referred to as the ``clock'' register.)
A Hamiltonian evolution, defined by the matrix $A$ and controlled by the value in the phase register, is then performed on the input register, as in~\autoref{eq:hhl_ham_evolution} where $N_p=2^{n_p}$.
\begin{equation}
    \label{eq:hhl_ham_evolution}
    \frac{1}{\sqrt{N_p}}\sum_{t=0}^{N_p-1}\ket{t}e^{-iAt}\ket{b}
\end{equation}

By decomposing the input state $\ket{b}=\sum_j\beta_j\ket{\mu_j}$ into the eigenbasis of $A$,
the state is then given by~\autoref{eq:hhl_before_iqft}:
\begin{equation}
    \label{eq:hhl_before_iqft}
    \sum_j\beta_j\frac{1}{\sqrt{N_p}}\sum_{t=0}^{N_p-1}\ket{t} e^{-i\lambda_jt}\ket{\mu_j},
\end{equation}
where $\lambda_j$ is the eigenvalue corresponding to $\ket{\mu_j}$.
The next step is an inverse QFT on the phase register, resulting in the state shown in~\autoref{eq:hhl_after_iqft}:
\begin{equation}
\label{eq:hhl_after_iqft}
\sum_{j=0}^{N-1}\beta_j\ket{\lambda_j}\ket{\mu_j}
\end{equation}

To effect a matrix inversion, an ancilla qubit is appended to the circuit and a rotation on the ancilla is performed, conditioned on the phase register.
This results in the state in~\autoref{eq:hhl_after_irot}, in which $\sin{\theta_j/2}=\frac{C}{\lambda_j}$ for constant $C\le\frac{1}{\kappa}$
and $\kappa$ is the condition number of $A$.
\begin{equation}
\label{eq:hhl_after_irot}
\sum_{j=0}^{N-1}\beta_j\ket{\lambda_j}\ket{\mu_j}\big(\cos{\theta_j/2}\ket{0}+\sin{\theta_j/2}\ket{1}\big)
\end{equation}

Following this, an inverse Quantum Phase Estimation (QPE$^{\dagger}$) is applied to the input and phase qubits, and measurement is performed on the qubits, conditioned on a measurement outcome of $\ket{1}$ in the ancilla.
The final state is proportional to the unknown vector $x$, as seen in~\autoref{eq:hhl_final}.
\begin{equation}
\label{eq:hhl_final}
\sum_j\frac{\beta_j}{\lambda_j}\ket{\mu_j}=\ket{x}
\end{equation}

The post-selection step is used to amplify the components of the quantum state that correspond to non-zero elements since the solution to the linear system is encoded in these non-zero amplitudes.


\subsection{The HHL Benchmark Implementation}
\label{subsec:hhl_scalability}

There are many ways that the HHL algorithm can be implemented, with various trade-offs~\cite{Liu_Xie_Liu_Zhao_2022,dervovic2018quantum}.
We outline here the design choices made to ensure a scalable and efficient implementation, tailored for execution as a benchmark on current quantum computing systems.

For scalability, we defined a set of problem instances represented by matrices $A$ and vectors $b$ for a range of phase register qubit numbers $n_p$ and input qubit numbers $n_b$.
The values for these instances are generated such that the matrices are sufficiently sparse and well-conditioned, making it possible to efficiently load the non-zero elements of $A$ onto a quantum register, and to prepare the input state $\ket{b}$.
Each instance is given by a quantum oracle corresponding to a $2$-sparse matrix $A$ with $\kappa\le4$.
The elements of $A$ are chosen as fractions that are exactly expressible with $n_p$ bits of precision.

The Hamiltonian evolution is implemented in the benchmarking code using the algorithm of~\cite{Childs2003} for sparse Hamiltonian simulation based on a quantum walk.
This algorithm makes use of an oracle $V$ acting on
two registers of size $n_b$ according to $V\ket{b_1}\ket{b_2}=\ket{b_1}\ket{a(b_1)\oplus b_2}$,
for all length-$n_b$ bitstrings $b_1$ and $b_2$.
Here, the binary vector $b_1$ indexes the rows of $A$
and $a(b_1)$ is the column index in which the $b_1$-th row of $A$ has an off-diagonal element.
For this reason, the Hamiltonian simulation part of the
HHL algorithm implementation requires $2n_b$ qubits.

To implement the controlled rotation step in~\autoref{eq:hhl_after_irot}, we use the algorithm for uniformly controlled rotations in~\cite{Mottonen2005}.
More generally, the rotation angle
$\theta_j=\sin^{-1}(2C/\lambda_j)$
could be computed efficiently from $\lambda_j$ using quantum arithmetic~\cite{Ruiz-Perez2017}.
However, for small problem sizes the constant overhead of quantum arithmetic is too large to
be practical for realistic near-term devices~\cite{yalovetzky2023}.
The uniformly controlled rotation requires a number of gates exponential in $n_p$,
so it is important that $n_p$ not scale with $n$ in order for the
benchmark to be scalable.

To explore volumetric profiles for the algorithm, we vary the values of $n_b$ and $n_p$ as a function of the total number of qubits under test, using~\autoref{eq:hhl_dim_function}.
The value for $n_b$ is multiplied by $2$ for the reason described above, and the addition of 1 represents the ancilla qubit used for inverse rotation.
\begin{equation}
\label{eq:hhl_dim_function}
n_{total} = 2 \times n_b + n_p + 1 
\end{equation}
In general, the value of $n_p$ should be capped at some upper limit
in order for the benchmark to be scalable,
however the results presented below use a modest
number of qubits and therefore $n_p$ is allowed to
vary over the full range consistent with~\autoref{eq:hhl_dim_function}.

The benchmark sweeps over a range of available qubit widths $[min\_qubits:max\_qubits]$.
For each width $n_{total}$, valid values of $n_b$ and $n_p$ are determined, random problem matrices $A$
and vectors $b$ are chosen, quantum circuits are generated, and the expected measurement distributions are calculated.
After executing each circuit, measurement results are analyzed, and normalized and Hellinger fidelity metrics are computed from the observed and ideal outcome distributions~\cite{lubinski2023_10061574}.
The benchmark returns the average of these metrics across all problem instances and combinations of $n_b$ and $n_p$.


\begin{figure}[t!]
    \includegraphics[width=0.88\columnwidth]{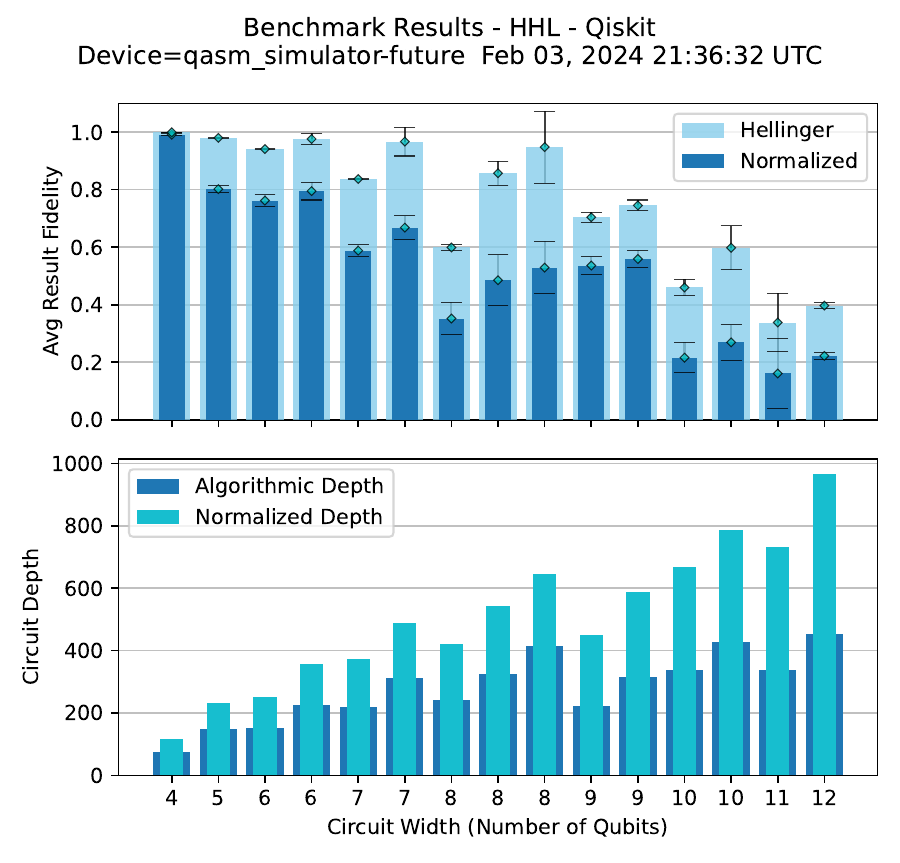}
    \caption{\textbf{Execution of HHL Benchmark up to 12 Qubits.} Here we show the results of running the HHL benchmark on the noisy quantum simulator from 4 to 12 qubits. The benchmark is configured to sweep several combinations of input and phase  register sizes at each total width executed (corresponding to $n\_total$ in~\autoref{eq:hhl_dim_function}). As a consequence, there are multiple execution groups at several of the circuit widths. This plot shows the average fidelities and circuit depth for each qubit width. For example, at 8 qubits, there are 3 valid combinations of input and phase size: (1,5), (2,3), and (3,1).}
    \label{fig:hhl_results_1}
\end{figure}

\begin{figure}[t!]
    \centering
    \includegraphics[width=0.92\columnwidth]{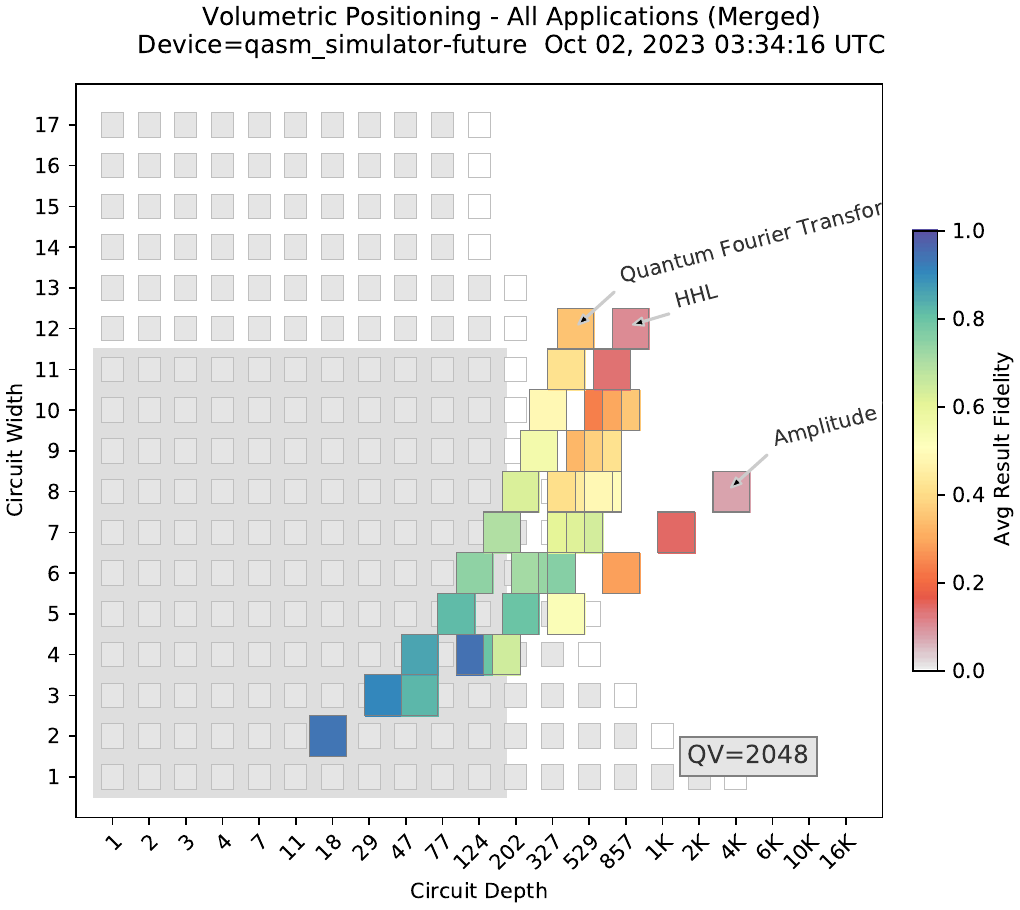}
    \caption{\textbf{Broadening Benchmark Coverage with HHL.} Here, the broadening of the application profile is visible in the volumetric positioning plot. The HHL benchmark fills the gap between the Quantum Fourier Transform (1) and Amplitude Estimation benchmarks. Each of the rectangles shown in the horizontal row for the HHL benchmark at a specific qubit width represents a variation in the algorithm's input and phase sizes. We sweep several of these to ensure more coverage of the width $\times$ height region.
    }
    \label{fig:hhl_results_2}
\end{figure}

\subsection{Results From Executing the HHL Benchmark}
\label{subsec:hhl_benchmark_and_results}

For this benchmark, we targeted the gap in~\autoref{fig:volumetric_application_profiles} between the Quantum Fourier Transform (QFT) (1) and the Amplitude Estimation volumetric profiles.
The sum of the circuit depths of the components of the HHL algorithm suggests that it should lie just to the right of the QFT (1).
Additionally, by executing the circuit multiple times while varying the width of the input and phase registers, it is possible to `broaden' the algorithm's profile and fill much of the targeted gap.

In~\autoref{fig:hhl_results_1}, we show the results of executing the HHL benchmark on a noisy quantum simulator from 4 to 12 qubits, using this broadened mode.
For a given maximum number of qubits, a valid combination of input and phase size is determined using~\autoref{eq:hhl_dim_function}.
In this case, for 12 qubits, valid values for the input and phase  size are $n_b = 4$ and $n_p = 3$.
From this, we generate a sweep over valid combinations of smaller phase and input sizes, producing circuit variants with a variety of depths at each circuit width. 
At the maximum and minimum, there is only a single valid combination, but in the middle range, there are multiple valid combinations.
The second figure,~\autoref{fig:hhl_results_2}, shows that by varying the width of the phase and input registers, the volumetric profile of the circuit is effectively broadened, increasing the coverage of the benchmark suite.

It is interesting to note the variation in fidelity and circuit depth for the qubit numbers in the middle of the profile. 
For example, at 8 qubits, there are 3 valid combinations of input and phase register size: (1,5), (2,3), and (3,1). 
Because of the implementation of the quantum phase estimation, the depth of the circuit grows faster with the number of input qubits than with the phase qubits, yet the fidelity of the execution is higher with fewer phase qubits.
We conjecture that this is due to the higher number
of two-qubit gates for circuits with more phase qubits. 
This example illustrates the way in which the benchmark framework can be useful in exploring the characteristics of quantum algorithms.
This approach could also be applied to other benchmarks in the suite, such as Amplitude Estimation, which similarly employs groups of qubits for specific functions, the size of which can be varied.

\begin{figure}[t!]
    \includegraphics[width=0.88\columnwidth]{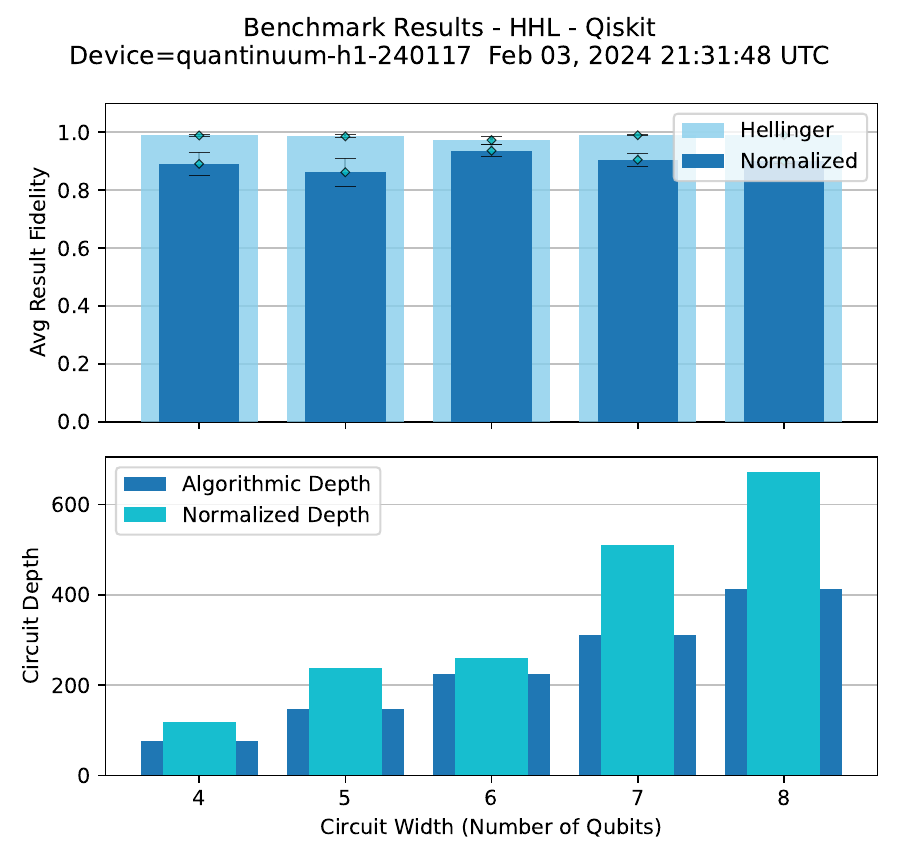}
    \caption{\textbf{Execution of HHL Benchmark on Quantinuum H1-1.}  Here we show the results of running the HHL benchmark on Quantinuum's H1-1 quantum computer. The benchmark was only run with $n_b=1$ but could be configured to sweep over a wider range to more fully cover the volumetric region. This plot shows the average fidelities and circuit depth for each qubit width. (\emph{Data provided by Quantinuum.})}
    \label{fig:hhl_results_quantinuum}
\end{figure}

\vspace{0.2cm}

In~\autoref{fig:hhl_results_quantinuum} we show the results from execution of the HHL benchmark on the Quantinuum H1-1 quantum computer with 1000 shots. H1-1 currently has 20 qubits, measured two-qubit gate fidelities >99.87\% from standard randomized benchmarking, and QV=$2^{19}$. H1-1 is also capable of mid-circuit measurement and conditional logic, which is required for this HHL benchmark. The HHL algorithm was run with $n_b=1$ and $n_p=1,2,3,4,5$ but in principle could be run up to $n_{total}=20$ with non-zero result fidelity expected up to $n_{total}\approx 15$ based on simulations. H1-1 ran the exact transpiled circuits from the QED-C suite, no further compilation was applied, although in principle could be used to further improve performance. The data in~\autoref{fig:hhl_results_quantinuum} shows that the HHL benchmark is possible to run on current hardware with small qubit number. We omit execution time data here to focus our analysis, as the execution times for the HHL algorithm closely track circuit depth and align with trends in earlier benchmarks~\cite{lubinski2023_10061574}.

\vspace{0.2cm}

The HHL benchmark is currently implemented to evaluate performance based on the fidelity of circuit execution as compared to ideal.
However, the benchmark could be enhanced to produce an application-specific metric based on how well the algorithm solves the linear equation that is given as its input.
We did not address this in our work, as there are many variables to consider in developing a valid figure of merit.
The approach would be to implement a method (2), similar to the other iterative benchmarks, which computes a figure of merit based on how accurately the algorithm solves the linear equation.
This work is left for future research.


\section{Benchmarking a Hydrogen Lattice Simulation}
\label{sec:hydrogen_lattice}

Quantum computing offers new and efficient ways to tackle classically challenging problems in chemistry and may be particularly relevant for the electronic structure problem, which can require exponential resources on classical machines~\cite{PhysRevX.7.031059,Cao_2023}. Although an exact solution to the electronic structure problem can be obtained with quantum phase estimation (QPE), the resource requirements of QPE are unattainable on today’s NISQ devices~\cite{Goings2022} due to high rates of error.
In contrast, the Variational Quantum Eigensolver (VQE) algorithm uses a hybrid quantum-classical approach with more attainable error requirements. As a result, it is thought to be a candidate algorithm that could be capable of addressing such electronic structure problems.

In this section, we introduce a VQE implementation of a Hydrogen Lattice simulation structured to make use of the QED-C benchmarking framework.
The VQE algorithm has been used previously as the basis for performance benchmarking of quantum computers~\cite{mccaskey2019quantum,yeteraydeniz2021benchmarking,dallairedemers2020application,sawaya2023hamlib,stair2020exploring}.
We found it necessary to constrain some characteristics of the problem to be addressed and carefully choose features of the benchmarking implementation for it to be practically useful for exploring various behaviors of the simulation as parameters and execution options are varied.
Our detailed analysis helps to quantify the performance of the algorithm on a target quantum computing system, with particular emphasis on the trade-off between the time taken to execute the application and the quality of the result obtained.


\subsection{Simulating a Hydrogen Lattice with VQE}
\label{sec:hydrogen_lattice_simulation_with_vqe}

\begin{figure}[t!]
\includegraphics[width=0.94\columnwidth]{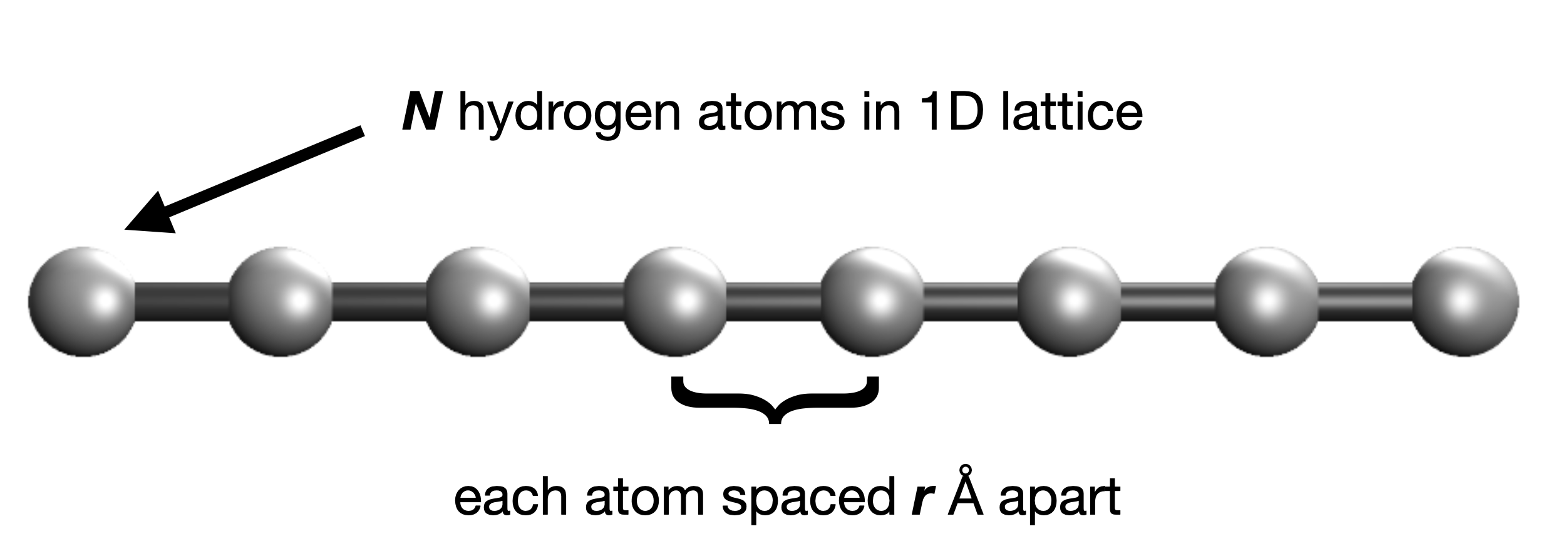}
\caption{\textbf{ Hydrogen Lattice Schematic.} The hydrogen lattice used in this work consists of one-dimensional chains of $N$ hydrogen atoms spaced equally $r$ \AA~apart. Here an 8-hydrogen chain at 1.0 \AA~interatomic spacing is depicted. More complex geometries, such as 2D or 3D arrangements in sheets or pyramids, may also be considered in future work.} 

\label{fig:hydrogen-lattice-schematic}
\end{figure}

\begin{figure}[t!]
\includegraphics[width=0.96\columnwidth]{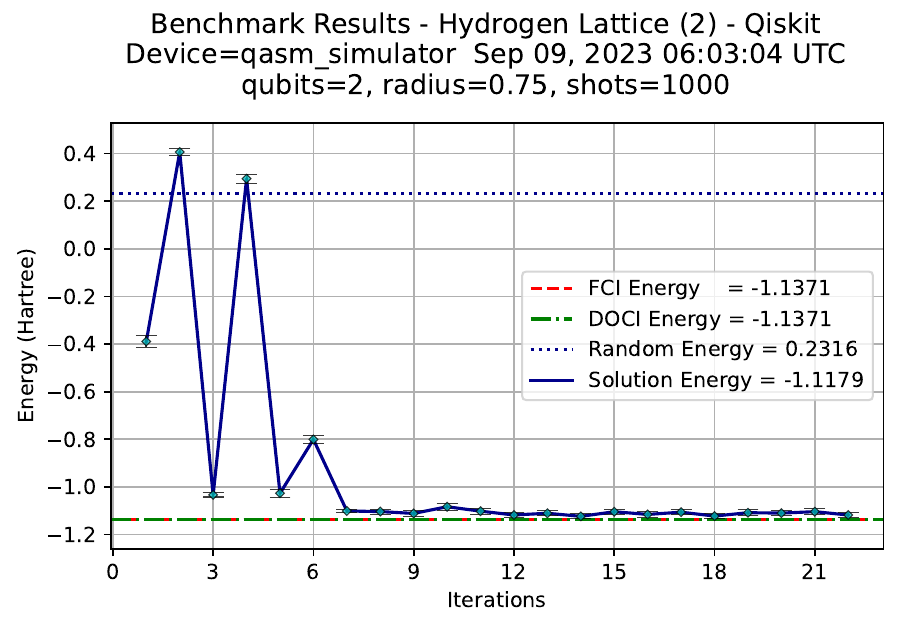}
\caption{\textbf{ Hydrogen Lattice Benchmark on 2 Qubits.} Results from executing this benchmark on a simulation of a simple chain of 2 Hydrogen atoms separated by 0.75 \AA. The benchmark implements a VQE algorithm using an upCCD ansatz with observables derived from the problem Hamiltonian.
A classical optimizer iteratively executes the ansatz, varying its parameters to find the ground state energy and converging to a final energy near to the classically computed exact FCI and DOCI energy. 
The benchmark was executed on a noisy quantum simulator with a quantum volume of 2048. 
The energy computed at the final iteration is displayed along with relevant reference energies.}
\label{fig:hydrogen_lattice_energy_2q}
\end{figure}

A simple example of the hydrogen lattice problem is depicted in~\autoref{fig:hydrogen-lattice-schematic}.
The challenge is to determine the radius at which the ground state energy of this chain of atoms is the lowest, by sweeping over a range of radii and computing the ground state energy at each.
Our benchmark is limited to evaluating the ability of a quantum computing system to calculate the ground state energy at any of a small set of pre-defined radii. 
We currently use hydrogen chains of varying lengths but plan to examine more complex geometries in the future (see Ref.~\cite{stair2020exploring} for alternatives.)

Using a variational optimization loop, the VQE algorithm can estimate the ground state energy of a given Hamiltonian, rendering it well-suited for quantum chemistry applications.
Our benchmark VQE algorithm implements a trial wave function using a unitary pair coupled-cluster doubles (upCCD) ansatz with observables derived from the problem Hamiltonian.
A classical optimizer iteratively executes the ansatz, varying its parameters in an attempt to find the lowest ground state energy. 
\autoref{fig:hydrogen-lattice-schematic} shows the progression of the benchmark over each iteration as it converges to a final energy near to the classically computed (and numerically exact) full configuration interaction (FCI) and doubly-occupied configuration interaction (DOCI) energies.
The final energy is displayed along with relevant reference energies.

The heuristic, but variational, nature of VQE allows it to operate with shallower circuits compared to common QPE algorithms~\cite{tilly2022variational}, and the iterative, hybrid quantum-classical approach allows for some noise tolerance as the noisy quantum function is classically minimized~\cite{McClean_2016}.
The algorithm is inherently approximate except in certain specific cases, and it may be susceptible to noise from insufficient statistical sampling, as it calculates an expectation value (energy). While other approaches exist for chemistry problems, such as Robust Amplitude Estimation~\cite{johnson2022reducing}, VQE’s blend of efficiency and robustness against noise makes it a compelling instrument for near-term quantum chemical calculations. 

Navigating the broad range of potential Hamiltonians for VQE benchmarking presents a significant challenge. Various Hamiltonian libraries, e.g. HamLib, have been implemented in~\cite{sawaya2023hamlib} and others, reflecting the wide variety of molecules and basis sets available.
Amid this complexity, we chose the hydrogen lattices for their simplicity and flexibility.
They are easily tunable systems, allowing for diverse physical scenarios to be explored. Moreover, these lattices are exactly solvable in the one-dimensional (1D) case using Density Matrix Renormalization Group (DMRG) methods \cite{white1992density}, offering a benchmark against which approximate methods can be assessed. This feature, along with their operation at half-filling with minimal basis sets — a trait shared by many “real-world” valence electron active spaces — enables the thorough investigation of various electronic correlation regimes.

Though no single model can fully encapsulate all possible chemical systems, hydrogen lattices serve as practical and representative models~\cite{HartreeFockGoogle2020,stair2020exploring,Motta_2017,PhysRevX.7.031059,Cao_2019}. For instance, in dealing with strongly correlated Hamiltonians, a similarly complex system can often be constructed using a hydrogen lattice. By judiciously selecting the lattice’s geometry, one can create Hamiltonian complexity that closely mirrors the original system, and VQE performance is correspondingly similar. Thus, hydrogen lattices offer a manageable and effective model, faithfully reflecting various facets of more complex systems, and validating their use for VQE benchmarking.

\vspace{0.3cm}

Practical implementation of VQE requires the selection of a parameterized trial circuit, or ansatz, that introduces additional layers of complexity for benchmarking. Given this, our work initially focuses on the upCCD ansatz, which treats electrons as pairs rather than as independent, single electrons. This approach is particularly effective for closed-shell systems, where electron pairing is a natural representation~\cite{elfving2021simulating,zhao2023orbital,OBrien2023-aa}. In other words, the upCCD ansatz intrinsically modifies the structure of the problem by allowing only paired double excitations. 

On account of the upCCD ansatz, we can represent quantum states efficiently by mapping to electron pairs~\cite{nam2020ground,elfving2021simulating}, which simplifies the problem significantly. A key benefit of this approach is that it cuts down the number of qubits we need by half compared to other methods like the Jordan-Wigner mapping. However, it's important to note that this method only works for certain types of systems, specifically those with paired electrons.

In addition to a reduced qubit requirement, the paired mapping provides another practical benefit. 
The upCCD Hamiltonian can be partitioned into three qubit-wise commuting groups, such that only three circuits need to be measured to evaluate the energy, regardless of problem size.
This is in contrast to other mappings, where energy measurements require a circuit count that can grow quartically with system size.

Assuming full connectivity, the upCCD method requires significantly fewer entangling gates than traditional unitary coupled cluster methods---about two orders of magnitude less~\cite{lee2018generalized}. Additionally, if gates can be applied in parallel, the time to execute the upCCD circuit is further reduced, potentially becoming linear with respect to system size. The upCCD approach, characterized by its less complex circuit and fewer measurements, is, therefore, a strong early option for NISQ applications. Though more complex ansatz will be explored in future work, we note that for cases where upCCD lacks accuracy, enhancements like orbital optimization and perturbation theory are viable options~\cite{zhao2023orbital, goings2023molecular}.


\subsection{The Hydrogen Lattice Benchmark Algorithm}
\label{subsec:hydrogen_lattice_benchmark_algorithm}

\begin{algorithm}[t!]
\caption{Benchmark Algorithm for VQE}\label{alg:cap}
\begin{algorithmic}[1]
\State $target \gets \textcolor{cyan}{backend\_id}$
\State initialize\_metrics()
\For{$size \gets \textcolor{cyan}{min\_size}, \textcolor{cyan}{max\_size}$}
    \State $circuit\_def \gets define\_problem(\textcolor{cyan}{problem}, size, \textcolor{cyan}{args})$ 
    \For{$restart\_id \gets 1, \textcolor{cyan}{max\_restarts}$}
            \State $cost\_function \gets define\_cost\_function(problem)$
            \State $circuit, num\_params \gets create\_circuit(circuit\_def)$
            \State $cached\_circuit \gets compile\_circuit(circuit)$
            \State $params[\bm{\alpha}] \gets random(num\_params)$
            \While{\textit{minimizer() not done} }\Comment{minimizing}

                \State $circuit \gets apply\_params(cached\_circuit, params)$
                \For{$pauli\_op \gets problem$}
                    \State $\textcolor{blue}{counts }\gets execute(target, circuit, \textcolor{cyan}{num\_shots})$
                \EndFor
                \State $energy, quality \gets cost\_function(\textcolor{blue}{counts})$
                \State $store\_iteration\_metrics(\textcolor{blue}{quality}, \textcolor{blue}{timing})$
                \State $params[\bm{\alpha}] \gets optimize(params[\bm{\alpha}])$
                \State $done \gets \textit{True if lowest(energy) found}$
                
            \EndWhile
            \State $compute\_and\_store\_restart\_metrics()$
    \EndFor
    \State $compute\_and\_store\_group\_metrics()$
\EndFor

\end{algorithmic}
\label{hydrogen_lattice_algo1}
\end{algorithm}

The implementation of the VQE algorithm, shown in Algorithm~\ref{hydrogen_lattice_algo1}, is modeled after the Max-Cut QAOA benchmark implementation from ~\cite{lubinski2023optimization}.
We describe the algorithm here, omitting some details for brevity.
As another type of iterative quantum algorithm, the new benchmark similarly optimizes by iterating over circuits executed with varying parameters, but brings additional complexity.
The benchmark can be configured to sweep over a range of qubit widths, executing the VQE algorithm at the problem size associated with that qubit width.
For each problem size, a limited set of inputs (radii in this case) can be selected, and the results at each radius collected and displayed. 
For each radius, however, execution is more involved and we focus here on several important distinctions.

One key difference is the addition of a loop over the three circuits with their appended basis rotations to measure the multiple non-commuting upCCD Hamiltonian terms of the hydrogen lattice simulation~\cite{elfving2021simulating}.
For every iteration within the VQE algorithm, we find the overall expectation value by summing over the expectation value of all these individual terms. This requires an additional for loop that goes through every Hamiltonian term to collect the measurement counts for each one.
The \texttt{\textcolor{blue}{$counts$}} variable represents an array of counts for each Pauli term, rather than just the single counts seen in each iteration of the Max-Cut benchmark.
Measurements of execution fidelity and run-time are similarly collected as arrays over the Pauli terms and aggregated in summary displays.

The execution parameters for the Hydrogen Lattice VQE benchmark are different from those of the MaxCut QAOA benchmark.
For the Hydrogen Lattice simulation, there is only a single cost-function parameter $\bm{\alpha}$ (rather than the $\bm{\beta}$ and $\bm{\gamma}$ parameters used in QAOA). This $\bm{\alpha}$ generally encompasses the parameter composition that a VQE ansatz might have. 
There is also the \texttt{\textcolor{cyan}{$radius$}} parameter, the spacing constant $r$ that is specific to the Hydrogen Lattice problem. 

Another complication to consider is related to the hydrogen lattice Hamiltonians over which we are iterating. For every system size, there are different shapes of the hydrogen lattice as well as the spacing constant $r$.
In future work, we will need to add an option to execute over different Hamiltonians that refer to different shapes of the hydrogen lattice (1D, 2D, 3D) in addition to the spacing constant.

\vspace{0.3cm}

As with our other benchmark algorithms, this benchmark is implemented with multiple methods.
The simpler of these, Method (1), gauges the fidelity of the VQE ansatz execution using the normalized Hellinger fidelity.
Executing the VQE ansatz at various sizes, we expect to acquire an ideal distribution over the outcome space.
For our analysis in the next section, we can state specifically that, for the VQE ansatz, as the problem size grows larger, the fidelity will generally decrease due to noise for the same number of shots.

The iterative, variational VQE algorithm discussed in this section is invoked as Method (2) of the Hydrogen Lattice Benchmark.
The entirety of the discussion in this manuscript is centered around the implementation and analysis of results from this Method (2) of the benchmark.


\subsection{Analysis of Benchmark Result Accuracy}
\label{subsec:hydrogen_lattice_results_analysis}

\begin{figure}[t!]
\includegraphics[width=0.99\columnwidth]{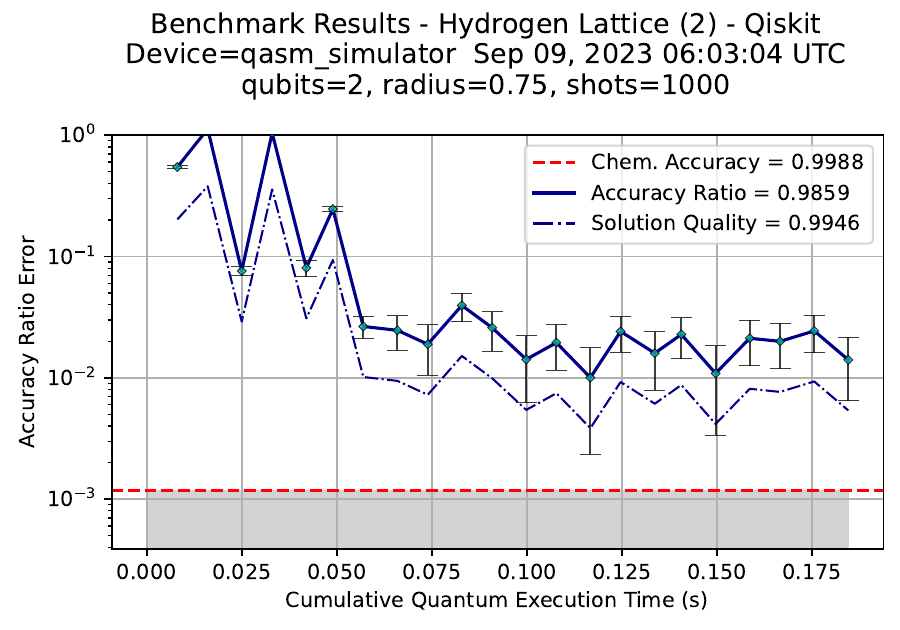}
\caption{\textbf{ Hydrogen Lattice Benchmark Results Analysis.} Here we present a detailed analysis of the data shown in~\autoref{fig:hydrogen_lattice_energy_2q}.
Two figures of merit, the ``Accuracy Ratio'' and ``Solution Quality'', are computed as ratios of the solution energy to the FCI energy, scaled relative to the energy produced from a random measurement distribution. At each iteration, the error in these metrics is displayed on a log scale to highlight results near optimal.
The grey band envelops energy values within the commonly accepted threshold for ``chemical accuracy'', or $1.6\times 10^{-3}$ Hartree, scaled similarly. Here, the algorithm converged to an energy about an order of magnitude from chemical accuracy. 
Results are plotted against the cumulative quantum execution time to capture the run-time cost of executing the algorithm.
}
\label{fig:hydrogen_lattice_analysis_2q}
\end{figure}
 
The final ground state energy that results from the execution of the VQE hydrogen lattice benchmark algorithm can be compared against a classically computed ``exact energy''.
While~\autoref{fig:hydrogen_lattice_energy_2q} shows the energy computed by the VQE algorithm along with the relevant reference energies, we introduce in~\autoref{fig:hydrogen_lattice_analysis_2q} several new figures of merit, or measures of accuracy, computed from the same benchmark data.
The derivation and significance of these is discussed in this section.

The simplest measure of the accuracy of a simulation result is the absolute energy difference $\Delta E$ between the application's computed result (solution energy) and the exact FCI energy.
This is the method used in much of the literature that explores the use of quantum computers for computational chemistry~\cite{dallairedemers2020application,mccaskey2019quantum,yeteraydeniz2021benchmarking}.
A given value of absolute energy difference is considered a good result if it falls within the commonly accepted threshold for ``chemical accuracy'', or $1.6\times 10^{-3}$ Hartree. 
In~\autoref{fig:hydrogen_lattice_energy_2q}, this $\Delta E$ is $0.0192$ or about $12$ times the acceptable distance for chemical accuracy.

However, it is challenging to gauge whether an absolute energy difference is good, mediocre, or really bad, as it depends greatly on the nature and size of the problem it addresses, making it difficult to use as a generalized performance benchmark.
Even more complicated is deciding on a metric to formalize the accuracy of the VQE algorithm at intermediate steps.
What might be considered a ``good" result is highly dependent on the application. For example, while large fluctuations in the energy at the start of the VQE may be acceptable, the same fluctuations towards the end of the iterations could be considered a failure of the algorithm to find a true ground state. There is also an issue of scale - because the VQE could result in a state close to but not exactly the ground state, it is difficult to define what a successful energy might be for different applications.

Another complication could be what to use as a ``correct" ground state energy to verify the VQE's estimated value. As an example, because the upCCD ansatz operates in a seniority-zero Hilbert subspace, the exact solution for the ground state energy is found via DOCI~\cite{elfving2021simulating}, which is the exact solution in the seniority-zero subspace. In other words, while FCI will calculate an exact solution in the full Hilbert space, the upCCD ansatz will converge instead to the DOCI energy. 

\vspace{0.3cm}

\begin{figure}[t!]
\includegraphics[width=0.99\columnwidth]{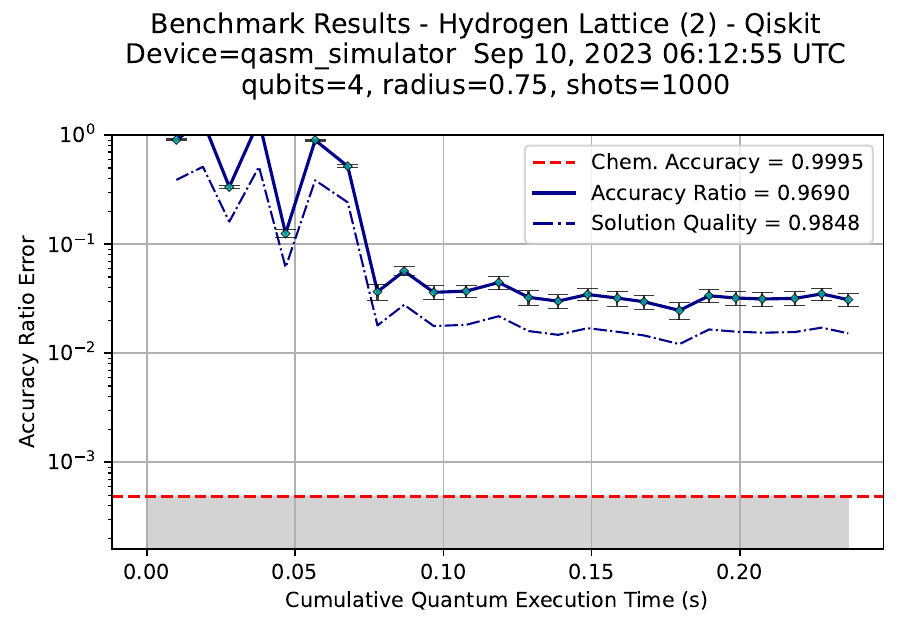}
\caption{\textbf{ Hydrogen Lattice Results Analysis on 4 Qubits.} This plot presents the error in two figures of merit, the Accuracy Ratio $AR$, and the Solution Quality $SQ$, for a 4-qubit Hydrogen Lattice simulation problem.
During initial iterations of the algorithm, the accuracy ratio metric takes on negative values, resulting in an error greater the $1.0$.
The solution quality is normalized so that its values always lie in the range $[0.0,1.0]$, ensuring that a full trace appears in the plot.
}
\label{fig:hydrogen_lattice_analysis_4q}
\end{figure}

For our purposes, we introduce a new figure of merit relevant for benchmarking, the ``Accuracy Ratio'' or $AR$.
Defined in~\autoref{eq:accuracy_ratio_def}, this metric is calculated as a ratio of the computed energy $E_{\rm solution}$ to the exact FCI energy $E_{\rm FCI}$, scaled relative to the energy that would be computed using a quantum system that returns completely random measurement values, generating a uniform distribution (the ``random'' energy) or $E_{\rm random}$.
\begin{equation}
    \label{eq:accuracy_ratio_def}
    AR = 1.0 - \frac{ \mid E_{\rm solution} - E_{\rm FCI} \mid } { \mid E_{\rm random} - E_{\rm FCI} \mid }
\end{equation}

The values of $E_{\rm FCI}$, $E_{\rm DOCI}$, $E_{\rm HF}E_{\rm FCI}$, and $E_{\rm random}$, are computed (classically) in advance for each problem that is addressed by the benchmark. With this approach, running the benchmark program requires only the execution of the algorithm under test, followed by a comparison against the previously computed reference values.
Note that $AR$ can take on a value that is outside of the range $\left[0, 1\right]$.
 
To bound our metric in the range $\left[0, 1\right]$ we define a second figure of merit, referred to as ``Solution Quality'' or $SQ$, defined in~\autoref{eq:solution_quality_def}.
\begin{equation}
    \label{eq:solution_quality_def}
    SQ = -\arctan ( \frac{ \frac{ E_{\rm solution} } { E_{\rm FCI} } \cdot p \cdot 2 } {\pi} )
\end{equation}

The metric $SQ$ is a measure of where the solution energy $E_{\rm solution}$ is positioned in the range from $E_{\rm FCI}$ to infinity. Since the most interesting computed energies lie closer to $E_{\rm FCI}$, we found it necessary to apply a conditioning function to the ratio $SQ$ to make it useful as a normalized metric. A heuristically determined ``precision'' factor $p$ (default $5.0$) determines the shape of the curve that is conditioned by the $arctan$ function and is chosen to best portray the results seen in this region.

The difference between these two figures of merit is illustrated in~\autoref{fig:hydrogen_lattice_analysis_4q}. The accuracy ratio $AR$ is seen to take on negative values, resulting in an error that is greater than $1.0$. The normalized value of solution quality $SQ$ is bounded in the range $[0.0,1.0]$. 
The value of $AR$ acts as a measure of the degree to which a quantum computing system is more effective than a random number generator. The value of $SQ$ is useful for visualizations that depend upon a bounded set of values for all to be visible. 
There are trade-offs between the two, and in this work, we primarily use $AR$, since the shape of the $SQ$ curve depends on precision $p$ which can vary based on the size of the problem.

The grey band seen in the figure envelops energy values that lie within the commonly accepted threshold for ``chemical accuracy'', or $1.6\times 10^{-3}$ Hartree, scaled for $AR$ using~\autoref{eq:accuracy_ratio_def}. In this 4-qubit case, the algorithm converged to an energy about two orders of magnitude from chemical accuracy, when viewed in the $AR$ accuracy ratio domain from  $E_{\rm FCI}$ to  $E_{\rm random}$.


\begin{figure}[t!]
\includegraphics[width=0.99\columnwidth]{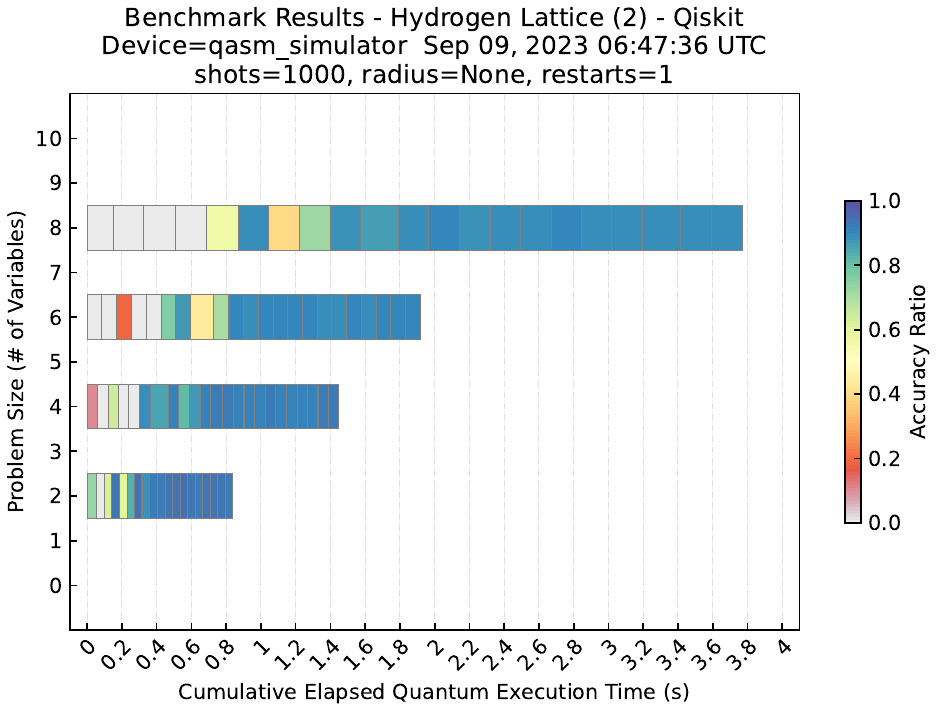}
\caption{\textbf{ Area Plot Shows Quality/Run-Time Trade-off.} In this ``area plot'', each horizontal row summarizes the data for the execution of the benchmark at a specific number of qubits. Each rectangle in the horizontal strip represents the elapsed quantum execution time. The elapsed time includes circuit compilation, transpilation, initialization and loading, and the final transfer of result data to the classical computer for classical processing. Elapsed time is a realistic representation of the time that a user will experience when executing a quantum circuit of this type.}
\label{fig:hydrogen_lattice_area_plots}
\end{figure}


\subsection{Quality of Solution vs Execution Run-Time}
\label{subsec:hydrogen_lattice_runtime_vs_quality}

In iterative algorithms, the quality of the results obtained is impacted by the length of time the algorithm is permitted to execute.
Here, we show how benchmarking results are presented in ways that permit some comparisons across system size, quality, and runtime.
We also describe the methods used to collect and analyze the metrics and illustrate the time versus quality trade-off for the Hydrogen Lattice VQE algorithm using a performance profile referred to as an ``area plot'', introduced in our previous work~\cite{lubinski2023optimization}.

An example of an area plot depicting the Quality/Time trade-off is shown in~\autoref{fig:hydrogen_lattice_area_plots}. It presents data collected for the benchmark on problems ranging from 2 to 8 qubits and executed on the same noisy quantum simulator used in prior plots. Each horizontal row represents successive iterations at each problem size (number of qubits), where position on the X-axis represents the cumulative elapsed quantum execution time, and color tracks the $AR$ computed after each classical optimizer iteration.
The elapsed time includes circuit compilation, transpilation, initialization and loading, and the transfer of result data to the classical computer for classical processing. It is a realistic representation of the time that a user will experience when executing a quantum circuit of this type. The time at which the quality of result reaches an acceptable level is visible in this plot and grows with circuit width.

\begin{figure}[t!]
\includegraphics[width=0.90\columnwidth]{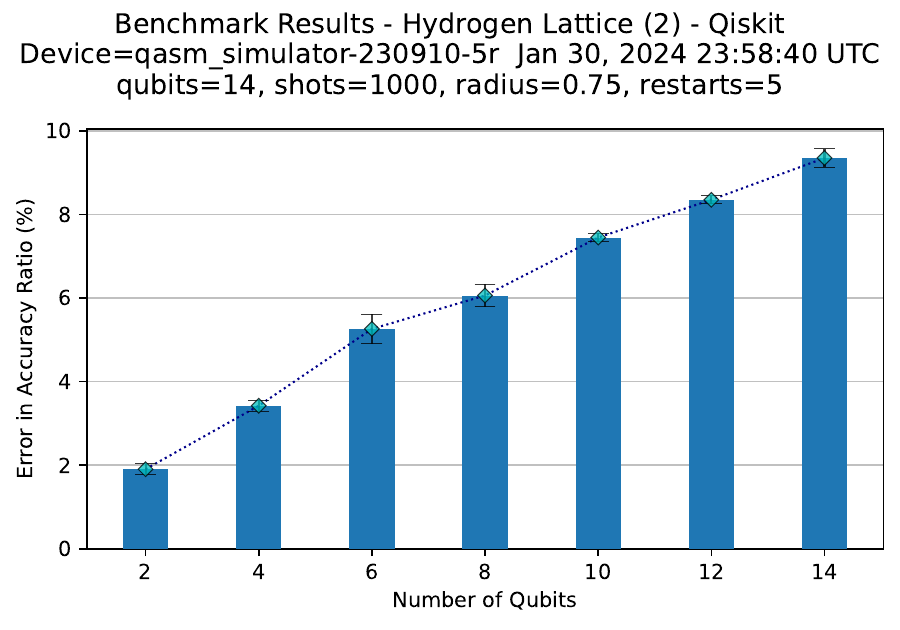}
\caption{\textbf{ Hydrogen Lattice Accuracy Ratio Error over a Range of Qubit Widths.} Here, the Hydrogen Lattice benchmark was executed on a noisy simulator over a range of qubit widths, from 2 to 14.  Each bar represents the error in the accuracy ratio computed in the final iteration at each qubit width, averaged over 5 restarts. Degradation in the solution quality is evident as the problem size increases. This is largely due to the decrease in result fidelity seen in method (1), correlated with the depth of the ansatz circuit. Other factors may also impact the results, such as the limited shot count or the random initialization of angles which can lead to barren plateaus.}
\label{fig:hydrogen_lattice_final_quality_bars}
\end{figure}

In~\autoref{fig:hydrogen_lattice_final_quality_bars}, we show the error in $AR$ (distance from optimal) seen in the final iteration of each run of the benchmark executed on our noisy quantum simulator. The benchmark sweeps over a range of problem sizes, implemented with qubit widths from 2 to 14, and executed 5 times to obtain variance. This plot illustrates how the quality of the result degrades as the problem size increases. This is roughly in line with expectations, as the depth of the ansatz increases in proportion to qubit width. The fidelities obtained with method (1) show a similar degradation with problem size.

Another important perspective is in how the execution time of the application changes as the problem size increases. Execution time on a classically implemented simulator is expected to increase much more rapidly than on a computing system implemented in hardware.

\begin{figure}[t!]
    \begin{subfigure}{0.46\textwidth}
        \includegraphics[width=\textwidth]{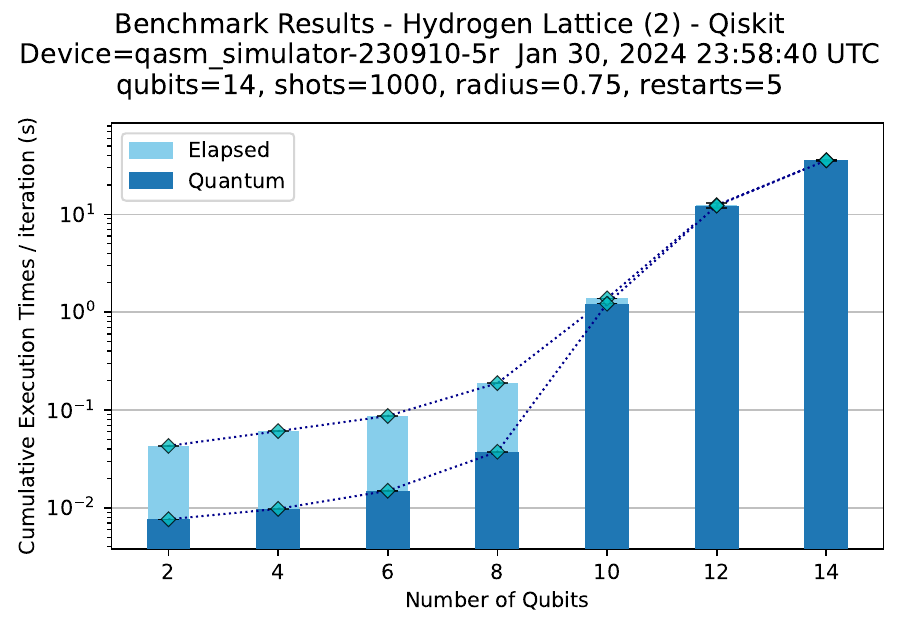}
        \caption{Execution on Quantum Simulator}
        \label{fig:hl_times_simulator}
    \end{subfigure}
    \begin{subfigure}{0.46\textwidth}
        \includegraphics[width=\textwidth]{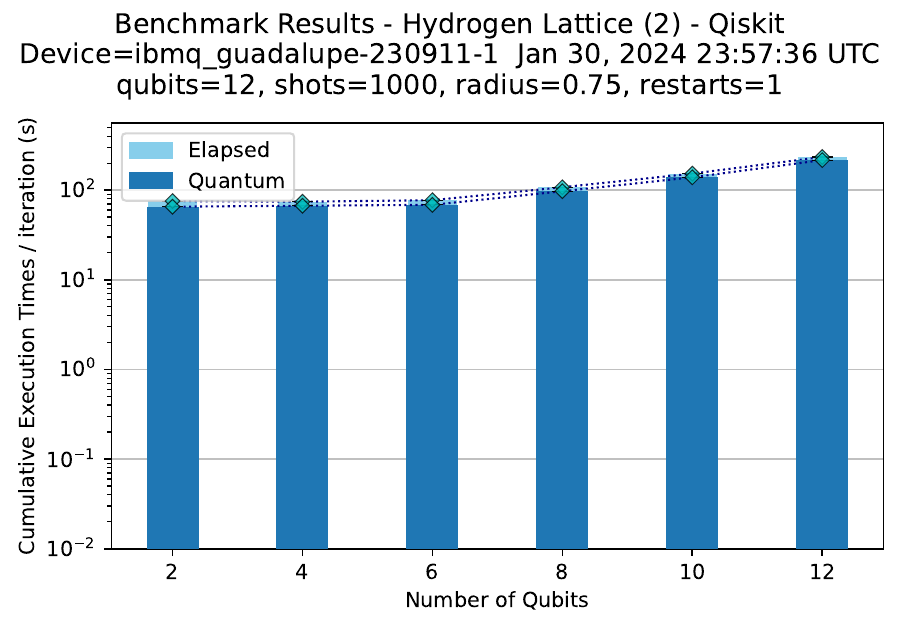}
        \caption{Execution on Hardware System}
        \label{fig:hl_times_guadalupe}
    \end{subfigure}
\caption{\textbf{ Average Execution and Elapsed Times per Iteration.} Here we show average execution and elapsed times per iteration of the objective function, at each qubit width. On a quantum simulator (a), the exponential growth of the quantum execution time is evident, while the elapsed time increment becomes less significant as the quantum time increases. On a physical hardware system, IBM Quantum $ibmq\_guadalupe$ (b), the quantum execution time increases more slowly, a defining characteristic of a quantum system.
(\emph{Data collected via cloud service}.)}
\label{fig:hydrogen_lattice_avg_exec_time_bars}
\end{figure}

A bar chart showing average execution and elapsed time per iteration of the objective function, at each qubit width, is shown in~\autoref{fig:hydrogen_lattice_avg_exec_time_bars}.
These averages, and associated error bars, are computed as a function of the array of time measurements collected as the iterative algorithm progresses. Executed on a noisy quantum simulator (a), the exponential growth of the quantum execution time is evident, while the increment in elapsed time is a smaller fraction of the total as the quantum time increases.
On a representative physical hardware system, e.g. ibmq\_guadalupe, (b), the quantum execution time starts at a larger value but increases much more slowly. The growth in execution time is primarily due to the increase in circuit depth which can be seen when running the method (1) benchmark.

The hydrogen lattice benchmark algorithm enables users to define both the classical optimizer function governing the iteration of the algorithm and the method by which angles are specified within the ansatz. 
While not explored in this manuscript, it is important to acknowledge that variations in these parameters can yield markedly different outcomes. Maintaining consistency in these parameters is imperative when utilizing the benchmarks to compare performance across backend systems.


\section{Program Optimization Techniques}
\label{program_optimization_techniques}

In this section, we review advancements made within the QED-C benchmark framework to allow the integration of custom program optimization functions into the execution pipeline (i.e. circuit creation, compilation, transpilation, execution, measurement processing).
We demonstrate the impact that such options can have on benchmark results, using optional third-party tools that have newly become available.

We analyze the impact of three program optimization techniques, using the new execution control features added to the benchmark framework. First, we consider state preparation and measurement error mitigation, employed through the ``Sampler'' primitive recently added to the Qiskit library~\cite{qiskit_sampler}.
Second, we explore approximate circuit resynthesis, implemented via the open-source toolkit TKET~\cite{Sivarajah_2020}.
Third, we consider a form of deterministic error suppression, available through Fire Opal and via native integration with Qiskit Runtime primitives~\cite{q-ctrl}.
For each of these, we evaluate the effect on the fidelity of the result data returned to the user and the impact on total application performance, including execution run-time.


\subsection{Error Mitigation with Qiskit Sampler}
\label{subsec:qiskit_error_mitigation}

Error mitigation has been shown to be an effective technique for improving the performance of quantum program execution~\cite{cai2023quantum,Qin_2023}.
However, there are limits to the effectiveness of these techniques~\cite{Takagi_2022}.
In particular, there is a trade-off in the improvement in result quality and the total time taken to execute all the required circuits, while the degree of improvement can depend on the device used and the circuit types tested~\cite{Cirstoiu2023volumetric}.

\begin{figure}[t!]
    \centering
    \begin{subfigure}[b]{\columnwidth}
        \includegraphics[width=0.80\columnwidth]{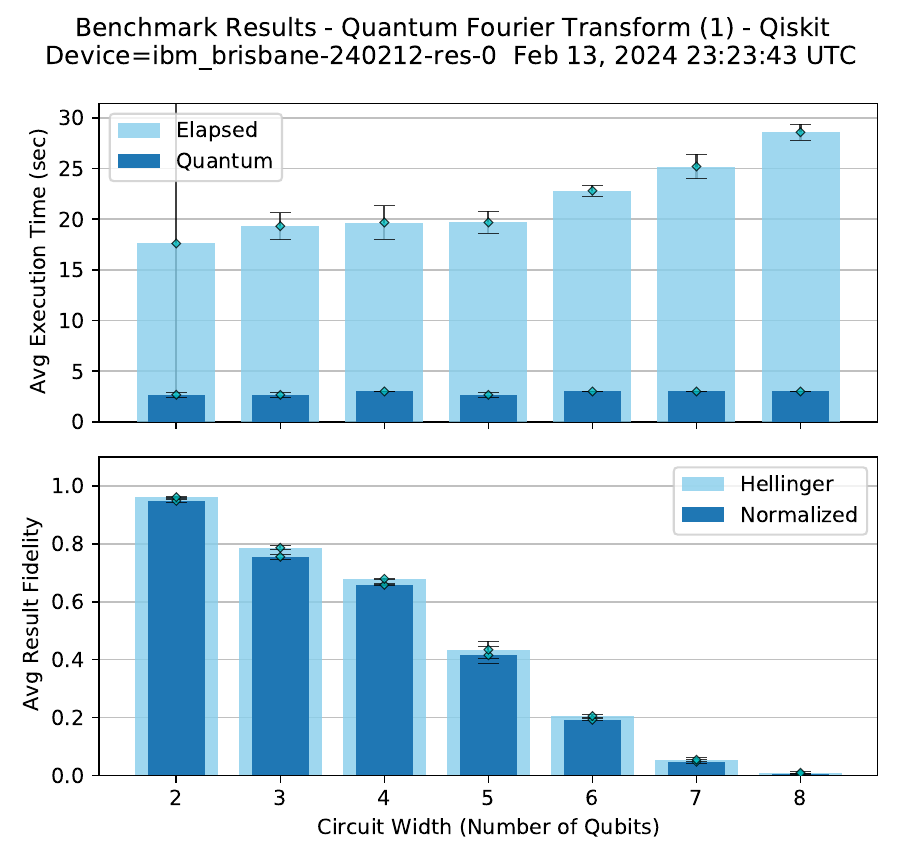}
        \caption{QFT (1) Execution  - Resilience Level 0}
    \end{subfigure}
    \begin{subfigure}[b]{\columnwidth}
        \includegraphics[width=0.80\columnwidth]{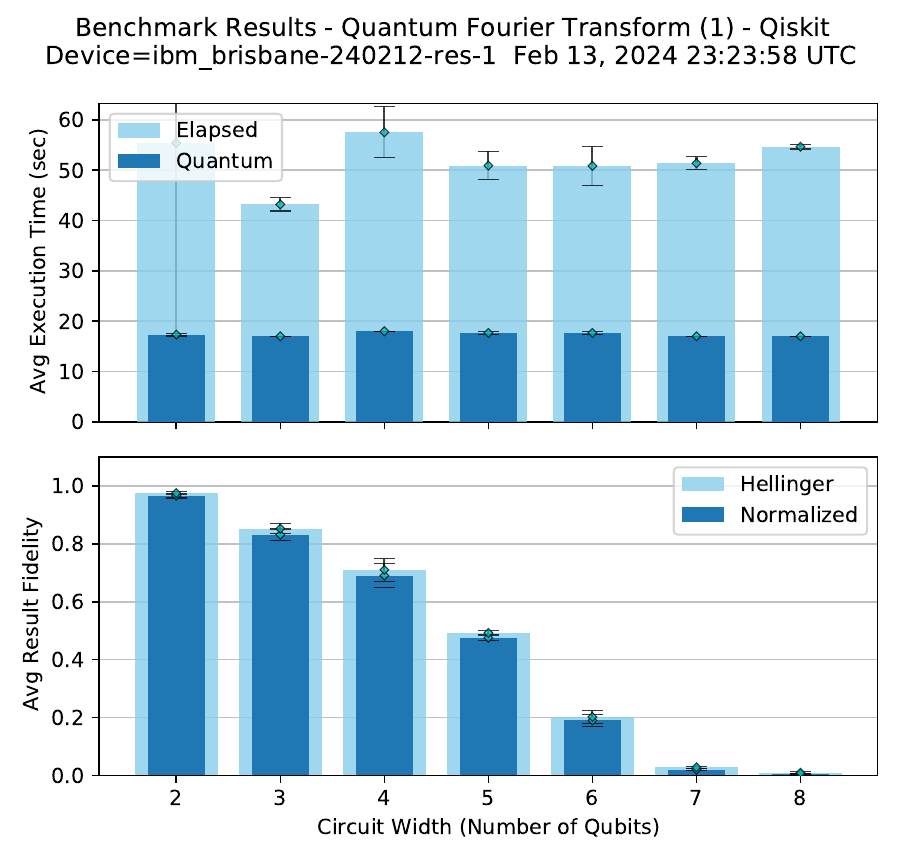}
        \caption{QFT (1) Execution - Resilience Level 1}
    \end{subfigure}
    \caption{\textbf{Execution of QFT (1) Benchmark using Sampler.} The elapsed and quantum execution times are shown, with fidelity results, from executing the QFT (1) benchmark using the Qiskit Sampler Primitive on the IBM Quantum $ibm\_brisbane$ backend (1000 shots). In (a), the circuits were executed using resilience level 0, in which no error mitigation is applied. In (b), the circuits were executed with error mitigation level 1, which applies dynamical decoupling to the circuits and mthree error mitigation automatically.
    With resilience level 1, fidelity is improved by up to 14\% at small qubit counts, but execution times are about 2X larger.
    (\emph{Data collected via cloud service}.)
    }
    \label{fig:sampler_error_mitigation_qft}
\end{figure}

In this section, we study the Qiskit Runtime Sampler Primitive~\cite{qiskit_sampler}, a feature of the Qiskit library that performs error mitigation automatically after circuit execution, and before data are returned to the caller.
The Sampler makes use of functions from a library called ``mthree''~\cite{Nation_2021,qiskit_mthree}, short for matrix-free measurement mitigation.
The mthree library uses calibrated measures of error rate variability in the components of a target quantum computing system to recover missing fidelity from measurement data. This can compensate for SPAM errors,
returning a quasi-probability distribution of bitstrings.
The Sampler may also perform multiple executions of the circuit and uses a heuristic algorithm in an attempt to achieve the highest fidelity of execution.

The Sampler provides a default error mitigation level, referred to as resilience level 1, which activates the mthree error mitigation functions.
Setting the resilience level to 0 disables error mitigation and returns the uncorrected measurement distribution.
We enable this optimization when executing the benchmarks by specifying an option that invokes the Sampler primitive with the desired resilience level.

We chose the Sampler for its ease of use in activating this error mitigation feature. 
Alternatively, the QED-C benchmarks can be programmatically configured to execute any user-defined post-processing function, such as mthree, immediately after executing each quantum circuit to apply corrections to the measured data and improve the quality of result.
Mthree has been studied previously with this alternate approach and shown to produce statistically significant improvements in the measured fidelity of the QED-C benchmarks~\cite{Nation_2023}.

\vspace{0.3cm}

In~\autoref{fig:sampler_error_mitigation_qft}, we contrast the quality of the results and execution times that are observed when the circuits of the benchmark suite are executed using the Sampler, with and without error mitigation configured. For one representative benchmark program, the QFT (1), the improvement in fidelity is visible for circuits of qubit widths ranging from 2 to 8 qubits. The average Hellinger fidelity, normalized fidelity, elapsed, and execution times for each circuit group are compared across the range of qubit widths using the provided benchmark plots. The first plot (a) shows the results of executing without enabling error mitigation, while the second plot (b) shows the improvement when mitigation is enabled.
For example, at 5 qubits, the normalized fidelity improves from 0.415 to 0.475 when the resilience level is set to 1, a gain of about 14\%.

\vspace{0.3cm}

The introduction of classical post-processing into the execution pipeline has a cost in terms of the total run-time for the execution of the circuit. 
Our benchmark framework provides a standard mechanism for collecting execution time on a backend quantum device and the total elapsed or wall-clock time which includes classical computation and data transfer time (described in our prior work~\cite{lubinski2023optimization}.)
For circuits executed with Qiskit, the elapsed time $t_{\rm elapsed}$ is the sum of circuit compilation time $t_{\rm compile}$, queue time $t_{\rm queue}$, time to load the circuit into the control system $t_{\rm load}$, execution time on the quantum backend $t_{\rm quantum}$, and time required to classically compute error mitigation $t_{\rm mitigate}$ as in~\autoref{eq:elapsed_mitigation_time}.
\begin{equation}
    \label{eq:elapsed_mitigation_time}
    t_{\rm elapsed} = t_{\rm compile} + t_{\rm queue} + t_{\rm load} + t_{\rm quantum} + t_{\rm mitigate}
\end{equation}

Quantum execution time $t_{\rm quantum}$ is the only metric that is reliably returned in the Qiskit result record, although the method of timing may differ across backend providers.
When the Sampler is used with resilience level 1, a portion of the error mitigation processing time is included in $t_{\rm quantum}$, while the remainder is captured by the benchmark framework in $t_{\rm elapsed}$.
As a result, both the elapsed time and the quantum execution time increase when error mitigation is applied.
Currently, we do not have a mechanism to partition the time into all of its components with more granularity than this.

In the QFT (1) example presented in~\autoref{fig:sampler_error_mitigation_qft}, the time for execution with error mitigation enabled is about 2-3X greater on average than without, depending on the qubit width of the circuit.
For example, at 6 qubits, the total elapsed time $t_{\rm elapsed}$ increases from $51s$ to $129s$ with resilience level set to 1.
At this width, the quantum execution time component $t_{\rm quantum}$ shows an increase from $18s$ to $42s$.

\begin{figure}[t!]
    \centering
    \begin{subfigure}[b]{\columnwidth}
        \includegraphics[width=0.80\columnwidth]{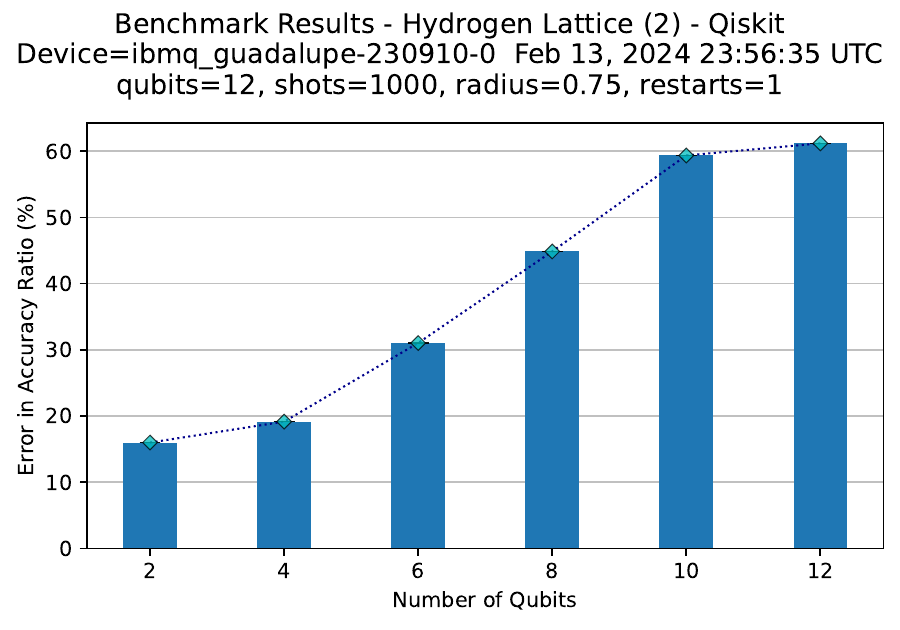}
        \caption{Hydrogen Lattice Accuracy Error - Resilience Level 0}
    \end{subfigure}
    \begin{subfigure}[b]{\columnwidth}
        \includegraphics[width=0.80\columnwidth]{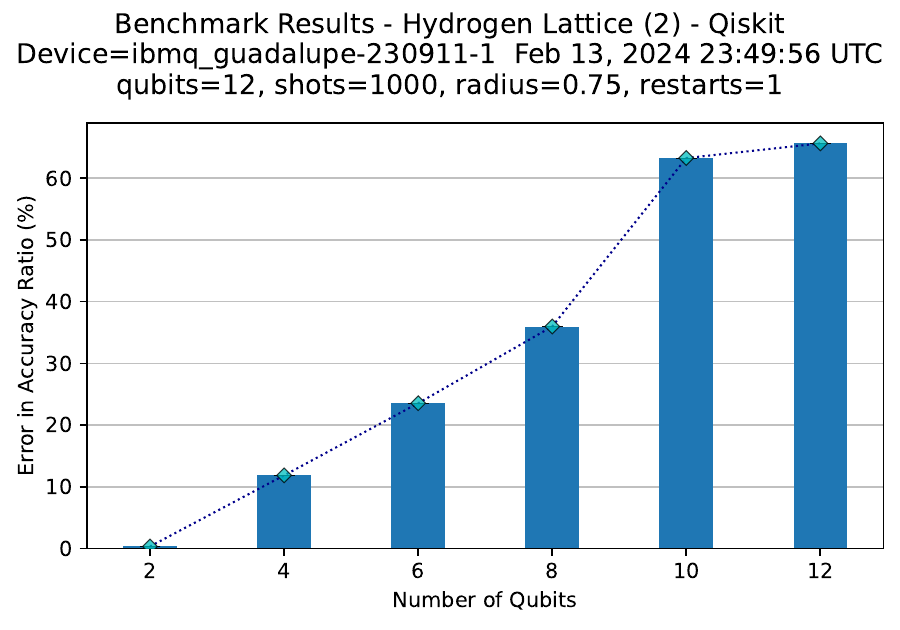}
        \caption{Hydrogen Lattice Accuracy Error - Resilience Level 1}
    \end{subfigure}
    \caption{\textbf{Accuracy Improvement for Hydrogen Lattice Benchmark using Sampler.} Here we display the percentage error in an application-specific metric, the accuracy ratio, that is computed in the final iteration of the Hydrogen Lattice benchmark when executed on the IBM Quantum $ibmq\_guadalupe$ system. In plot (a), the error is 16\% and 19\% at 2 and 4 qubit widths, respectively. In plot (b), the error has been reduced with error mitigation to about 1\% and 11\%. The level of improvement appears to diminish at larger problem sizes
    (\emph{Data collected via cloud service}.)
    }
    \label{fig:sampler_error_mitigation_hydro}
\end{figure}

As another example,~\autoref{fig:sampler_error_mitigation_hydro} illustrates an improvement in the accuracy ratio obtained when the Hydrogen Lattice benchmark (\autoref{sec:hydrogen_lattice}) is run with error mitigation enabled via the Sampler.
This benchmark uses a hybrid algorithm that executes a set of quantum circuits repeatedly,
wherein error mitigation is applied to the measurement results after each iteration.
In (a), we observe that without error mitigation, the error in the accuracy ratio ranges from 16\% to 30\% for small qubit widths.
However, in (b), we see that the error in accuracy ratio is reduced to a range of 1\% to 23\% for small problem sizes, ranging from 2 to 6 qubits.
At larger qubit widths, there is no noticeable improvement in the accuracy ratio.


\subsection{Quantum Circuit Transformation with TKET}
\label{transformations_with_pytket}

The QED-C benchmark circuits are typically executed using default methods for compiling and mapping them to target systems, sometimes with API-specific options, such as the resilience level discussed in an earlier section.
In this section, we illustrate an enhancement made to the benchmark execution pipeline that enables the exploration of custom techniques for preparing the circuits before their execution.
A new benchmark API option permits a user to invoke a custom \emph{transformer} function on each circuit before it is executed.

Here, we make use of an open-source toolkit, TKET (accessed via the python package pytket)~\cite{Sivarajah_2020}, to perform program transformations on the benchmark circuits prior to execution.
This permits us to explore the impact that such transformations have on benchmark algorithms that have differing characteristics.
This exercise revealed weaknesses in some of the benchmarks, where our algorithmic benchmark examples turned out to compile to much fewer gates than the standard algorithmic circuits. We mention several examples below and plan to investigate more complicated circuits to address this.

\begin{figure}[t!]
    \centering
    \begin{subfigure}[b]{\columnwidth}
        \includegraphics[width=0.80\columnwidth]{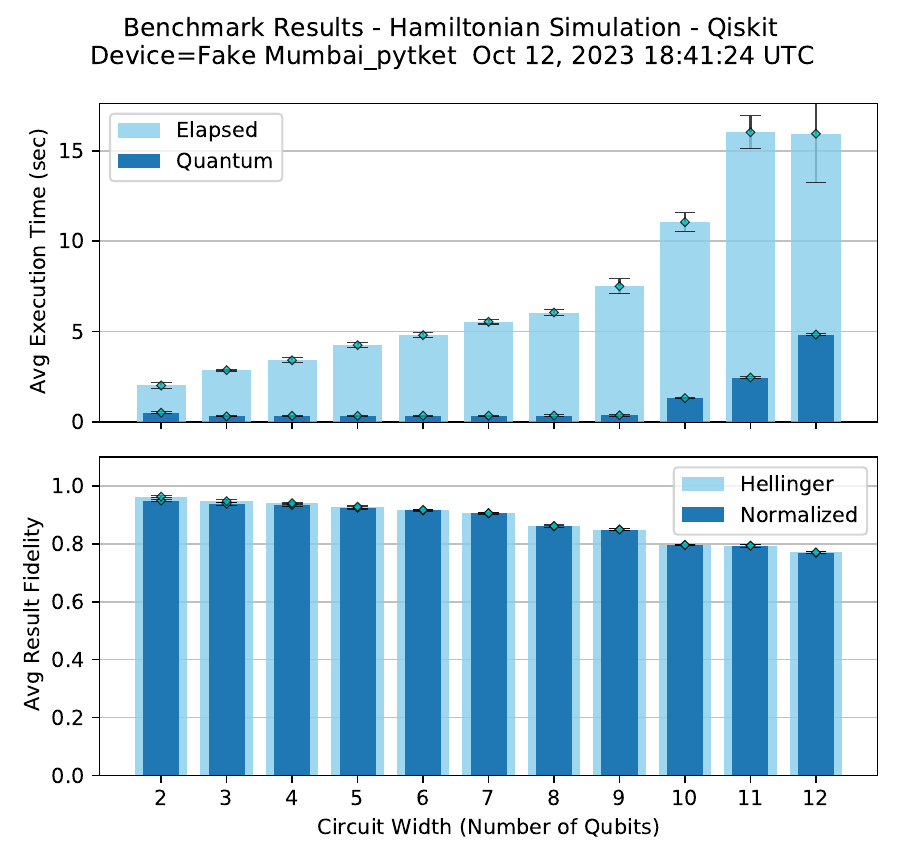}
        \caption{Hamiltonian Simulation with TKET Approximate Resynthesis}
    \end{subfigure}
    \begin{subfigure}[b]{\columnwidth}
        \includegraphics[width=0.80\columnwidth]{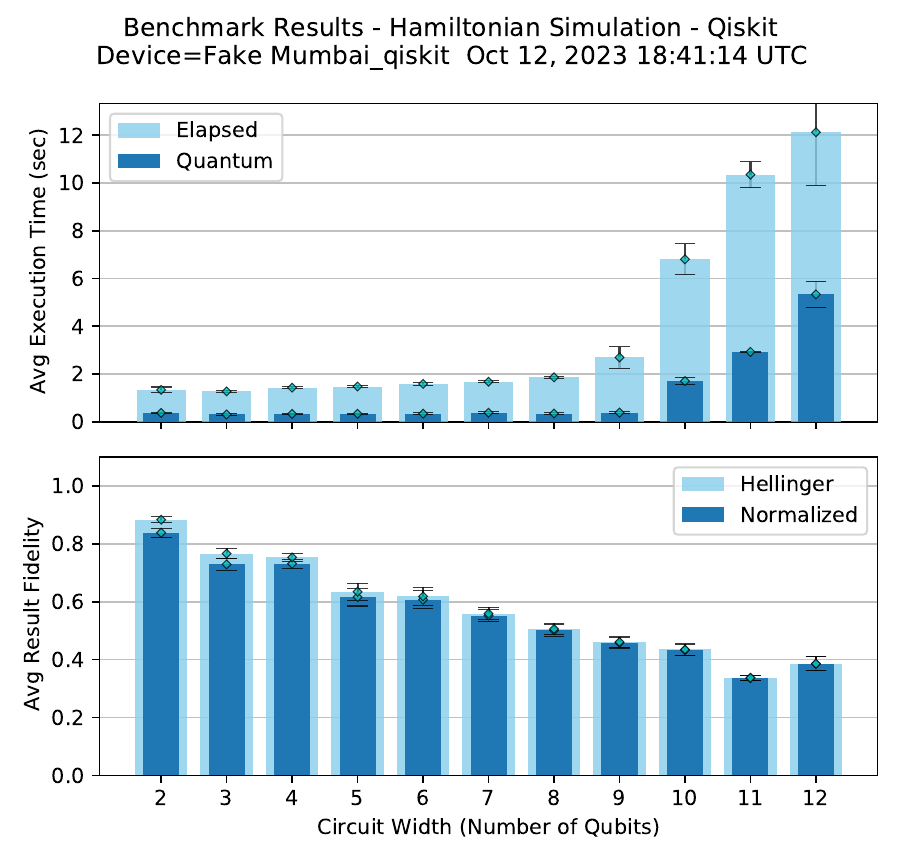}
        \caption{Hamiltonian Simulation with default optimization level 3}
    \end{subfigure}
    \caption{\textbf{Approximate Resynthesis TKET Circuit Optimization applied to Hamiltonian Simulation.} Hellinger and normalized fidelities are displayed over circuit width for the Hamiltonian Simulation benchmark circuits executed on the Qiskit Aer simulator, using a Fake Melbourne noise model. A significant increase in execution fidelity is visible in plot (a) as the approximate resynthesis algorithm produces circuits of shorter lengths, compensating for a decrease in the accuracy of the gate representation of the problem.}
    \label{fig:tket ham sim approx}
\end{figure}

\begin{figure}[t!]
    \centering
    \begin{subfigure}[b]{\columnwidth}
        \includegraphics[width=0.80\columnwidth]{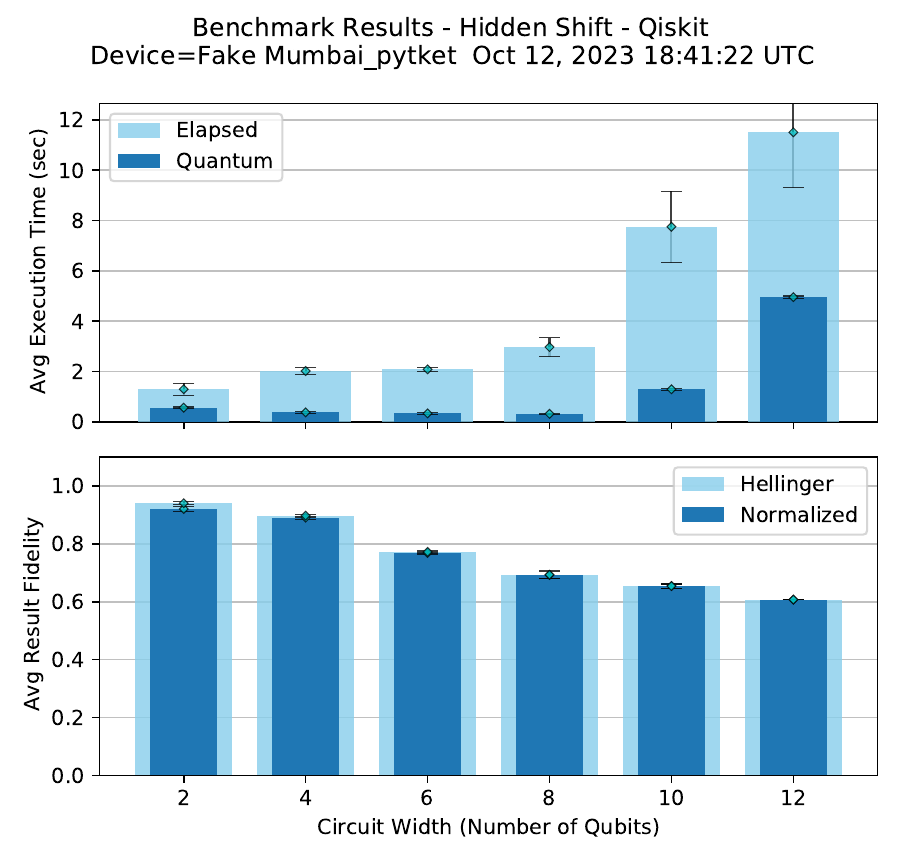}
        \caption{Hidden Shift with TKET Approximate Resynthesis.}
    \end{subfigure}
    \begin{subfigure}[b]{\columnwidth}
        \includegraphics[width=0.80\columnwidth]{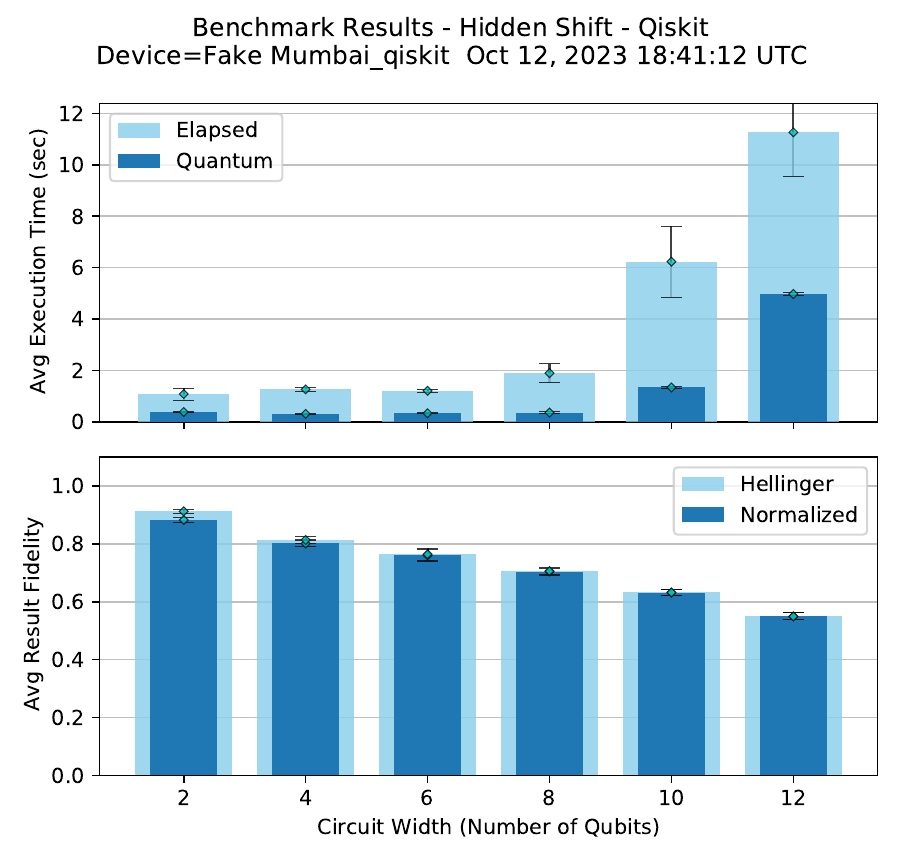}
        \caption{Hidden Shift with default optimization level 3}
    \end{subfigure}
    \caption{\textbf{Approximate Resynthesis TKET Circuit Optimization applied to Hidden Shift.} Hellinger and normalized fidelities are displayed over circuit width for the  Hidden Shift benchmark circuits executed on the Qiskit Aer simulator, using a Fake Melbourne noise model. No increase in execution fidelity is seen in plot (a) as the approximate resynthesis algorithm is unable to regroup any of the small number of gates in the Hidden Shift circuit, making no change to circuit depth, resulting in no improvement in fidelity.}
    \label{fig:tket hidden shift}
\end{figure}

It is expected that transformation passes on benchmark circuits will affect the performance when running application-oriented benchmarks~\cite{Mills_2021}, as the resulting circuits may be more or less vulnerable to noise.
Limits on the optimization techniques permitted when benchmarking have been explored~\cite{amico2023defining}, and prevent the use of some techniques. Here, we demand only that the optimization passes utilized are clearly reported in the results of these benchmarks, that the optimizations used are classically efficient to implement, and that barriers in the circuit are respected.

In the remainder of this section, we examine the impact the transformation technique ``approximate circuit resynthesis''.
Approximate resynthesis proceeds by merging sequences of two-qubit gates acting on the same two qubits into a single two-qubit unitary, and resynthesizing the resulting unitary. Two-qubit operations can always be synthesized using at most three CX gates and single-qubit gates, by a result known as the KAK or Cartan decomposition~\cite{tucci2005introduction}. The set of unitaries that can be generated by two CX gates and single-qubit gates is smaller than the set of all 2 qubit unitaries, but when such a circuit implements a unitary that is ``close enough'' to the desired one it may be preferable to use that \emph{approximate decomposition} instead of the perfect one~\cite{Cross_2019}. This is true if the error that results from using an approximate decomposition is greater than the error that would be introduced from implementing an additional CX gate on a noisy quantum computer. Approximate synthesis typically improves performance, since shorter depth circuits compensate for inaccuracies introduced by altering the unitary considered (by removing gates). Here we use the \texttt{KAKDecomposition} pass, available through pytket~\cite{Sivarajah_2020} as a means of performing such an approximate resynthesis.

A function is defined to implement the approximate resynthesis transformation on an input circuit and return a modified circuit. This function is specified via the \texttt{exec\_options} argument of a benchmark's \texttt{run} method. 
When the benchmark is executed, the transformation is applied to each circuit prior to its execution.
Both the fidelity score achieved on the benchmark as well as the execution time are metrics of concern. 

\vspace{0.2cm}

In~\autoref{fig:tket ham sim approx} and~\autoref{fig:tket hidden shift}, we show the fidelities and execution times obtained for two of the benchmarks, Hamiltonian Simulation and Hidden Shift, when executed over widths of 2 to 12 qubits, with approximate resynthesis applied. 
The Hamiltonian Simulation results are significantly improved while the Hidden Shift is essentially unchanged.

The Hamiltonian simulation implementation benefits the most from the approximate resynthesis since all two-qubit gates correspond to small angle ($\pi/300$) ZZ interactions. In current-generation quantum computers implementing such small interactions actually adds more errors to the overall circuit than simply skipping the gates. This is seen by comparing the average fidelity between the small angle ZZ interactions and the identity gate to the fidelity of typical two-qubit gates.
The use of approximate resynthesis is generically applicable, not just to Hamiltonian Simulation, but the advantage is particularly notable in this case due to the small angle gates.
Other QED-C benchmarks that benefit include QFT (1) and (2), quantum phase estimation, as well as HHL and the Hydrogen Lattice examples discussed above. Conducting such transformations comes at an additional, but reasonable time cost, as expected. In future work, we plan to implement other variants of Hamiltonian simulation circuits (and other benchmark circuits) that may limit the impact of such techniques.

Conversely, there is little change in performance in the case of the Hidden shift class of circuits, as seen in~\autoref{fig:tket hidden shift}. This particular class of circuits contains two-qubit sub-circuits separated by barriers. These two-qubit subcircuits contain only a single two-qubit gate between barriers, and therefore do not benefit from resynthesis passes leaving the fidelity unchanged.

\begin{figure}[t!]
    \centering
    \begin{subfigure}[b]{\columnwidth}
        \includegraphics[width=0.95\columnwidth]{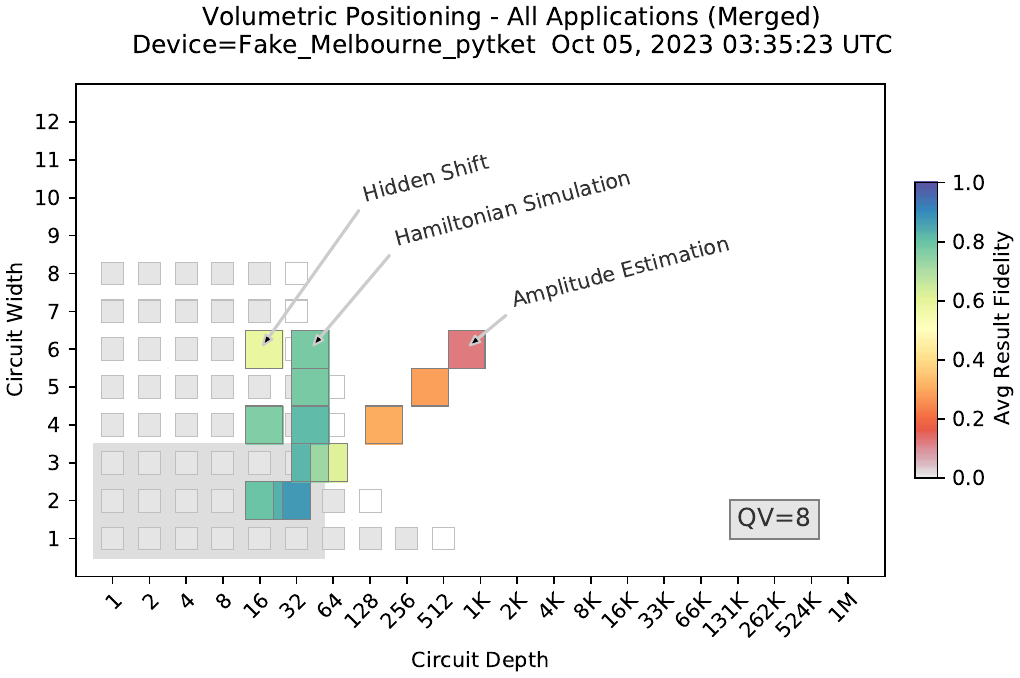}
        \caption{TKET circuit optimization - Approximate resynthesis.}
    \end{subfigure}
    \begin{subfigure}[b]{\columnwidth}
        \includegraphics[width=0.95\columnwidth]{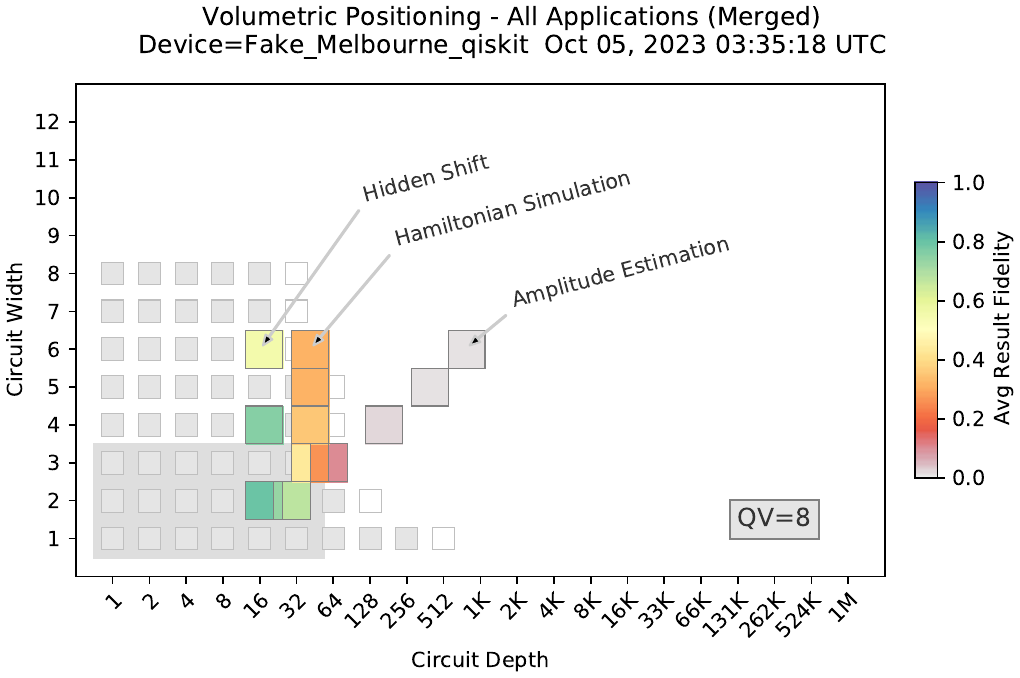}
        \caption{Default Qiskit circuit optimization - level 3.}
    \end{subfigure}
    \caption{\textbf{Volumetric plot of several benchmark circuits} These circuits were executed on the Qiskit Aer simulator, using the Fake Melbourne noise model. Quantum volume on the Melbourne system (now retired) was measured to be 8, suggesting that circuits wider than 3 qubits will have a low probability of success. Here, the impact of using the TKET approximate resynthesis method is visible for both the Hamiltonian Simulation and the Amplitude Estimation benchmark.}
    \label{fig:tket approx}
\end{figure}

In~\autoref{fig:tket approx}, we present a volumetric plot of the results from executing three of the benchmarks with and without the use of the approximate resynthesis transformation.
The improvement in the Hamiltonian Simulation and the lack of improvement in Hidden Shift fidelities is visible between the two. 
While we do not include the detailed bar charts here, our analysis shows that Amplitude Estimation (plotted in~\autoref{fig:tket approx}) and Monte Carlo (not plotted) also benefit from resynthesis across three-qubit sub-circuits (also available in pytket). In those tests, the target state is created only over two qubits with several subcircuits that consist of controlled operations between the target state and the ancilla qubit (three total qubits). Three qubit resynthesis significantly reduces the two-qubit gate count in these cases. In future work, we will scale the target state size in these examples with the number of qubits to avoid such a high reduction.


\subsection{Custom Execution Pipeline with Fire Opal}
\label{sec:optimization_fire_opal}
 
In this section, we discuss another enhancement developed for this work, a custom ``executor'' function that permits a user to take complete control of the execution pipeline.
This custom function receives an un-processed benchmark circuit object, processes and executes it under user control, and returns measurement results post-processed as desired.

To illustrate this enhancement, we developed a custom executor function that employs the Fire Opal package from Q-CTRL~\cite{q-ctrl} to perform transpilation, optimization, execution, and mitigation of errors during the execution of each benchmark circuit. 
The Fire Opal package is intended to be agnostic to the hardware backend and to the type of algorithm executed.
Given an arbitrary algorithm and hardware properties, such as device topology, connectivity, and backend data on qubit quality,
Fire Opal uses a deterministic approach to improve quantum hardware performance and suppress non-Markovian noise in the system without the use of sampling or randomization methods (described in~\cite{mundada_2022}.)

\vspace{0.3cm}

Here, we present an analysis of the results of executing two of the QED-C benchmarks within the Q-CTRL environment, configured to use this custom executor function.
To illustrate the types of operations that can be performed in a custom executor function, we describe the error-suppressing workflow implemented for Fire Opal, and performed on each benchmark circuit:

\begin{enumerate}
    \item A quantum circuit (Qiskit QuantumCircuit object), along with execution parameters, is passed to the function, where a Fire Opal front-end compiler reduces the circuit depth and transpiles it to the backend device topology using a sequence of mathematical identities. Fire Opal also accepts OpenQASM circuit definition.
    \item An error-aware hardware mapping function determines the best circuit layout to maximize performance on the physical device, using knowledge of qubit coherences and gate errors that typically vary across the device.
    Dynamical decoupling pulses are embedded to suppress crosstalk using a context-aware ranking protocol.
    \item If the gate fidelities are lower than the coherence limit as calculated by the $T_1$ and $T_2$ times provided by the backend, then optimal pulse control (\cite{Baum2021, Carvalho2021} is autonomously deployed to replace the hardware-level instructions for the specified gate.
    \item The user circuit is converted to the API supported by the specified backend (i.e. Pyquil, Qiskit pulse, AWS Braket, etc). The circuit is then executed and the results are returned for further processing. 
    \item In post-processing, measurement errors from the obtained probability distribution are eliminated using a protocol that utilizes a set of pre-measured confusion matrices. The details of this error mitigation protocol are available in Appendix C-4 of Ref.\cite{mundada_2022}
\end{enumerate}

\begin{figure}[t!]
    \centering
    \begin{subfigure}[b]{\columnwidth}
        \includegraphics[width=0.86\columnwidth]{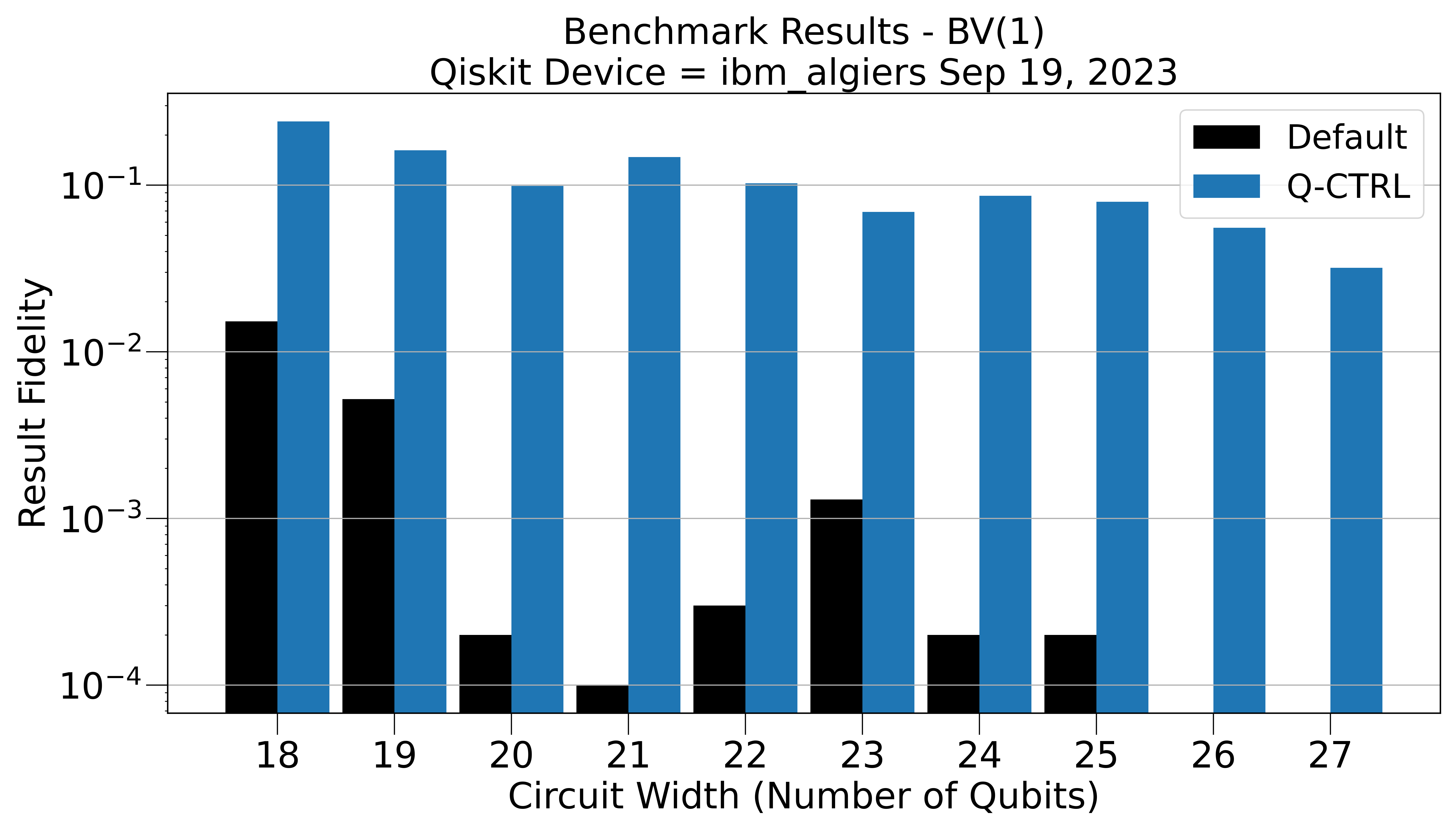}
    \end{subfigure}
    \begin{subfigure}[b]{\columnwidth}
        \includegraphics[width=0.86\columnwidth]{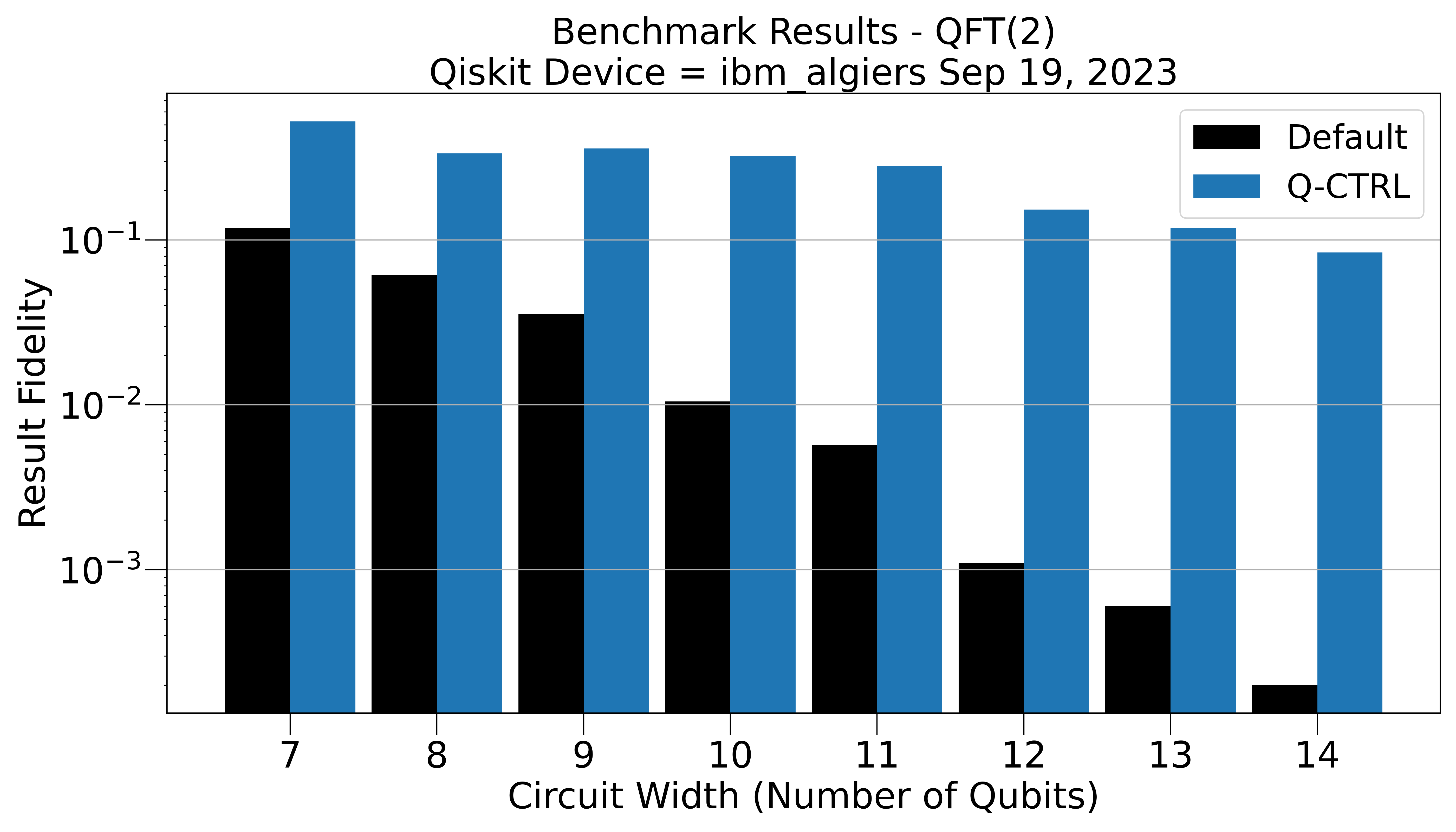}
    \end{subfigure}
    \caption{\textbf{Benchmark execution on the 27Q IBM Quantum Algiers backend.} (In logscale) (a) Normalized fidelity of the BV (1) circuits executed using default sampler primitive (black) is below 1\% beyond 18Q and is 0\% for 26Q and 27Q. Using the error suppression in Q-CTRL Fire Opal (blue) gives fidelities ranging from 25\% down to 5\% for 27 qubits. (b) Normalized fidelities for QFT (2) using default sample (black) shows an exponential decay to 0.01\% for 14Q. Using Fire Opal (blue) gives fidelities ranging from 50\% (for 7Q) to 9\% (for 14Q).
    (\emph{Data provided by Q-CTRL}.)
    }
    \label{fig:qctrl_bv_qft}
\end{figure}

To understand the impact of error-suppression techniques on performance,
we compare the result fidelity of the Bernstein-Vazirani BV (1) and the QFT (2) benchmarks run on the IBM quantum processor \textit{ibm\_algiers} via the default sampler primitive (black) and Q-CTRL's Fire Opal software (blue) in~\autoref{fig:qctrl_bv_qft}. In this experiment, we focus specifically on circuits with increasing qubit widths, where the result fidelity obtained with default settings is very low. We highlight the improvement in result fidelity that can be obtained in this region of circuit widths when the Fire Opal executor function is invoked.
The default sampler primitive includes an \textit{optimization level} = 3 and \textit{resilience level} = 1. We pass exactly the same input circuits to both methods.

For BV (1), we examine results from 18 qubits to 27 qubits.
We modified this benchmark to use a target string of all ``1''s, instead of random inputs, as it results in the largest depth circuits. 
The fidelity of the BV (1) circuits executed using the default sampler primitive (black) is below 1\% beyond 18Q and we never observe the target bitstring for 26Q and 27Q over 10,000 shots. Using the error suppression in Q-CTRL Fire Opal (blue) gives improved fidelities ranging from 25\% to 5\% for 27 qubits.

For QFT (2), we examine results from 7 to 14 qubits, where the default result fidelity is below 10\% and as low as 
0.01\% for 14Q.
Fire Opal (blue) gives improved fidelities ranging from 50\% (for 7Q) down to 9\% (for 14Q). It is important to note that the magnitude of improvement in fidelity varies with the type and size of circuits. Typically, benefits are expected to increase with circuit width and depth~\cite{mundada_2022}.

This demonstrates how the QED-C benchmark framework allows users to test additional error suppression features like Fire Opal to improve result fidelity.
However, in this exercise, we do not present measurements of the run-time cost of this additional processing.
We plan to do this in a future work.




\section{Benchmarking Machine Learning Applications}
\label{machine_learning_benchmark}

Applications in machine learning have widespread utility, revolutionizing data analysis and decision-making across various domains, and recently being extended to creative tasks as well. However, the heavy computational requirements of classical machine learning make it an attractive area for the development of quantum approaches.
While early quantum machine learning algorithms required large numbers of qubits and gates which are beyond the capabilities of current generation quantum computers, much recent algorithmic development has focused on near-term heuristic approaches which are possible to test on real quantum hardware.

In this section, we present a framework for benchmarking these emerging quantum machine learning algorithms, using a representative quantum algorithm for binary classification of images within a large dataset.
While there have been many benchmark frameworks developed for quantum machine learning applications~\cite{Nguyen_2019,West_2023,kiwit2023applicationoriented,Benedetti_2019},
our work generalizes an approach to be consistent with our benchmarking of other classes of applications.
This is designed to provide consistency and validation of results across application domains and to enable performance-driven exploration of new types of machine learning applications.


\subsection{Quantum Machine Learning}
\label{sec:quantum_machine_learning}

Machine learning involves analyzing existing data to make predictions when presented with new information. Some commonly encountered examples of machine learning are image and speech recognition, product recommendation, and anomaly detection. Machine learning can be supervised, that is, based on a labeled dataset, or unsupervised, where the dataset is unlabeled. A further distinction can be made between discriminative, which seeks to maximize the ability of an algorithm to assign a class to unseen data, and generative learning, which seeks to produce new data that could plausibly belong to the original dataset. 

Quantum computers provide a new paradigm for machine learning algorithms. Near-term quantum machine learning generally relies on using parameterized circuits to efficiently capture the correlations between different variables in a dataset which can be optimized according to a desired outcome. 
Theoretical work suggests favorable generalization properties for such quantum models~\cite{peters2022generalization, caro2022outofdistribution, Caro2022-ae} and arguments for quantum advantage are presented based on the expressivity of these models~\cite{Lukin2022, bowles2023contextuality,PhysRevResearch.4.043092}.

Notable experimental results in quantum machine learning include training quantum-enhanced generative adversarial networks (GANs) on a variety of image generation and multivariate problems~\cite{Rudolph20,PhysRevApplied.16.024051,silver2023mosaiq,PhysRevResearch.4.043092,zhu2022copulabased},
and using quantum algorithms, transformers, or ensembles to perform image classification tasks of varying complexity~\cite{Johri20,cherrat2022quantum,Silver_Patel_Tiwari_2022}.
In addition to the potential for quantum advantage, theoretical and numerical work has established the possibility of effective training of certain classes of parameterized quantum circuits, including Quantum Convolutional Neural Networks~\cite{PhysRevX.11.041011} and orthogonal quantum circuits~\cite{kerenidis2022classical}.


\subsection{Image Recognition Quantum Algorithm}

For our benchmark, we selected a simple binary image classification problem that uses a database consisting of images of two digits, 7 or 9, as the source of data.
Each image is labeled to identify the class, 7 or 9, to which it belongs. 
The challenge is to execute and benchmark a quantum algorithm that can recognize the class of an unknown image. 
To accomplish this, we first use a subset of the images in `training' mode by repeatedly encoding each of these images into an ansatz circuit and executing a variational algorithm to search for parameters that maximize the classification accuracy of the images.
These parameters can later be used to identify any unknown encoded image by re-executing the circuit with these parameters.

In the algorithm used, each data point (image) is uploaded to the quantum computer one at a time, after which it is acted upon by the parameterized quantum circuit. After measurement, the output is classically processed and submitted to a loss function. This procedure effectively encodes a Fourier series in the data, whose coefficients depend on the parameterized circuit and frequencies on the data encoding procedure~\cite{Schuld_data_encoding}. Since Fourier functions are known to be universal function approximators quantum models can be used to model any kind of data. 

Specifically, the classification of the $k$-th image with pixel values $\{i, p^{(k)}_i\}$, where $i\in[1,n_0]$ is the pixel index and $p^{(k)}_i$ is the pixel value which lies between 0 and 1, proceeds as follows:
\begin{enumerate}[(i)]
\item
Compressing the image using Principal Component Analysis to a vector of size equal to the number $n$ of available qubits: 
$\{p^{(k)}_i\}\rightarrow\{q^{(k)}_j\}$ with $j\in[1,n]$ for $n<n_0$.
\item
Loading into a quantum state with $n$ qubits using product state encoding, $\ket{\Psi_0(\{q_j^{(k)}\})}=\prod_{j=1}^n \exp(i2\pi q^{(k)}_jX_j)\ket{0\ldots 0}$
\item
Acting on it with a parameterized unitary $U(\theta)$ based on the quantum convolutional neural network~\cite{Cong2019,Hur_2022} with the circuit tapering towards the $0$-th qubit to give the state
$U(\vec{\theta})\ket{\Psi(\{q_j^{(k)}\})}$
\item
Measuring operator $Z_0$ (i.e. Pauli $Z$ acting on the 0-th qubit) on the resulting state and converting that measurement into a prediction for the image class:
$m_k=f\bigg(\bra{\Psi_0(\{q_j^{(k)}\})} U(\vec{\theta})^{\dagger} Z_0 U(\vec{\theta})\ket{\Psi_0(\{q_j^{(k)}\})}\bigg)$, where $f$ is a simple classical function
\end{enumerate}

While training the classifier, the prediction is used to calculate a mean square loss function for image $k$ against the provided label $y_k$, $l_k=|y_k-m_k|^2$ which is input to a mean loss for the whole training dataset $L=\frac{1}{N_{\text{train}}}\sum_{k=1}^{N_{\text{train}}} l_k$, where $N_{\text{train}}$ is the size of the training dataset. This loss function is used by an optimizer to iteratively change the parameters to minimize the loss. While in validation (also known as testing) mode, the prediction is used to calculate how well the classifier performs. Finally, the machine learning algorithm can be deployed in inference mode on new data.


\subsection{Image Recognition Benchmark Implementation}

This image recognition benchmark makes use of a publicly available MNIST image database, `mnist\_784'~\cite{dbmnist784}. The digits in this database are size-normalized and centered in a fixed-size 28x28 pixel image.
We use the OpenML `sklearn' python package~\cite{sklearn} to load the database and extract only the images for the digits 7 and 9 along with their labels.
From a specified number of images to process, 
we select 80\% for a training set, with the remaining 20\% used for testing the effectiveness of the training.  

Once the images have been loaded, the benchmark can be executed in three different modes, or `methods', each exercising a different aspect of the benchmark.
All three modes produce a series of plots displaying the benchmarking results which are described below.

Method (1) is designed to characterize the result fidelity of ansatz execution, as with the other iterative algorithms in the suite (e.g. maxcut, hydrogen-lattice).
The quantum circuit that encodes a single image for classification is executed with random input parameters and its output is evaluated against the output from an ideal simulator. This can be done for different numbers of qubits, and it produces results similar to the non-iterative benchmarks, including a volumetric plot and bar charts displaying execution times, circuit depths, and fidelities.

\begin{algorithm}[t!]
\caption{Benchmark Algorithm for Quantum Image Recognition (Training Mode, method 2)}\label{alg:cap-ir}
\begin{algorithmic}[1]
\State $target \gets \textcolor{cyan}{backend\_id}$
\State initialize\_metrics()
\State $image\_data, image\_label \gets load\_mnist\_data()$
\State $image\_data\_batch, image\_label\_batch \gets create\_batch(\textcolor{cyan}{batch\_size}, image\_data, image\_label)$
\For{$qubit\_count \gets \textcolor{cyan}{min\_count}, \textcolor{cyan}{max\_count}$}
    
    \State $params[\bm{\alpha}] \gets random(num\_params)$ 
    \While{\textit{num$\_$iteration not done} }\Comment{minimizing}
        \For{$data\_point \gets {image\_data\_batch}$}
            \State $feature\_map\_circuit \gets create\_feature\_map(qubit\_count, data\_point)$
            \State $parameterized\_circuit \gets create\_ansatz\_circuit(qubit\_count)$
            \State $merged_circuit \gets compose\_circuits(feature\_map\_circuit,  parameterized\_circuit)$
            \State $cached\_circuit \gets compile\_circuit(merged_circuit)$
            \State $circuit \gets apply\_params(cached\_circuit, params)$
            \State $\textcolor{blue}{counts }\gets execute(target, circuit, \textcolor{cyan}{num\_shots})$
            \State $prediction  \gets calculate\_prediction(counts)$
            \State $prediction\_batch  \gets append(prediction)$
        \EndFor
            
        \State $loss, accuracy \gets loss\_function(prediction\_batch, image\_data\_batch)$
        \State $store\_iteration\_metrics(loss, accuracy)$
        \State $params[\bm{\alpha}] \gets optimize(params[\bm{\alpha}])$
        \State $image\_data\_batch, image\_label\_batch \gets update\_batch(batch\_size, image\_data, image\_label)$ \Comment{no repetition}
        \State $done \gets \textit{True if iterations completed}$

    \EndWhile

    \State compute\_and\_store\_restart\_metrics()
    \State compute\_and\_store\_group\_metrics()
\EndFor

\end{algorithmic}
\label{alg:image recognition}
\end{algorithm}

\begin{figure*}[!t]
    \centering
    \begin{minipage}{\textwidth}
        \includegraphics[width=0.84\linewidth]{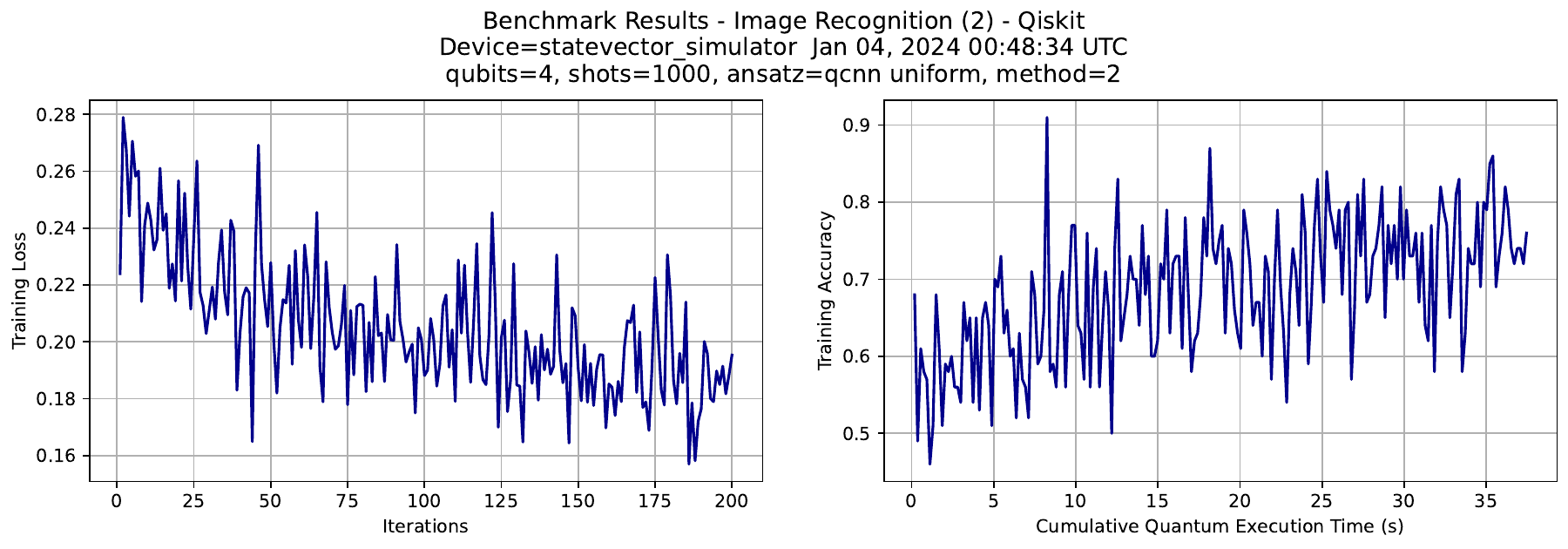}
    \end{minipage}
    \vspace{1cm} 
    \begin{minipage}{\textwidth}
        \includegraphics[width=0.84\linewidth]{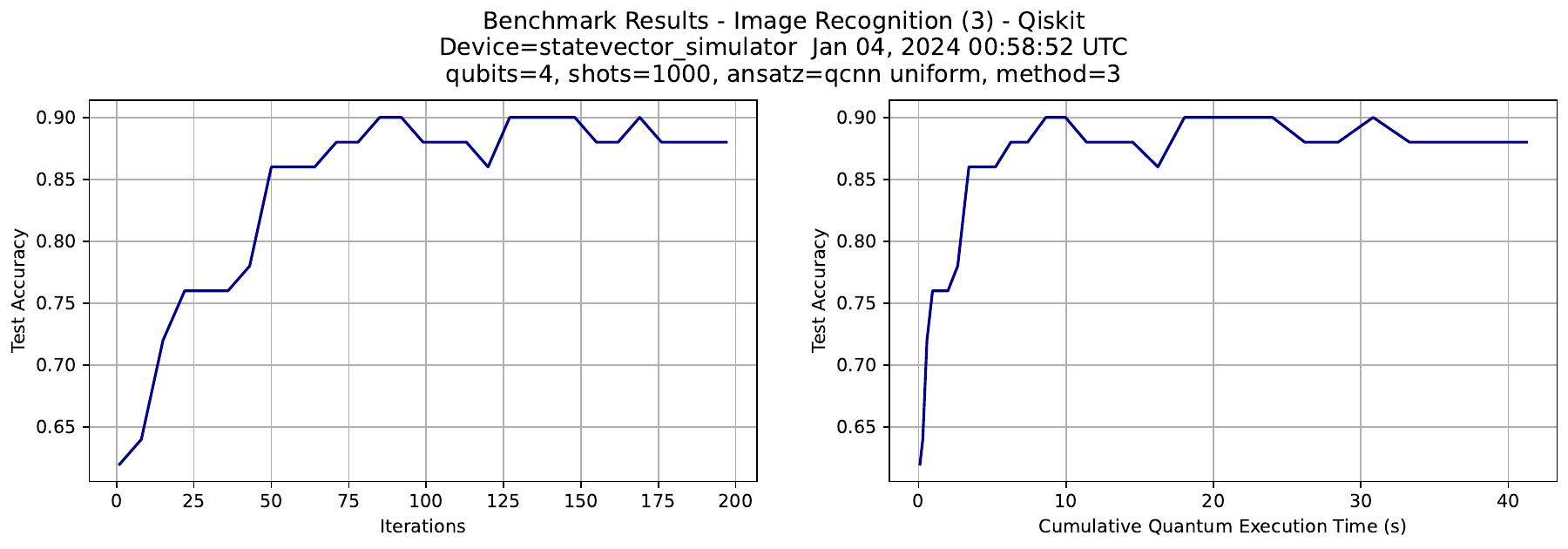}     
    \end{minipage}
    \vspace{-.03cm}
    \caption{\textbf{Progression of Training and Test Accuracies.} Here we show results from executing the Image Recognition benchmark methods 2 and 3, the training pass and the testing pass, respectively.
    As the training progresses, the classification accuracy increases, and the loss decreases for both the training and test datasets. The benchmarking suite evaluates the mean training loss after every iteration of the optimizer. The test loss can be calculated after the training at parameter values that are stored periodically after a set number of iterations. Here, 200 images are used for training and 50 for testing.}
    \label{fig:image_accuracy_loss_q_4}
\end{figure*}

Method (2) is the `training pass', in which the images in the training subset are processed by the variational algorithm to find optimal parameters for the given problem dataset.
Algorithm~\ref{alg:image recognition} outlines the pseudocode for the operation of this training mode.

The benchmark loops over a specified range of qubit widths, as in other benchmarks. 
For each width, and under the control of a classical optimizer, the training images undergo classical pre-processing and are compressed so that they can be encoded into a small number of qubits at that width.
For each image, an ansatz circuit, with image features encoded, is created and executed with trial parameters.
An aggregate loss, expressed as a distance from the expected classifications, is calculated for all the images in the set.
This loss value is returned to the optimizer which attempts to find the parameters for the ansatz that result in the lowest loss. The benchmark tracks the progress of the algorithm as it attempts to converge on the lowest loss value.
 
The Simultaneous Perturbation Stochastic Approximation (SPSA) optimizer is chosen for its ability to perform multi-parameter optimization where the number of function calls is independent of the number of parameters.
This is important here since many images need to be evaluated at each optimizer iteration.  In principle, the loss function could be evaluated as an average over all the training images at each iteration. However, this would make training very expensive. In practice, we find that a stochastic sampling technique, where a number of images equal to a fixed batch size are evaluated at each iteration of the optimizer works just as well.
In this work, we use a default batch size of 50.

Method (3) is the ‘validation’ pass, in which the benchmark performs inference on the remaining images, known as the test dataset. The ansatz is executed once for each of the images, using the parameters obtained during the training pass.
The distance of the resulting classifications from the expected classifications results in a loss that acts as a measure of how well this quantum classification algorithm works when trained.  This value can be compared against classical solutions using the same problem set.

While the overall framework remains the same as for other variational algorithms, some key differences distinguish this benchmark. First, the classical data, in this case images, need to be effectively compressed before loading into the quantum computer. Secondly, a large number of images need to be processed at each iteration corresponding to which a large number of quantum circuits need to be run on the quantum computer or simulator. Finally, even validating the efficacy of the training involves running the classification algorithm on a test dataset that itself requires many calls to the quantum computer or simulator.


\subsection{Simulation Results and Analysis}

\begin{figure}[!t]
    \includegraphics[width=0.84\columnwidth]{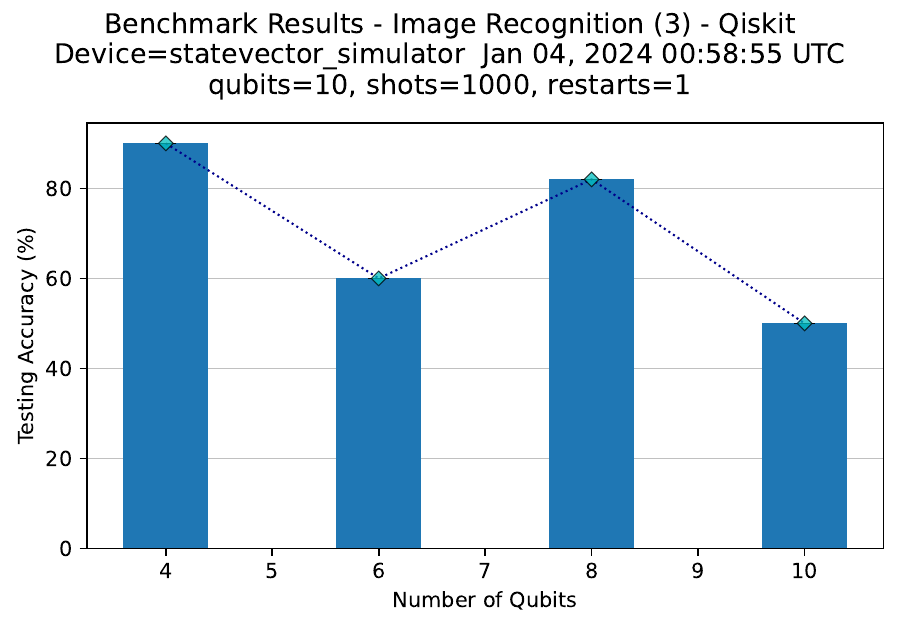}
    \caption{\textbf{Maximum Training Accuracies.} Here we show the accuracy of the algorithm on the training dataset as a function of the number of qubits when the number of training iterations is fixed at 200 and the size of the training dataset is 200.} 
    \label{fig:image_train_accuracy_q_4_8}
\end{figure}

\begin{figure}[!t]
    \includegraphics[width=0.84\columnwidth]{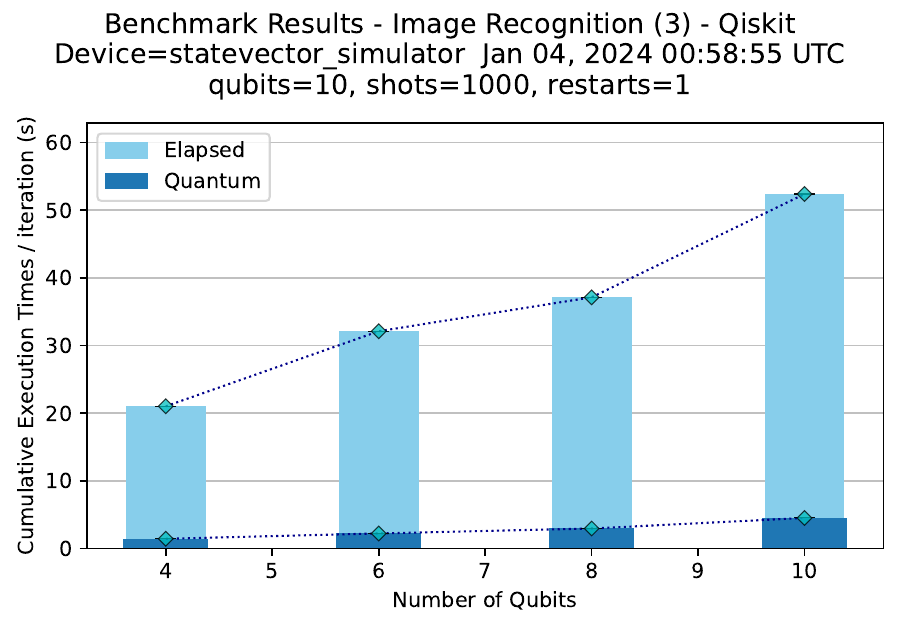}
    \caption{\textbf{Training Times.} Here we show the execution time of the algorithm on the training dataset as a function of the number of qubits when the number of training iterations is fixed at 200 and the size of the training dataset is 200.} 
    \label{fig:image_train_time_q_4_8}
\end{figure}

The benchmark was tested in simulation using the Qiskit statevector simulator. Some of the output graphs are shown in~\autoref{fig:image_accuracy_loss_q_4} -~\autoref{fig:image_train_time_q_4_8}.
In~\autoref{fig:image_accuracy_loss_q_4} (top), the measured loss function decreases as the training of the classifier circuit progress. At the same time, the accuracy of the circuit in classifying images of the training dataset increases. The corresponding behavior is also seen on the test dataset (bottom). The test accuracy is measured less frequently than the training accuracy since it is not essential for training, but rather a measure of how it is progressing. We see that the curves oscillate around an increasing or decreasing overall trend. This oscillation is due to the stochastic nature of the batch construction as well as the use of the SPSA optimizer.

\autoref{fig:image_train_accuracy_q_4_8} shows the training accuracy that is achieved for 4 through 10 qubits after 200 iterations. Here the maximum training accuracy is defined as an average of the 5 highest accuracies that are achieved on the training dataset during the training. The accuracy overall trends downwards as the number of qubits increases, indicating that models with a larger number of qubits take longer to train. Additionally, the models with 4 and 8 qubits have increased accuracy which is a result of using the quantum convolutional neural network, which has a symmetry when the number of qubits is a power of 2.

\autoref{fig:image_train_time_q_4_8} shows the training times for 4 through 10 qubits for executing 200 iterations of the optimizer. It is notable that for execution on the simulator, the elapsed time is considerably longer than the quantum time, indicating that more optimized software for handling the construction and execution of parameterized quantum circuits could have a significant impact on the feasibility of quantum machine learning which involves running batches of such circuits.

\begin{figure}
    \centering
    \begin{minipage}{\columnwidth}
        \includegraphics[width=0.84\linewidth]{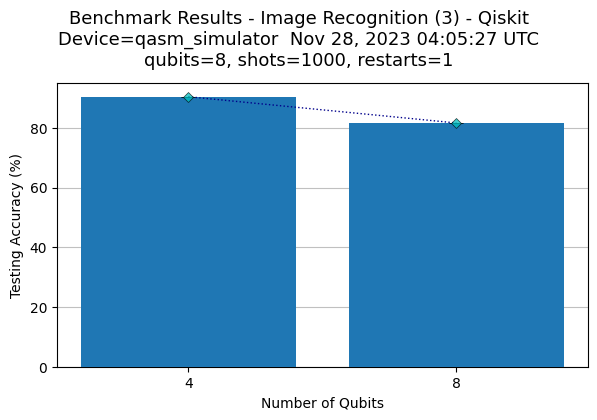}
    \end{minipage}
    \begin{minipage}{\columnwidth}
        \includegraphics[width=0.84\linewidth]{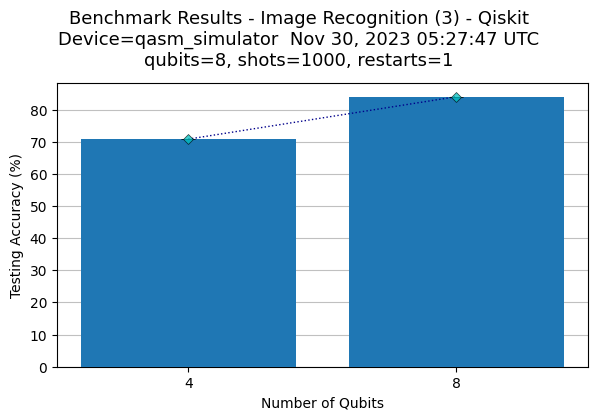}
    \end{minipage}
    \begin{minipage}{\columnwidth}
        \includegraphics[width=0.84\linewidth]{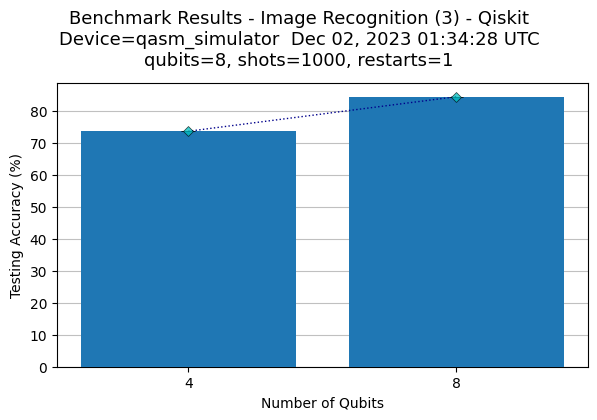}
    \end{minipage}
\caption{\textbf{Effect of Training Dataset Size.} Here is an example of the hyperparameter exploration that can be done with the benchmarking framework. The top panel shows the accuracy on the test dataset when $N_{\text{train}}=200$, while the middle panel shows the results when $N_{\text{train}}=14201$. For both the top and middle panels, the number of iterations is 200. The bottom panel shows the results when $N_{\text{train}}=14201$ and number of iterations is 400.}
\label{fig:training_size}
\end{figure}

The benchmarking framework allows us to test the effect of different hyperparameters on the training. For instance,~\autoref{fig:training_size} shows the effect of the size of the training dataset. The top panel shows the accuracy on the test dataset when the number of training images is fixed at 200, while the middle one shows the results when all the available images in MNIST for our binary classification problem that are not in the test dataset, numbering 14201, are used. We see that for the smaller dataset size, the test accuracies are larger for 4 qubits, while the accuracy of the 8 qubit ansatz stays about the same in both cases. In the bottom panel, we see that the accuracy for the larger dataset for 4 qubits increases when we increase the number of iterations, indicating that the larger dataset size makes it harder for the optimizer to converge at a fixed number of iterations.


\section{Summary and Conclusions}
\label{sec:summary-and-conclusions}

Cloud-accessible quantum computers are attracting a wide audience of potential users, but the challenges in understanding their capabilities create a significant barrier to the adoption of quantum computing. The QED-C Application-Oriented Benchmark suite makes it easy for new users to assess a machine's ability to implement applications, and its volumetric visualization of the results is designed to be simple and intuitive.
In this manuscript, we described several enhancements made to this benchmark suite.
This is an ongoing effort, established under the direction of the QED-C (Quantum Economic Development Consortium), with work organized and managed by QED-C members within the broad community of quantum computing system providers and quantum software developers.

In enhancing the suite and developing new benchmarks, we have expanded the framework to offer greater control over the properties and configuration of the applications used as benchmarks. 
It makes it possible, not only to increase coverage of the execution landscape but to explore various algorithmic variations.
One reason to do this is to determine whether there are certain applications or variants of them that perform better on a specific class of hardware.
We anticipate that our work will facilitate the adoption of quantum computing, and encourage economic development within the industry.

\vspace{0.3cm}

In this work, we introduced several new benchmarks, the first based on a scalable version of the HHL linear equation solver that illustrates how variations on the algorithm impact its volumetric profile and add coverage to the benchmark suite.
The second evaluates the performance of a VQE algorithm that finds the ground state energy of a hydrogen lattice simulation, with a new methodology for analyzing the quality/run-time trade-off and two new normalized measures of chemical accuracy, the accuracy ratio, and the solution quality.

This was followed by a review of new options for inserting custom circuit preparation, execution, and error mitigation procedures. We show how these enable the use of new tools such as the Qiskit Sampler, TKET, and Q-CTRL's Fire Opal to improve the results obtained from running the benchmark results.
Lastly, we presented an early-stage version of a machine learning algorithm, a simple image classification problem, that defines another application-specific measure and executes many more circuits iteratively than does the VQE benchmark.

The updated QED-C benchmarks provide extensive support for measuring the run-time costs associated with the execution of both simple and variational algorithms.
In evaluating the performance of quantum computers, a precise understanding of the practical trade-offs in run-time cost versus quality of solution is essential to evaluating the total cost of ownership for a solution technology.

\vspace{0.3cm}

A primary goal of this effort is to identify ways in which this highly flexible framework could be made available and extended easily to new quantum applications.
Surfacing these benchmarks early offers potential users and investors the assurance that progress is being made in advancing the technology. 
To truly develop confidence in the technology, users must be able to execute simple programs easily and gauge the performance of these programs as the machine improves in performance.
That is the purpose of this suite of benchmarks and the effort described in this paper.

Looking to the future, we envision proposals for additional efforts to facilitate the exploration of algorithmic options and their impact on performance in this continually evolving benchmarking framework.

\section*{Code Availability}
\label{sec:data_and_code_availability}

The code for the benchmark suite described in this work is available at 
\href{https://github.com/SRI-International/QC-App-Oriented-Benchmarks}{https://github.com/SRI-International/QC-App-Oriented-Benchmarks}.
Detailed instructions are provided in the repository. 


\section*{Acknowledgement}

The Quantum Economic Development Consortium (QED-C), a group of commercial organizations, government institutions, and academia formed a Technical Advisory Committee (TAC) to study the landscape of standards development in quantum technologies and to identify ways to encourage economic development through standards. In this context, the Standards TAC undertook to create the suite of Application-Oriented Performance Benchmarks for Quantum Computing as an open source project, with contributions from many members of the QED-C involved in Quantum Computing.
We thank the many members of the QED-C for their valuable input in reviewing and enhancing this work.

We acknowledge the use of IBM Quantum services for this work. The views expressed are those of the authors and do not reflect the official policy or position of IBM or the IBM Quantum team.
IBM Quantum. https://quantum-computing.ibm.com/, 2023.
We acknowledge Quantinuum for contributing the results from their commercial H1-1 hardware.
We thank Q-CTRL for supplying the environment and performing the execution of several of the QED-C benchmarks.


\clearpage
\appendix
\clearpage





\clearpage

\bibliographystyle{unsrtnat}  
\bibliography{references}

\begin{thebibliography}{102}
\providecommand{\natexlab}[1]{#1}
\providecommand{\url}[1]{\texttt{#1}}
\expandafter\ifx\csname urlstyle\endcsname\relax
  \providecommand{\doi}[1]{doi: #1}\else
  \providecommand{\doi}{doi: \begingroup \urlstyle{rm}\Url}\fi

\bibitem[Knill et~al.(2008)Knill, Leibfried, Reichle, Britton, Blakestad, Jost,
  Langer, Ozeri, Seidelin, and Wineland]{PhysRevA.77.012307}
E.~Knill, D.~Leibfried, R.~Reichle, J.~Britton, R.~B. Blakestad, J.~D. Jost,
  C.~Langer, R.~Ozeri, S.~Seidelin, and D.~J. Wineland.
\newblock Randomized benchmarking of quantum gates.
\newblock \emph{Phys. Rev. A}, 77:\penalty0 012307, Jan 2008.
\newblock \doi{10.1103/PhysRevA.77.012307}.
\newblock URL \url{https://link.aps.org/doi/10.1103/PhysRevA.77.012307}.

\bibitem[Magesan et~al.(2011)Magesan, Gambetta, and
  Emerson]{PhysRevLett.106.180504}
Easwar Magesan, J.~M. Gambetta, and Joseph Emerson.
\newblock Scalable and robust randomized benchmarking of quantum processes.
\newblock \emph{Phys. Rev. Lett.}, 106:\penalty0 180504, May 2011.
\newblock \doi{10.1103/PhysRevLett.106.180504}.
\newblock URL \url{https://link.aps.org/doi/10.1103/PhysRevLett.106.180504}.

\bibitem[Blume-Kohout et~al.(2017)Blume-Kohout, Gamble, Nielsen, Rudinger,
  Mizrahi, Fortier, and Maunz]{Blume-Kohout2017-no}
Robin Blume-Kohout, John~King Gamble, Erik Nielsen, Kenneth Rudinger, Jonathan
  Mizrahi, Kevin Fortier, and Peter Maunz.
\newblock Demonstration of qubit operations below a rigorous fault tolerance
  threshold with gate set tomography.
\newblock \emph{Nat. Commun.}, 8:\penalty0 14485, February 2017.
\newblock \doi{10.1038/ncomms14485}.
\newblock URL \url{https://www.nature.com/articles/ncomms14485}.

\bibitem[Cross et~al.(2019)Cross, Bishop, Sheldon, Nation, and
  Gambetta]{Cross_2019}
Andrew~W. Cross, Lev~S. Bishop, Sarah Sheldon, Paul~D. Nation, and Jay~M.
  Gambetta.
\newblock Validating quantum computers using randomized model circuits.
\newblock \emph{Physical Review A}, 100\penalty0 (3), sep 2019.
\newblock \doi{10.1103/physreva.100.032328}.
\newblock URL \url{https://doi.org/10.1103%2Fphysreva.100.032328}.

\bibitem[Boixo et~al.(2018)Boixo, Isakov, Smelyanskiy, Babbush, Ding, Jiang,
  Bremner, Martinis, and Neven]{Boixo_2018}
Sergio Boixo, Sergei~V. Isakov, Vadim~N. Smelyanskiy, Ryan Babbush, Nan Ding,
  Zhang Jiang, Michael~J. Bremner, John~M. Martinis, and Hartmut Neven.
\newblock Characterizing quantum supremacy in near-term devices.
\newblock \emph{Nature Physics}, 14\penalty0 (6):\penalty0 595–600, Apr 2018.
\newblock ISSN 1745-2481.
\newblock \doi{10.1038/s41567-018-0124-x}.
\newblock URL \url{http://dx.doi.org/10.1038/s41567-018-0124-x}.

\bibitem[Proctor et~al.(2020)Proctor, Rudinger, Young, Nielsen, and
  Blume-Kohout]{proctor2020measuring}
Timothy Proctor, Kenneth Rudinger, Kevin Young, Erik Nielsen, and Robin
  Blume-Kohout.
\newblock Measuring the capabilities of quantum computers, 2020.

\bibitem[Wack et~al.(2021)Wack, Paik, Javadi-Abhari, Jurcevic, Faro, Gambetta,
  and Johnson]{wack_clops_2021}
Andrew Wack, Hanhee Paik, Ali Javadi-Abhari, Petar Jurcevic, Ismael Faro,
  Jay~M. Gambetta, and Blake~R. Johnson.
\newblock Quality, speed, and scale: three key attributes to measure the
  performance of near-term quantum computers, 2021.
\newblock URL \url{https://arxiv.org/abs/2110.14108}.

\bibitem[Quetschlich et~al.(2023)Quetschlich, Burgholzer, and
  Wille]{Quetschlich2023mqtbench}
Nils Quetschlich, Lukas Burgholzer, and Robert Wille.
\newblock {MQT} {B}ench: {B}enchmarking {S}oftware and {D}esign {A}utomation
  {T}ools for {Q}uantum {C}omputing.
\newblock \emph{{Quantum}}, 7:\penalty0 1062, July 2023.
\newblock ISSN 2521-327X.
\newblock \doi{10.22331/q-2023-07-20-1062}.
\newblock URL \url{https://doi.org/10.22331/q-2023-07-20-1062}.

\bibitem[Tomesh et~al.(2022)Tomesh, Gokhale, Omole, Ravi, Smith, Viszlai, Wu,
  Hardavellas, Martonosi, and Chong]{tomesh2022supermarq}
Teague Tomesh, Pranav Gokhale, Victory Omole, Gokul~Subramanian Ravi,
  Kaitlin~N. Smith, Joshua Viszlai, Xin-Chuan Wu, Nikos Hardavellas,
  Margaret~R. Martonosi, and Frederic~T. Chong.
\newblock Supermarq: A scalable quantum benchmark suite, 2022.

\bibitem[Mesman et~al.(2022)Mesman, {Al-Ars}, and M{\"o}ller]{Mesman2022}
Koen Mesman, Zaid {Al-Ars}, and Matthias M{\"o}ller.
\newblock {{QPack}}: {{Quantum Approximate Optimization Algorithms}} as
  universal benchmark for quantum computers, April 2022.
\newblock URL \url{https://arxiv.org/abs/2103.17193}.

\bibitem[Donkers et~al.(2022)Donkers, Mesman, Al-Ars, and Möller]{Donkers2022}
Huub Donkers, Koen Mesman, Zaid Al-Ars, and Matthias Möller.
\newblock Qpack scores: Quantitative performance metrics for
  application-oriented quantum computer benchmarking, 2022.
\newblock URL \url{https://arxiv.org/abs/2205.12142}.

\bibitem[Finzgar et~al.(2022)Finzgar, Ross, Holscher, Klepsch, and
  Luckow]{Finzgar_2022}
Jernej~Rudi Finzgar, Philipp Ross, Leonhard Holscher, Johannes Klepsch, and
  Andre Luckow.
\newblock {QUARK}: A framework for quantum computing application benchmarking.
\newblock In \emph{2022 {IEEE} International Conference on Quantum Computing
  and Engineering ({QCE})}. {IEEE}, sep 2022.
\newblock \doi{10.1109/qce53715.2022.00042}.
\newblock URL \url{https://doi.org/10.1109%2Fqce53715.2022.00042}.

\bibitem[Kiwit et~al.(2023)Kiwit, Marso, Ross, Riofrío, Klepsch, and
  Luckow]{kiwit2023applicationoriented}
Florian~J. Kiwit, Marwa Marso, Philipp Ross, Carlos~A. Riofrío, Johannes
  Klepsch, and Andre Luckow.
\newblock Application-oriented benchmarking of quantum generative learning
  using quark, 2023.
\newblock URL \url{https://arxiv.org/abs/2308.04082}.

\bibitem[Lubinski et~al.(2023{\natexlab{a}})Lubinski, Johri, Varosy, Coleman,
  Zhao, Necaise, Baldwin, Mayer, and Proctor]{lubinski2023_10061574}
Thomas Lubinski, Sonika Johri, Paul Varosy, Jeremiah Coleman, Luning Zhao,
  Jason Necaise, Charles~H. Baldwin, Karl Mayer, and Timothy Proctor.
\newblock Application-oriented performance benchmarks for quantum computing.
\newblock \emph{IEEE Transactions on Quantum Engineering}, 4:\penalty0 1--32,
  2023{\natexlab{a}}.
\newblock \doi{10.1109/TQE.2023.3253761}.

\bibitem[Lubinski et~al.(2023{\natexlab{b}})Lubinski, Coffrin, McGeoch, Sathe,
  Apanavicius, and Neira]{lubinski2023optimization}
Thomas Lubinski, Carleton Coffrin, Catherine McGeoch, Pratik Sathe, Joshua
  Apanavicius, and David E.~Bernal Neira.
\newblock Optimization applications as quantum performance benchmarks,
  2023{\natexlab{b}}.
\newblock URL \url{https://arxiv.org/abs/2302.02278}.

\bibitem[spe(2021)]{spec_org}
{Standard Performance Evaluation Corporation}, 2021.
\newblock URL \url{https://spec.org/}.
\newblock {SPEC Benchmark Suite, accessed 2021-05-28}.

\bibitem[Hennessy and Patterson(2019)]{hennessy_patterson_2019_all}
John~L. Hennessy and Patterson.
\newblock \emph{{Computer Architecture: a Quantitative Approach}}.
\newblock Morgan Kaufmann, 2019.

\bibitem[Peruzzo et~al.(2014)Peruzzo, McClean, Shadbolt, Yung, Zhou, Love,
  {Aspuru-Guzik}, and O'Brien]{peruzzoVariationalEigenvalueSolver2014}
Alberto Peruzzo, Jarrod McClean, Peter Shadbolt, Man-Hong Yung, Xiao-Qi Zhou,
  Peter~J. Love, Al{\'a}n {Aspuru-Guzik}, and Jeremy~L. O'Brien.
\newblock A variational eigenvalue solver on a photonic quantum processor.
\newblock \emph{Nature Communications}, 5\penalty0 (1):\penalty0 1--7, July
  2014.
\newblock ISSN 2041-1723.
\newblock \doi{10.1038/ncomms5213}.

\bibitem[Farhi et~al.(2014)Farhi, Goldstone, and Gutmann]{farhi2014quantum}
Edward Farhi, Jeffrey Goldstone, and Sam Gutmann.
\newblock A quantum approximate optimization algorithm, 2014.

\bibitem[LaRose and Coyle(2020)]{larose2020robust}
Ryan LaRose and Brian Coyle.
\newblock Robust data encodings for quantum classifiers.
\newblock \emph{Phys. Rev. A}, 102:\penalty0 032420, Sep 2020.
\newblock \doi{10.1103/PhysRevA.102.032420}.
\newblock URL \url{https://link.aps.org/doi/10.1103/PhysRevA.102.032420}.

\bibitem[Chen et~al.(2023)Chen, Nielsen, Ebert, Inlek, Wright, Chaplin,
  Maksymov, Páez, Poudel, Maunz, and Gamble]{chen2023benchmarking}
Jwo-Sy Chen, Erik Nielsen, Matthew Ebert, Volkan Inlek, Kenneth Wright,
  Vandiver Chaplin, Andrii Maksymov, Eduardo Páez, Amrit Poudel, Peter Maunz,
  and John Gamble.
\newblock Benchmarking a trapped-ion quantum computer with 29 algorithmic
  qubits, 2023.

\bibitem[Harrow et~al.(2009)Harrow, Hassidim, and Lloyd]{Harrow2009}
Aram~W. Harrow, Avinatan Hassidim, and Seth Lloyd.
\newblock Quantum algorithm for linear systems of equations.
\newblock \emph{Phys. Rev. Lett.}, 103:\penalty0 150502, Oct 2009.
\newblock \doi{10.1103/PhysRevLett.103.150502}.
\newblock URL \url{https://link.aps.org/doi/10.1103/PhysRevLett.103.150502}.

\bibitem[Gambetta et~al.(2012)Gambetta, C{\'o}rcoles, Merkel, Johnson, Smolin,
  Chow, Ryan, Rigetti, Poletto, Ohki, et~al.]{gambetta2012characterization}
Jay~M Gambetta, AD~C{\'o}rcoles, Seth~T Merkel, Blake~R Johnson, John~A Smolin,
  Jerry~M Chow, Colm~A Ryan, Chad Rigetti, S~Poletto, Thomas~A Ohki, et~al.
\newblock Characterization of addressability by simultaneous randomized
  benchmarking.
\newblock \emph{Phys. Rev. Lett.}, 109\penalty0 (24):\penalty0 240504, 2012.
\newblock URL
  \url{https://journals.aps.org/prl/abstract/10.1103/PhysRevLett.109.240504}.

\bibitem[Sarovar et~al.(2020)Sarovar, Proctor, Rudinger, Young, Nielsen, and
  Blume-Kohout]{sarovar2019detecting}
Mohan Sarovar, Timothy Proctor, Kenneth Rudinger, Kevin Young, Erik Nielsen,
  and Robin Blume-Kohout.
\newblock Detecting crosstalk errors in quantum information processors.
\newblock \emph{Quantum}, 4:\penalty0 321, 2020.
\newblock URL \url{https://quantum-journal.org/papers/q-2020-09-11-321/}.

\bibitem[Proctor et~al.(2022{\natexlab{a}})Proctor, Seritan, Rudinger, Nielsen,
  Blume-Kohout, and Young]{Proctor_2022}
Timothy Proctor, Stefan Seritan, Kenneth Rudinger, Erik Nielsen, Robin
  Blume-Kohout, and Kevin Young.
\newblock Scalable randomized benchmarking of quantum computers using mirror
  circuits.
\newblock \emph{Physical Review Letters}, 129\penalty0 (15), October
  2022{\natexlab{a}}.
\newblock ISSN 1079-7114.
\newblock \doi{10.1103/physrevlett.129.150502}.
\newblock URL \url{http://dx.doi.org/10.1103/PhysRevLett.129.150502}.

\bibitem[Team(2021)]{qiskit_measuring_quantum_volume}
The~Qiskit Team.
\newblock Measuring quantum volume, Aug 2021.
\newblock URL
  \url{https://qiskit.org/textbook/ch-quantum-hardware/measuring-quantum-volume.html}.

\bibitem[Blume-Kohout and Young(2020)]{BlumeKohout2020volumetricframework}
Robin Blume-Kohout and Kevin~C. Young.
\newblock A volumetric framework for quantum computer benchmarks.
\newblock \emph{{Quantum}}, 4:\penalty0 362, November 2020.
\newblock ISSN 2521-327X.
\newblock \doi{10.22331/q-2020-11-15-362}.
\newblock URL \url{https://doi.org/10.22331/q-2020-11-15-362}.

\bibitem[Baldwin et~al.(2022)Baldwin, Mayer, Brown, Ryan-Anderson, and
  Hayes]{Baldwin2022reexaminingquantum}
Charles~H. Baldwin, Karl Mayer, Natalie~C. Brown, Ciar{\'{a}}n Ryan-Anderson,
  and David Hayes.
\newblock Re-examining the quantum volume test: {I}deal distributions, compiler
  optimizations, confidence intervals, and scalable resource estimations.
\newblock \emph{{Quantum}}, 6:\penalty0 707, May 2022.
\newblock ISSN 2521-327X.
\newblock \doi{10.22331/q-2022-05-09-707}.
\newblock URL \url{https://doi.org/10.22331/q-2022-05-09-707}.

\bibitem[Pelofske et~al.(2022)Pelofske, Bartschi, and Eidenbenz]{Pelofske_2022}
Elijah Pelofske, Andreas Bartschi, and Stephan Eidenbenz.
\newblock Quantum volume in practice: What users can expect from {NISQ}
  devices.
\newblock \emph{{IEEE} Transactions on Quantum Engineering}, 3:\penalty0 1--19,
  2022.
\newblock \doi{10.1109/tqe.2022.3184764}.
\newblock URL \url{https://doi.org/10.1109%2Ftqe.2022.3184764}.

\bibitem[Proctor et~al.(2022{\natexlab{b}})Proctor, Seritan, Nielsen, Rudinger,
  Young, Blume-Kohout, and Sarovar]{proctor2022trust}
Timothy Proctor, Stefan Seritan, Erik Nielsen, Kenneth Rudinger, Kevin Young,
  Robin Blume-Kohout, and Mohan Sarovar.
\newblock Establishing trust in quantum computations, 2022{\natexlab{b}}.
\newblock URL \url{https://arxiv.org/abs/2204.07568}.

\bibitem[qc-(2015)]{qc-proto-benchmarks}
{Application-Oriented Performance Benchmarks for Quantum Computing}, 2015.
\newblock URL
  \url{https://github.com/SRI-International/QC-App-Oriented-Benchmarks}.

\bibitem[Aaronson(2015)]{Aaronson2015}
Scott Aaronson.
\newblock Read the fine print.
\newblock \emph{Nature Physics}, 11\penalty0 (4):\penalty0 291--293, Apr 2015.
\newblock ISSN 1745-2481.
\newblock \doi{10.1038/nphys3272}.
\newblock URL \url{https://doi.org/10.1038/nphys3272}.

\bibitem[Liu et~al.(2022)Liu, Xie, Liu, and Zhao]{Liu_Xie_Liu_Zhao_2022}
Xiaonan Liu, Haoshan Xie, Zhengyu Liu, and Chenyan Zhao.
\newblock Survey on the improvement and application of hhl algorithm.
\newblock \emph{Journal of Physics: Conference Series}, 2333\penalty0
  (1):\penalty0 012023, 2022.
\newblock \doi{10.1088/1742-6596/2333/1/012023}.

\bibitem[Dervovic et~al.(2018)Dervovic, Herbster, Mountney, Severini, Usher,
  and Wossnig]{dervovic2018quantum}
Danial Dervovic, Mark Herbster, Peter Mountney, Simone Severini, Naïri Usher,
  and Leonard Wossnig.
\newblock Quantum linear systems algorithms: a primer, 2018.

\bibitem[au2 et~al.(2023)au2, Zaman, and Wong]{morrell2023stepbystep}
Hector Jose Morrell~Jr au2, Anika Zaman, and Hiu~Yung Wong.
\newblock Step-by-step hhl algorithm walkthrough to enhance the understanding
  of critical quantum computing concepts, 2023.

\bibitem[Cao et~al.(2012)Cao, Daskin, Frankel, and Kais]{Cao_2012}
Yudong Cao, Anmer Daskin, Steven Frankel, and Sabre Kais.
\newblock Quantum circuit design for solving linear systems of equations.
\newblock \emph{Molecular Physics}, 110\penalty0 (15-16):\penalty0 1675--1680,
  aug 2012.
\newblock \doi{10.1080/00268976.2012.668289}.
\newblock URL \url{https://doi.org/10.1080%2F00268976.2012.668289}.

\bibitem[Lee et~al.(2019)Lee, Joo, and Lee]{Lee2019_HHL_IBMQ}
Yonghae Lee, Jaewoo Joo, and Soojoon Lee.
\newblock Hybrid quantum linear equation algorithm and its experimental test on
  ibm quantum experience.
\newblock \emph{Scientific Reports}, 9, 03 2019.
\newblock \doi{10.1038/s41598-019-41324-9}.

\bibitem[Martin et~al.(2023)Martin, Ibarrondo, and Sanz]{Martin_2023}
Ana Martin, Ruben Ibarrondo, and Mikel Sanz.
\newblock Digital-analog co-design of the harrow-hassidim-lloyd algorithm.
\newblock \emph{Physical Review Applied}, 19\penalty0 (6), jun 2023.
\newblock \doi{10.1103/physrevapplied.19.064056}.
\newblock URL \url{https://doi.org/10.1103%2Fphysrevapplied.19.064056}.

\bibitem[Childs et~al.(2003)Childs, Cleve, Deotto, Farhi, Gutmann, and
  Spielman]{Childs2003}
Andrew~M. Childs, Richard Cleve, Enrico Deotto, Edward Farhi, Sam Gutmann, and
  Daniel~A. Spielman.
\newblock Exponential algorithmic speedup by a quantum walk.
\newblock In \emph{Proceedings of the Thirty-Fifth Annual ACM Symposium on
  Theory of Computing}, STOC '03, page 59–68, New York, NY, USA, 2003.
  Association for Computing Machinery.
\newblock ISBN 1581136749.
\newblock \doi{10.1145/780542.780552}.
\newblock URL \url{https://doi.org/10.1145/780542.780552}.

\bibitem[M\"{o}tt\"{o}nen et~al.(2005)M\"{o}tt\"{o}nen, Vartiainen, Bergholm,
  and Salomaa]{Mottonen2005}
Mikko M\"{o}tt\"{o}nen, Juha~J. Vartiainen, Ville Bergholm, and Martti~M.
  Salomaa.
\newblock Transformation of quantum states using uniformly controlled
  rotations.
\newblock 5\penalty0 (6):\penalty0 467–473, sep 2005.
\newblock ISSN 1533-7146.

\bibitem[Ruiz-Perez and Garcia-Escartin(2017)]{Ruiz-Perez2017}
Lidia Ruiz-Perez and Juan~Carlos Garcia-Escartin.
\newblock Quantum arithmetic with the quantum fourier transform.
\newblock \emph{Quantum Information Processing}, 16\penalty0 (6):\penalty0 152,
  Apr 2017.
\newblock ISSN 1573-1332.
\newblock \doi{10.1007/s11128-017-1603-1}.
\newblock URL \url{https://doi.org/10.1007/s11128-017-1603-1}.

\bibitem[Yalovetzky et~al.(2023)Yalovetzky, Minssen, Herman, and
  Pistoia]{yalovetzky2023}
Romina Yalovetzky, Pierre Minssen, Dylan Herman, and Marco Pistoia.
\newblock Hybrid hhl with dynamic quantum circuits on real hardware, 2023.

\bibitem[Motta et~al.(2017{\natexlab{a}})Motta, Ceperley, Chan, Gomez, Gull,
  Guo, Jim\'enez-Hoyos, Lan, Li, Ma, Millis, Prokof'ev, Ray, Scuseria, Sorella,
  Stoudenmire, Sun, Tupitsyn, White, Zgid, and Zhang]{PhysRevX.7.031059}
Mario Motta, David~M. Ceperley, Garnet Kin-Lic Chan, John~A. Gomez, Emanuel
  Gull, Sheng Guo, Carlos~A. Jim\'enez-Hoyos, Tran~Nguyen Lan, Jia Li, Fengjie
  Ma, Andrew~J. Millis, Nikolay~V. Prokof'ev, Ushnish Ray, Gustavo~E. Scuseria,
  Sandro Sorella, Edwin~M. Stoudenmire, Qiming Sun, Igor~S. Tupitsyn, Steven~R.
  White, Dominika Zgid, and Shiwei Zhang.
\newblock Towards the solution of the many-electron problem in real materials:
  Equation of state of the hydrogen chain with state-of-the-art many-body
  methods.
\newblock \emph{Phys. Rev. X}, 7:\penalty0 031059, Sep 2017{\natexlab{a}}.
\newblock \doi{10.1103/PhysRevX.7.031059}.
\newblock URL \url{https://link.aps.org/doi/10.1103/PhysRevX.7.031059}.

\bibitem[Cao et~al.(2023)Cao, Sun, Yuan, Hu, Pham, and Lv]{Cao_2023}
Changsu Cao, Jinzhao Sun, Xiao Yuan, Han-Shi Hu, Hung~Q. Pham, and Dingshun Lv.
\newblock Ab initio quantum simulation of strongly correlated materials with
  quantum embedding.
\newblock \emph{npj Computational Materials}, 9\penalty0 (1), may 2023.
\newblock \doi{10.1038/s41524-023-01045-0}.
\newblock URL \url{https://doi.org/10.1038%2Fs41524-023-01045-0}.

\bibitem[Goings et~al.(2022)Goings, White, Lee, Tautermann, Degroote, Gidney,
  Shiozaki, Babbush, and Rubin]{Goings2022}
Joshua~J. Goings, Alec White, Joonho Lee, Christofer~S. Tautermann, Matthias
  Degroote, Craig Gidney, Toru Shiozaki, Ryan Babbush, and Nicholas~C. Rubin.
\newblock Reliably assessing the electronic structure of cytochrome p450 on
  today's classical computers and tomorrow's quantum computers.
\newblock \emph{Proceedings of the National Academy of Sciences}, 119\penalty0
  (38), sep 2022.
\newblock \doi{10.1073/pnas.2203533119}.
\newblock URL \url{https://doi.org/10.1073%2Fpnas.2203533119}.

\bibitem[McCaskey et~al.(2019)McCaskey, Parks, Jakowski, Moore, Morris, Humble,
  and Pooser]{mccaskey2019quantum}
Alexander~J. McCaskey, Zachary~P. Parks, Jacek Jakowski, Shirley~V. Moore,
  T.~Morris, Travis~S. Humble, and Raphael~C. Pooser.
\newblock Quantum chemistry as a benchmark for near-term quantum computers,
  2019.
\newblock URL \url{https://doi.org/10.1038/s41534-019-0209-0}.

\bibitem[Yeter-Aydeniz et~al.(2021)Yeter-Aydeniz, Gard, Jakowski, Majumder,
  Barron, Siopsis, Humble, and Pooser]{yeteraydeniz2021benchmarking}
Kübra Yeter-Aydeniz, Bryan~T. Gard, Jacek Jakowski, Swarnadeep Majumder,
  George~S. Barron, George Siopsis, Travis Humble, and Raphael~C. Pooser.
\newblock Benchmarking quantum chemistry computations with variational,
  imaginary time evolution, and krylov space solver algorithms, 2021.

\bibitem[Dallaire-Demers et~al.(2020)Dallaire-Demers, Stęchły, Gonthier,
  Bashige, Romero, and Cao]{dallairedemers2020application}
Pierre-Luc Dallaire-Demers, Michał Stęchły, Jerome~F. Gonthier,
  Ntwali~Toussaint Bashige, Jonathan Romero, and Yudong Cao.
\newblock An application benchmark for fermionic quantum simulations, 2020.

\bibitem[Sawaya et~al.(2023)Sawaya, Marti-Dafcik, Ho, Tabor, Bernal, Magann,
  Premaratne, Dubey, Matsuura, Bishop, de~Jong, Benjamin, Parekh, Tubman,
  Klymko, and Camps]{sawaya2023hamlib}
Nicolas~PD Sawaya, Daniel Marti-Dafcik, Yang Ho, Daniel~P Tabor, David Bernal,
  Alicia~B Magann, Shavindra Premaratne, Pradeep Dubey, Anne Matsuura, Nathan
  Bishop, Wibe~A de~Jong, Simon Benjamin, Ojas~D Parekh, Norm Tubman, Katherine
  Klymko, and Daan Camps.
\newblock Hamlib: A library of hamiltonians for benchmarking quantum algorithms
  and hardware, 2023.

\bibitem[Stair and Evangelista(2020)]{stair2020exploring}
Nicholas~H. Stair and Francesco~A. Evangelista.
\newblock Exploring hilbert space on a budget: Novel benchmark set and
  performance metric for testing electronic structure methods in the regime of
  strong correlation, 2020.
\newblock URL \url{https://doi.org/10.1063/5.0014928}.

\bibitem[Tilly et~al.(2022)Tilly, Chen, Cao, Picozzi, Setia, Li, Grant,
  Wossnig, Rungger, Booth, et~al.]{tilly2022variational}
Jules Tilly, Hongxiang Chen, Shuxiang Cao, Dario Picozzi, Kanav Setia, Ying Li,
  Edward Grant, Leonard Wossnig, Ivan Rungger, George~H Booth, et~al.
\newblock The variational quantum eigensolver: a review of methods and best
  practices.
\newblock \emph{Physics Reports}, 986:\penalty0 1--128, 2022.

\bibitem[McClean et~al.(2016)McClean, Romero, Babbush, and
  Aspuru-Guzik]{McClean_2016}
Jarrod~R McClean, Jonathan Romero, Ryan Babbush, and Al{\'{a}}n Aspuru-Guzik.
\newblock The theory of variational hybrid quantum-classical algorithms.
\newblock \emph{New Journal of Physics}, 18\penalty0 (2):\penalty0 023023, feb
  2016.
\newblock \doi{10.1088/1367-2630/18/2/023023}.
\newblock URL \url{https://doi.org/10.1088/1367-2630/18/2/023023}.

\bibitem[Johnson et~al.(2022)Johnson, Kunitsa, Gonthier, Radin, Buda, Doskocil,
  Abuan, and Romero]{johnson2022reducing}
Peter~D. Johnson, Alexander~A. Kunitsa, Jérôme~F. Gonthier, Maxwell~D. Radin,
  Corneliu Buda, Eric~J. Doskocil, Clena~M. Abuan, and Jhonathan Romero.
\newblock Reducing the cost of energy estimation in the variational quantum
  eigensolver algorithm with robust amplitude estimation, 2022.
\newblock URL \url{https://arxiv.org/abs/2203.07275}.

\bibitem[White(1992)]{white1992density}
Steven~R White.
\newblock Density matrix formulation for quantum renormalization groups.
\newblock \emph{Physical review letters}, 69\penalty0 (19):\penalty0 2863,
  1992.

\bibitem[Arute et~al.(2020)Arute, Arya, Babbush, Bacon, Bardin, Barends, Boixo,
  Broughton, Buckley, Buell, Burkett, Bushnell, Chen, Chen, Chiaro, Collins,
  Courtney, Demura, Dunsworth, Farhi, Fowler, Foxen, Gidney, Giustina, Graff,
  Habegger, Harrigan, Ho, Hong, Huang, Huggins, Ioffe, Isakov, Jeffrey, Jiang,
  Jones, Kafri, Kechedzhi, Kelly, Kim, Klimov, Korotkov, Kostritsa, Landhuis,
  Laptev, Lindmark, Lucero, Martin, Martinis, McClean, McEwen, Megrant, Mi,
  Mohseni, Mruczkiewicz, Mutus, Naaman, Neeley, Neill, Neven, Niu, O'Brien,
  Ostby, Petukhov, Putterman, Quintana, Roushan, Rubin, Sank, Satzinger,
  Smelyanskiy, Strain, Sung, Szalay, Takeshita, Vainsencher, White, Wiebe, Yao,
  Yeh, and Zalcman]{HartreeFockGoogle2020}
Frank Arute, Kunal Arya, Ryan Babbush, Dave Bacon, Joseph~C. Bardin, Rami
  Barends, Sergio Boixo, Michael Broughton, Bob~B. Buckley, David~A. Buell,
  Brian Burkett, Nicholas Bushnell, Yu~Chen, Zijun Chen, Benjamin Chiaro,
  Roberto Collins, William Courtney, Sean Demura, Andrew Dunsworth, Edward
  Farhi, Austin Fowler, Brooks Foxen, Craig Gidney, Marissa Giustina, Rob
  Graff, Steve Habegger, Matthew~P. Harrigan, Alan Ho, Sabrina Hong, Trent
  Huang, William~J. Huggins, Lev Ioffe, Sergei~V. Isakov, Evan Jeffrey, Zhang
  Jiang, Cody Jones, Dvir Kafri, Kostyantyn Kechedzhi, Julian Kelly, Seon Kim,
  Paul~V. Klimov, Alexander Korotkov, Fedor Kostritsa, David Landhuis, Pavel
  Laptev, Mike Lindmark, Erik Lucero, Orion Martin, John~M. Martinis, Jarrod~R.
  McClean, Matt McEwen, Anthony Megrant, Xiao Mi, Masoud Mohseni, Wojciech
  Mruczkiewicz, Josh Mutus, Ofer Naaman, Matthew Neeley, Charles Neill, Hartmut
  Neven, Murphy~Yuezhen Niu, Thomas~E. O'Brien, Eric Ostby, Andre Petukhov,
  Harald Putterman, Chris Quintana, Pedram Roushan, Nicholas~C. Rubin, Daniel
  Sank, Kevin~J. Satzinger, Vadim Smelyanskiy, Doug Strain, Kevin~J. Sung,
  Marco Szalay, Tyler~Y. Takeshita, Amit Vainsencher, Theodore White, Nathan
  Wiebe, Z.~Jamie Yao, Ping Yeh, and Adam Zalcman.
\newblock Hartree-fock on a superconducting qubit quantum computer.
\newblock \emph{Science}, 369\penalty0 (6507):\penalty0 1084--1089, aug 2020.
\newblock \doi{10.1126/science.abb9811}.
\newblock URL \url{https://doi.org/10.1126%2Fscience.abb9811}.

\bibitem[Motta et~al.(2017{\natexlab{b}})Motta, Ceperley, Chan, Gomez, Gull,
  Guo, Jim{\'{e} }nez-Hoyos, Lan, Li, Ma, Millis, Prokof'ev, Ray, Scuseria,
  Sorella, Stoudenmire, Sun, Tupitsyn, White, Zgid, and and]{Motta_2017}
Mario Motta, David~M. Ceperley, Garnet Kin-Lic Chan, John~A. Gomez, Emanuel
  Gull, Sheng Guo, Carlos~A. Jim{\'{e} }nez-Hoyos, Tran~Nguyen Lan, Jia Li,
  Fengjie Ma, Andrew~J. Millis, Nikolay~V. Prokof'ev, Ushnish Ray, Gustavo~E.
  Scuseria, Sandro Sorella, Edwin~M. Stoudenmire, Qiming Sun, Igor~S. Tupitsyn,
  Steven~R. White, Dominika Zgid, and Shiwei~Zhang and.
\newblock Towards the solution of the many-electron problem in real materials:
  Equation of state of the hydrogen chain with state-of-the-art many-body
  methods.
\newblock \emph{Physical Review X}, 7\penalty0 (3), sep 2017{\natexlab{b}}.
\newblock \doi{10.1103/physrevx.7.031059}.
\newblock URL \url{https://doi.org/10.1103%2Fphysrevx.7.031059}.

\bibitem[Cao et~al.(2019)Cao, Romero, Olson, Degroote, Johnson, Kieferová,
  Kivlichan, Menke, Peropadre, Sawaya, and et~al.]{Cao_2019}
Yudong Cao, Jonathan Romero, Jonathan~P. Olson, Matthias Degroote, Peter~D.
  Johnson, Mária Kieferová, Ian~D. Kivlichan, Tim Menke, Borja Peropadre,
  Nicolas P.~D. Sawaya, and et~al.
\newblock Quantum chemistry in the age of quantum computing.
\newblock \emph{Chemical Reviews}, 119\penalty0 (19):\penalty0 10856–10915,
  Aug 2019.
\newblock ISSN 1520-6890.
\newblock \doi{10.1021/acs.chemrev.8b00803}.
\newblock URL \url{http://dx.doi.org/10.1021/acs.chemrev.8b00803}.

\bibitem[Elfving et~al.(2021)Elfving, Millaruelo, G{\'a}mez, and
  Gogolin]{elfving2021simulating}
Vincent~E Elfving, Marta Millaruelo, Jos{\'e}~A G{\'a}mez, and Christian
  Gogolin.
\newblock Simulating quantum chemistry in the seniority-zero space on
  qubit-based quantum computers.
\newblock \emph{Physical Review A}, 103\penalty0 (3):\penalty0 032605, 2021.

\bibitem[Zhao et~al.(2023)Zhao, Goings, Shin, Kyoung, Fuks, Kevin~Rhee, Rhee,
  Wright, Nguyen, Kim, et~al.]{zhao2023orbital}
Luning Zhao, Joshua Goings, Kyujin Shin, Woomin Kyoung, Johanna~I Fuks,
  June-Koo Kevin~Rhee, Young~Min Rhee, Kenneth Wright, Jason Nguyen, Jungsang
  Kim, et~al.
\newblock Orbital-optimized pair-correlated electron simulations on trapped-ion
  quantum computers.
\newblock \emph{npj Quantum Information}, 9\penalty0 (1):\penalty0 60, 2023.

\bibitem[O'Brien et~al.(2023)O'Brien, Anselmetti, Gkritsis, Elfving, Polla,
  Huggins, Oumarou, Kechedzhi, Abanin, Acharya, Aleiner, Allen, Andersen,
  Anderson, Ansmann, Arute, Arya, Asfaw, Atalaya, Bardin, Bengtsson, Bortoli,
  Bourassa, Bovaird, Brill, Broughton, Buckley, Buell, Burger, Burkett,
  Bushnell, Campero, Chen, Chiaro, Chik, Cogan, Collins, Conner, Courtney,
  Crook, Curtin, Debroy, Demura, Drozdov, Dunsworth, Erickson, Faoro, Farhi,
  Fatemi, Ferreira, Flores~Burgos, Forati, Fowler, Foxen, Giang, Gidney,
  Gilboa, Giustina, Gosula, Grajales~Dau, Gross, Habegger, Hamilton, Hansen,
  Harrigan, Harrington, Heu, Hoffmann, Hong, Huang, Huff, Ioffe, Isakov,
  Iveland, Jeffrey, Jiang, Jones, Juhas, Kafri, Khattar, Khezri, Kieferov{\'a},
  Kim, Klimov, Klots, Korotkov, Kostritsa, Kreikebaum, Landhuis, Laptev, Lau,
  Laws, Lee, Lee, Lester, Lill, Liu, Livingston, Locharla, Malone, Mandr{\`a},
  Martin, Martin, McClean, McCourt, McEwen, Mi, Mieszala, Miao, Mohseni,
  Montazeri, Morvan, Movassagh, Mruczkiewicz, Naaman, Neeley, Neill, Nersisyan,
  Newman, Ng, Nguyen, Nguyen, Niu, Omonije, Opremcak, Petukhov, Potter,
  Pryadko, Quintana, Rocque, Roushan, Saei, Sank, Sankaragomathi, Satzinger,
  Schurkus, Schuster, Shearn, Shorter, Shutty, Shvarts, Skruzny, Smith, Somma,
  Sterling, Strain, Szalay, Thor, Torres, Vidal, Villalonga,
  Vollgraff~Heidweiller, White, Woo, Xing, Yao, Yeh, Yoo, Young, Zalcman,
  Zhang, Zhu, Zobrist, Bacon, Boixo, Chen, Hilton, Kelly, Lucero, Megrant,
  Neven, Smelyanskiy, Gogolin, Babbush, and Rubin]{OBrien2023-aa}
T~E O'Brien, G~Anselmetti, F~Gkritsis, V~E Elfving, S~Polla, W~J Huggins,
  O~Oumarou, K~Kechedzhi, D~Abanin, R~Acharya, I~Aleiner, R~Allen, T~I
  Andersen, K~Anderson, M~Ansmann, F~Arute, K~Arya, A~Asfaw, J~Atalaya, J~C
  Bardin, A~Bengtsson, G~Bortoli, A~Bourassa, J~Bovaird, L~Brill, M~Broughton,
  B~Buckley, D~A Buell, T~Burger, B~Burkett, N~Bushnell, J~Campero, Z~Chen,
  B~Chiaro, D~Chik, J~Cogan, R~Collins, P~Conner, W~Courtney, A~L Crook,
  B~Curtin, D~M Debroy, S~Demura, I~Drozdov, A~Dunsworth, C~Erickson, L~Faoro,
  E~Farhi, R~Fatemi, V~S Ferreira, L~Flores~Burgos, E~Forati, A~G Fowler,
  B~Foxen, W~Giang, C~Gidney, D~Gilboa, M~Giustina, R~Gosula, A~Grajales~Dau,
  J~A Gross, S~Habegger, M~C Hamilton, M~Hansen, M~P Harrigan, S~D Harrington,
  P~Heu, M~R Hoffmann, S~Hong, T~Huang, A~Huff, L~B Ioffe, S~V Isakov,
  J~Iveland, E~Jeffrey, Z~Jiang, C~Jones, P~Juhas, D~Kafri, T~Khattar,
  M~Khezri, M~Kieferov{\'a}, S~Kim, P~V Klimov, A~R Klots, A~N Korotkov,
  F~Kostritsa, J~M Kreikebaum, D~Landhuis, P~Laptev, K-M Lau, L~Laws, J~Lee,
  K~Lee, B~J Lester, A~T Lill, W~Liu, W~P Livingston, A~Locharla, F~D Malone,
  S~Mandr{\`a}, O~Martin, S~Martin, J~R McClean, T~McCourt, M~McEwen, X~Mi,
  A~Mieszala, K~C Miao, M~Mohseni, S~Montazeri, A~Morvan, R~Movassagh,
  W~Mruczkiewicz, O~Naaman, M~Neeley, C~Neill, A~Nersisyan, M~Newman, J~H Ng,
  A~Nguyen, M~Nguyen, M~Y Niu, S~Omonije, A~Opremcak, A~Petukhov, R~Potter, L~P
  Pryadko, C~Quintana, C~Rocque, P~Roushan, N~Saei, D~Sank, K~Sankaragomathi,
  K~J Satzinger, H~F Schurkus, C~Schuster, M~J Shearn, A~Shorter, N~Shutty,
  V~Shvarts, J~Skruzny, W~C Smith, R~D Somma, G~Sterling, D~Strain, M~Szalay,
  D~Thor, A~Torres, G~Vidal, B~Villalonga, C~Vollgraff~Heidweiller, T~White,
  B~W~K Woo, C~Xing, Z~J Yao, P~Yeh, J~Yoo, G~Young, A~Zalcman, Y~Zhang, N~Zhu,
  N~Zobrist, D~Bacon, S~Boixo, Y~Chen, J~Hilton, J~Kelly, E~Lucero, A~Megrant,
  H~Neven, V~Smelyanskiy, C~Gogolin, R~Babbush, and N~C Rubin.
\newblock Purification-based quantum error mitigation of pair-correlated
  electron simulations.
\newblock \emph{Nat. Phys.}, pages 1--6, October 2023.

\bibitem[Nam et~al.(2020)Nam, Chen, Pisenti, Wright, Delaney, Maslov, Brown,
  Allen, Amini, Apisdorf, et~al.]{nam2020ground}
Yunseong Nam, Jwo-Sy Chen, Neal~C Pisenti, Kenneth Wright, Conor Delaney,
  Dmitri Maslov, Kenneth~R Brown, Stewart Allen, Jason~M Amini, Joel Apisdorf,
  et~al.
\newblock Ground-state energy estimation of the water molecule on a trapped-ion
  quantum computer.
\newblock \emph{npj Quantum Information}, 6\penalty0 (1):\penalty0 33, 2020.

\bibitem[Lee et~al.(2018)Lee, Huggins, Head-Gordon, and
  Whaley]{lee2018generalized}
Joonho Lee, William~J Huggins, Martin Head-Gordon, and K~Birgitta Whaley.
\newblock Generalized unitary coupled cluster wave functions for quantum
  computation.
\newblock \emph{Journal of chemical theory and computation}, 15\penalty0
  (1):\penalty0 311--324, 2018.

\bibitem[Goings et~al.(2023)Goings, Zhao, Jakowski, Morris, and
  Pooser]{goings2023molecular}
Joshua Goings, Luning Zhao, Jacek Jakowski, Titus Morris, and Raphael Pooser.
\newblock Molecular symmetry in vqe: A dual approach for trapped-ion
  simulations of benzene.
\newblock \emph{arXiv preprint arXiv:2308.00667}, 2023.

\bibitem[qis(2022{\natexlab{a}})]{qiskit_sampler}
{Qiskit Runtime Sampler Primitive}.
\newblock
  \url{https://qiskit.org/ecosystem/ibm-runtime/stubs/qiskit_ibm_runtime.Sampler.html},
  2022{\natexlab{a}}.
\newblock {IBM Quantum Lab}.

\bibitem[Sivarajah et~al.(2020)Sivarajah, Dilkes, Cowtan, Simmons, Edgington,
  and Duncan]{Sivarajah_2020}
Seyon Sivarajah, Silas Dilkes, Alexander Cowtan, Will Simmons, Alec Edgington,
  and Ross Duncan.
\newblock tket: a retargetable compiler for {NISQ} devices.
\newblock \emph{Quantum Science and Technology}, 6\penalty0 (1):\penalty0
  014003, nov 2020.
\newblock \doi{10.1088/2058-9565/ab8e92}.
\newblock URL \url{https://doi.org/10.1088/2058-9565/ab8e92}.

\bibitem[q-c(2023)]{q-ctrl}
{Q-CTRL} web site.
\newblock \url{https://q-ctrl.com/}, 2023.
\newblock {Q-CTRL File Opal}.

\bibitem[Cai et~al.(2023)Cai, Babbush, Benjamin, Endo, Huggins, Li, McClean,
  and O'Brien]{cai2023quantum}
Zhenyu Cai, Ryan Babbush, Simon~C. Benjamin, Suguru Endo, William~J. Huggins,
  Ying Li, Jarrod~R. McClean, and Thomas~E. O'Brien.
\newblock Quantum error mitigation, 2023.

\bibitem[Qin et~al.(2023)Qin, Chen, and Li]{Qin_2023}
Dayue Qin, Yanzhu Chen, and Ying Li.
\newblock Error statistics and scalability of quantum error mitigation
  formulas.
\newblock \emph{npj Quantum Information}, 9\penalty0 (1), apr 2023.
\newblock \doi{10.1038/s41534-023-00707-7}.
\newblock URL \url{https://doi.org/10.1038%2Fs41534-023-00707-7}.

\bibitem[Takagi et~al.(2022)Takagi, Endo, Minagawa, and Gu]{Takagi_2022}
Ryuji Takagi, Suguru Endo, Shintaro Minagawa, and Mile Gu.
\newblock Fundamental limits of quantum error mitigation.
\newblock \emph{npj Quantum Information}, 8\penalty0 (1), sep 2022.
\newblock \doi{10.1038/s41534-022-00618-z}.
\newblock URL \url{https://doi.org/10.1038%2Fs41534-022-00618-z}.

\bibitem[Cirstoiu et~al.(2023)Cirstoiu, Dilkes, Mills, Sivarajah, and
  Duncan]{Cirstoiu2023volumetric}
Cristina Cirstoiu, Silas Dilkes, Daniel Mills, Seyon Sivarajah, and Ross
  Duncan.
\newblock Volumetric {B}enchmarking of {E}rror {M}itigation with {Q}ermit.
\newblock \emph{{Quantum}}, 7:\penalty0 1059, July 2023.
\newblock ISSN 2521-327X.
\newblock \doi{10.22331/q-2023-07-13-1059}.
\newblock URL \url{https://doi.org/10.22331/q-2023-07-13-1059}.

\bibitem[Nation et~al.(2021)Nation, Kang, Sundaresan, and
  Gambetta]{Nation_2021}
Paul~D. Nation, Hwajung Kang, Neereja Sundaresan, and Jay~M. Gambetta.
\newblock Scalable mitigation of measurement errors on quantum computers.
\newblock \emph{{PRX} Quantum}, 2\penalty0 (4), nov 2021.
\newblock \doi{10.1103/prxquantum.2.040326}.
\newblock URL \url{https://doi.org/10.1103%2Fprxquantum.2.040326}.

\bibitem[qis(2022{\natexlab{b}})]{qiskit_mthree}
{Mthree Error Mitigation}.
\newblock \url{https://qiskit.org/ecosystem/mthree/}, 2022{\natexlab{b}}.
\newblock {IBM Quantum Lab}.

\bibitem[Nation and Treinish(2023)]{Nation_2023}
Paul~D. Nation and Matthew Treinish.
\newblock Suppressing quantum circuit errors due to system variability.
\newblock \emph{{PRX} Quantum}, 4\penalty0 (1), mar 2023.
\newblock \doi{10.1103/prxquantum.4.010327}.
\newblock URL \url{https://doi.org/10.1103%2Fprxquantum.4.010327}.

\bibitem[Mills et~al.(2021)Mills, Sivarajah, Scholten, and Duncan]{Mills_2021}
Daniel Mills, Seyon Sivarajah, Travis~L. Scholten, and Ross Duncan.
\newblock Application-motivated, holistic benchmarking of a full quantum
  computing stack.
\newblock \emph{Quantum}, 5:\penalty0 415, Mar 2021.
\newblock ISSN 2521-327X.
\newblock \doi{10.22331/q-2021-03-22-415}.
\newblock URL \url{http://dx.doi.org/10.22331/q-2021-03-22-415}.

\bibitem[Amico et~al.(2023)Amico, Zhang, Jurcevic, Bishop, Nation, Wack, and
  McKay]{amico2023defining}
Mirko Amico, Helena Zhang, Petar Jurcevic, Lev~S. Bishop, Paul Nation, Andrew
  Wack, and David~C. McKay.
\newblock Defining standard strategies for quantum benchmarks, 2023.
\newblock URL \url{https://arxiv.org/abs/2303.02108}.

\bibitem[Tucci(2005)]{tucci2005introduction}
Robert~R. Tucci.
\newblock An introduction to cartan's kak decomposition for qc programmers,
  2005.
\newblock URL \url{https://arxiv.org/abs/quant-ph/0507171}.

\bibitem[Mundada et~al.(2023)Mundada, Barbosa, Maity, Wang, Merkh, Stace,
  Nielson, Carvalho, Hush, Biercuk, and Baum]{mundada_2022}
Pranav~S. Mundada, Aaron Barbosa, Smarak Maity, Yulun Wang, Thomas Merkh, T.M.
  Stace, Felicity Nielson, Andre~R.R. Carvalho, Michael Hush, Michael~J.
  Biercuk, and Yuval Baum.
\newblock Experimental benchmarking of an automated deterministic
  error-suppression workflow for quantum algorithms.
\newblock \emph{Phys. Rev. Appl.}, 20:\penalty0 024034, Aug 2023.
\newblock \doi{10.1103/PhysRevApplied.20.024034}.
\newblock URL \url{https://link.aps.org/doi/10.1103/PhysRevApplied.20.024034}.

\bibitem[Baum et~al.(2021)Baum, Amico, Howell, Hush, Liuzzi, Mundada, Merkh,
  Carvalho, and Biercuk]{Baum2021}
Yuval Baum, Mirko Amico, Sean Howell, Michael Hush, Maggie Liuzzi, Pranav
  Mundada, Thomas Merkh, Andre~R.R. Carvalho, and Michael~J. Biercuk.
\newblock Experimental deep reinforcement learning for error-robust gate-set
  design on a superconducting quantum computer.
\newblock \emph{PRX Quantum}, 2:\penalty0 040324, Nov 2021.
\newblock \doi{10.1103/PRXQuantum.2.040324}.
\newblock URL \url{https://link.aps.org/doi/10.1103/PRXQuantum.2.040324}.

\bibitem[Carvalho et~al.(2021)Carvalho, Ball, Biercuk, Hush, and
  Thomsen]{Carvalho2021}
Andre R.~R. Carvalho, Harrison Ball, Michael~J. Biercuk, Michael~R. Hush, and
  Felix Thomsen.
\newblock Error-robust quantum logic optimization using a cloud quantum
  computer interface.
\newblock \emph{Phys. Rev. Applied}, 15:\penalty0 064054, Jun 2021.
\newblock \doi{10.1103/PhysRevApplied.15.064054}.
\newblock URL \url{https://link.aps.org/doi/10.1103/PhysRevApplied.15.064054}.

\bibitem[Nguyen et~al.(2019)Nguyen, Behrman, Moustafa, and Steck]{Nguyen_2019}
Nam~H. Nguyen, E.~C. Behrman, Mohamed~A. Moustafa, and J.~E. Steck.
\newblock Benchmarking neural networks for quantum computations.
\newblock \emph{IEEE Transactions on Neural Networks and Learning Systems},
  page 1–10, 2019.
\newblock ISSN 2162-2388.
\newblock \doi{10.1109/tnnls.2019.2933394}.
\newblock URL \url{http://dx.doi.org/10.1109/TNNLS.2019.2933394}.

\bibitem[West et~al.(2023)West, Erfani, Leckie, Sevior, Hollenberg, and
  Usman]{West_2023}
Maxwell~T. West, Sarah~M. Erfani, Christopher Leckie, Martin Sevior, Lloyd
  C.~L. Hollenberg, and Muhammad Usman.
\newblock Benchmarking adversarially robust quantum machine learning at scale.
\newblock \emph{Physical Review Research}, 5\penalty0 (2), June 2023.
\newblock ISSN 2643-1564.
\newblock \doi{10.1103/physrevresearch.5.023186}.
\newblock URL \url{http://dx.doi.org/10.1103/PhysRevResearch.5.023186}.

\bibitem[Benedetti et~al.(2019)Benedetti, Garcia-Pintos, Perdomo,
  Leyton-Ortega, Nam, and Perdomo-Ortiz]{Benedetti_2019}
Marcello Benedetti, Delfina Garcia-Pintos, Oscar Perdomo, Vicente
  Leyton-Ortega, Yunseong Nam, and Alejandro Perdomo-Ortiz.
\newblock A generative modeling approach for benchmarking and training shallow
  quantum circuits.
\newblock \emph{npj Quantum Information}, 5\penalty0 (1), May 2019.
\newblock ISSN 2056-6387.
\newblock \doi{10.1038/s41534-019-0157-8}.
\newblock URL \url{http://dx.doi.org/10.1038/s41534-019-0157-8}.

\bibitem[Peters and Schuld(2023)]{peters2022generalization}
Evan Peters and Maria Schuld.
\newblock Generalization despite overfitting in quantum machine learning
  models.
\newblock \emph{Quantum}, 7:\penalty0 1210, December 2023.
\newblock ISSN 2521-327X.
\newblock \doi{10.22331/q-2023-12-20-1210}.
\newblock URL \url{http://dx.doi.org/10.22331/q-2023-12-20-1210}.

\bibitem[Caro et~al.(2023)Caro, Huang, Ezzell, Gibbs, Sornborger, Cincio,
  Coles, and Holmes]{caro2022outofdistribution}
Matthias~C. Caro, Hsin-Yuan Huang, Nicholas Ezzell, Joe Gibbs, Andrew~T.
  Sornborger, Lukasz Cincio, Patrick~J. Coles, and Zoë Holmes.
\newblock Out-of-distribution generalization for learning quantum dynamics.
\newblock \emph{Nature Communications}, 14\penalty0 (1), July 2023.
\newblock ISSN 2041-1723.
\newblock \doi{10.1038/s41467-023-39381-w}.
\newblock URL \url{http://dx.doi.org/10.1038/s41467-023-39381-w}.

\bibitem[Caro et~al.(2022)Caro, Huang, Cerezo, Sharma, Sornborger, Cincio, and
  Coles]{Caro2022-ae}
Matthias~C Caro, Hsin-Yuan Huang, M~Cerezo, Kunal Sharma, Andrew Sornborger,
  Lukasz Cincio, and Patrick~J Coles.
\newblock Generalization in quantum machine learning from few training data.
\newblock \emph{Nature Communications}, 13\penalty0 (1):\penalty0 4919, aug
  2022.

\bibitem[Gao et~al.(2022)Gao, Anschuetz, Wang, Cirac, and Lukin]{Lukin2022}
Xun Gao, Eric~R. Anschuetz, Sheng-Tao Wang, J.~Ignacio Cirac, and Mikhail~D.
  Lukin.
\newblock Enhancing generative models via quantum correlations.
\newblock \emph{Phys. Rev. X}, 12:\penalty0 021037, May 2022.
\newblock \doi{10.1103/PhysRevX.12.021037}.

\bibitem[Bowles et~al.(2023)Bowles, Wright, Farkas, Killoran, and
  Schuld]{bowles2023contextuality}
Joseph Bowles, Victoria~J Wright, Máté Farkas, Nathan Killoran, and Maria
  Schuld.
\newblock Contextuality and inductive bias in quantum machine learning, 2023.

\bibitem[Zhu et~al.(2022{\natexlab{a}})Zhu, Johri, Bacon, Esencan, Kim, Muir,
  Murgai, Nguyen, Pisenti, Schouela, Sosnova, and
  Wright]{PhysRevResearch.4.043092}
Elton~Yechao Zhu, Sonika Johri, Dave Bacon, Mert Esencan, Jungsang Kim, Mark
  Muir, Nikhil Murgai, Jason Nguyen, Neal Pisenti, Adam Schouela, Ksenia
  Sosnova, and Ken Wright.
\newblock Generative quantum learning of joint probability distribution
  functions.
\newblock \emph{Phys. Rev. Res.}, 4:\penalty0 043092, Nov 2022{\natexlab{a}}.
\newblock \doi{10.1103/PhysRevResearch.4.043092}.

\bibitem[Rudolph et~al.(2020)Rudolph, Toussaint, Katabarwa, Johri, Peropadre,
  and Perdomo-Ortiz]{Rudolph20}
Manuel~S. Rudolph, Ntwali~Bashige Toussaint, Amara Katabarwa, Sonika Johri,
  Borja Peropadre, and Alejandro Perdomo-Ortiz.
\newblock Generation of high-resolution handwritten digits with an ion-trap
  quantum computer.
\newblock 2020.

\bibitem[Huang et~al.(2021)Huang, Du, Gong, Zhao, Wu, Wang, Li, Liang, Lin, Xu,
  Yang, Liu, Hsieh, Deng, Rong, Peng, Lu, Chen, Tao, Zhu, and
  Pan]{PhysRevApplied.16.024051}
He-Liang Huang, Yuxuan Du, Ming Gong, Youwei Zhao, Yulin Wu, Chaoyue Wang,
  Shaowei Li, Futian Liang, Jin Lin, Yu~Xu, Rui Yang, Tongliang Liu, Min-Hsiu
  Hsieh, Hui Deng, Hao Rong, Cheng-Zhi Peng, Chao-Yang Lu, Yu-Ao Chen, Dacheng
  Tao, Xiaobo Zhu, and Jian-Wei Pan.
\newblock Experimental quantum generative adversarial networks for image
  generation.
\newblock \emph{Phys. Rev. Appl.}, 16:\penalty0 024051, Aug 2021.
\newblock \doi{10.1103/PhysRevApplied.16.024051}.
\newblock URL \url{https://link.aps.org/doi/10.1103/PhysRevApplied.16.024051}.

\bibitem[Silver et~al.(2023)Silver, Patel, Cutler, Ranjan, Gandhi, and
  Tiwari]{silver2023mosaiq}
Daniel Silver, Tirthak Patel, William Cutler, Aditya Ranjan, Harshitta Gandhi,
  and Devesh Tiwari.
\newblock Mosaiq: Quantum generative adversarial networks for image generation
  on nisq computers, 2023.

\bibitem[Zhu et~al.(2022{\natexlab{b}})Zhu, Shen, Giani, Majumder, Neculaes,
  and Johri]{zhu2022copulabased}
Daiwei Zhu, Weiwei Shen, Annarita Giani, Saikat~Ray Majumder, Bogdan Neculaes,
  and Sonika Johri.
\newblock Copula-based risk aggregation with trapped ion quantum computers.
\newblock 2022{\natexlab{b}}.
\newblock URL \url{https://arxiv.org/abs/2206.11937}.

\bibitem[Johri et~al.(2021)Johri, Debnath, Mocherla, SINGK, Prakash, Kim, and
  Kerenidis]{Johri20}
Sonika Johri, Shantanu Debnath, Avinash Mocherla, Alexandros SINGK, Anupam
  Prakash, Jungsang Kim, and Iordanis Kerenidis.
\newblock Nearest centroid classification on a trapped ion quantum computer.
\newblock \emph{npj Quantum Information}, 7\penalty0 (1):\penalty0 122, Aug
  2021.
\newblock ISSN 2056-6387.
\newblock \doi{10.1038/s41534-021-00456-5}.

\bibitem[Cherrat et~al.(2022)Cherrat, Kerenidis, Mathur, Landman, Strahm, and
  Li]{cherrat2022quantum}
El~Amine Cherrat, Iordanis Kerenidis, Natansh Mathur, Jonas Landman, Martin
  Strahm, and Yun~Yvonna Li.
\newblock Quantum vision transformers.
\newblock 2022.
\newblock URL \url{https://arxiv.org/abs/2209.08167}.

\bibitem[Silver et~al.(2022)Silver, Patel, and
  Tiwari]{Silver_Patel_Tiwari_2022}
Daniel Silver, Tirthak Patel, and Devesh Tiwari.
\newblock Quilt: Effective multi-class classification on quantum computers
  using an ensemble of diverse quantum classifiers.
\newblock \emph{Proceedings of the AAAI Conference on Artificial Intelligence},
  36\penalty0 (8):\penalty0 8324--8332, Jun. 2022.
\newblock \doi{10.1609/aaai.v36i8.20807}.
\newblock URL \url{https://ojs.aaai.org/index.php/AAAI/article/view/20807}.

\bibitem[Pesah et~al.(2021)Pesah, Cerezo, Wang, Volkoff, Sornborger, and
  Coles]{PhysRevX.11.041011}
Arthur Pesah, M.~Cerezo, Samson Wang, Tyler Volkoff, Andrew~T. Sornborger, and
  Patrick~J. Coles.
\newblock Absence of barren plateaus in quantum convolutional neural networks.
\newblock \emph{Phys. Rev. X}, 11:\penalty0 041011, Oct 2021.
\newblock \doi{10.1103/PhysRevX.11.041011}.

\bibitem[Kerenidis et~al.(2022)Kerenidis, Landman, and
  Mathur]{kerenidis2022classical}
Iordanis Kerenidis, Jonas Landman, and Natansh Mathur.
\newblock Classical and quantum algorithms for orthogonal neural networks.
\newblock 2022.
\newblock URL \url{https://arxiv.org/abs/2106.07198}.

\bibitem[Schuld et~al.(2021)Schuld, Sweke, and Meyer]{Schuld_data_encoding}
Maria Schuld, Ryan Sweke, and Johannes~Jakob Meyer.
\newblock Effect of data encoding on the expressive power of variational
  quantum-machine-learning models.
\newblock \emph{Phys. Rev. A}, 103:\penalty0 032430, Mar 2021.
\newblock \doi{10.1103/PhysRevA.103.032430}.
\newblock URL \url{https://link.aps.org/doi/10.1103/PhysRevA.103.032430}.

\bibitem[Cong and Choi(2019)]{Cong2019}
Iris Cong and Mikhail~D. Choi, Soonwonand~Lukin.
\newblock Quantum convolutional neural networks.
\newblock \emph{Nature Physics}, 15\penalty0 (12):\penalty0 1273--1278, Dec
  2019.
\newblock ISSN 1745-2481.
\newblock \doi{10.1038/s41567-019-0648-8}.
\newblock URL \url{https://arxiv.org/abs/1810.03787}.

\bibitem[Hur et~al.(2022)Hur, Kim, and Park]{Hur_2022}
Tak Hur, Leeseok Kim, and Daniel~K. Park.
\newblock Quantum convolutional neural network for classical data
  classification.
\newblock \emph{Quantum Machine Intelligence}, 4\penalty0 (1), February 2022.
\newblock ISSN 2524-4914.
\newblock \doi{10.1007/s42484-021-00061-x}.
\newblock URL \url{http://dx.doi.org/10.1007/s42484-021-00061-x}.

\bibitem[dbm(2023)]{dbmnist784}
Mnist database 784.
\newblock
  \url{https://www.openml.org/search?type=data&sort=runs&id=554&status=active},
  2023.
\newblock Handwritten Digit Database.

\bibitem[skl(2023)]{sklearn}
Sci-kit sklearn.
\newblock \url{https://scikit-learn.org/stable/}, 2023.
\newblock Machine Learning Package in Python.

\end{thebibliography}

\end{document}